\documentclass[twocolumn]{aastex7}
\usepackage{gensymb}

\shorttitle{Cloud-Haze Interactions}
\begin{document}

\title{Wet Removal and Cloud Enhancement: The Microphysics of Cloud-Haze Interactions on Sub-Neptunes}

\author[orcid=0000-0001-5909-4433,sname='Nagpal']{Vighnesh Nagpal}
\altaffiliation{NSF Graduate Research Fellow}
\affiliation{Department of Astronomy \& Astrophysics, the University of Chicago}
\email[show]{vnagpal@uchicago.edu}  

\author[orcid=0000-0001-8342-1895, sname='Steinrueck']{Maria Steinrueck} 
\altaffiliation{51 Pegasi b Fellow}
\affiliation{Department of Astronomy \& Astrophysics, the University of Chicago}
\affiliation{Max Planck Institute for Astronomy, 69117 Heidelberg, Germany}
\email{msteinrueck@uchicago.edu}

\author[orcid=0000-0002-4250-0957, sname='Powell']{Diana Powell}
\affiliation{Department of Astronomy \& Astrophysics, the University of Chicago}
\email{diana.powell@uchicago.edu}

\author[orcid=0000-0002-8518-9601, sname='Gao']{Peter Gao}
\affiliation{Earth \& Planets Laboratory, Carnegie Institution for Science, 5241 Broad Branch Road NW, Washington, DC 20015, USA}
\email{pgao@carnegiescience.edu}

\author[orcid=0000-0002-8956-2047, sname='Samra']{Dominic Samra}
\affiliation{Department of Astronomy \& Astrophysics, the University of Chicago}
\email{pgao@carnegiescience.edu}

\author[orcid=0000-0002-8658-3811, sname='Cukier']{Wolf Cukier}
\affiliation{Department of Astronomy \& Astrophysics, the University of Chicago}
\email{wcukier@uchicago.edu}

% \author[orcid=0000-0001-5909-4433,sname='Nagpal']{Vighnesh Nagpal}
% \altaffiliation{NSF Graduate Research Fellow}
% \affiliation{Department of Astronomy \& Astrophysics, the University of Chicago}

% \author[orcid=0000-0001-5909-4433,sname='Steinrueck']{Maria Steinrueck}
% \affiliation{Department of Astronomy \& Astrophysics, the University of Chicago}

\begin{abstract}

Aerosols are a near-ubiquitous feature of sub-Neptune atmospheres, yet their microphysical nature remains poorly understood. Both condensate clouds and photochemical hazes have been proposed to explain observations, but have largely been studied in isolation. Here we present a new bin-scheme microphysical model, adapted from the Community Aerosol and Radiation Model for Atmospheres (\texttt{CARMA}), that explicitly couples cloud and haze formation through heterogeneous nucleation---the dominant mode of cloud formation on Earth and throughout the Solar System---in which haze particles act as cloud condensation nuclei (CCN). Applying this model to KCl clouds on GJ~1214~b-like warm sub-Neptunes, we find that the microphysical contact angle $\theta$ between cloud and haze particles governs distinct regimes of aerosol behavior: at moderate contact angles ($25\degr \lessapprox \theta \lessapprox 70\degr$), hazes are efficiently removed from the upper atmosphere through ``wet removal'' as they seed gravitationally-settling clouds; at small contact angles ($\theta \lessapprox 25\degr$), heterogeneous nucleation instead produces an enhanced population of mixed cloud-haze particles at high altitudes, dramatically increasing aerosol optical depth (``cloud enhancement''). These structural changes produce differences of up to four scale heights in transmission spectra, with strong effects at optical and near-infrared wavelengths relevant to instruments such as JWST NIRISS/SOSS, while wavelengths beyond $\sim$3~$\mu$m remain comparatively unaffected. We map these effects across orders of magnitude in metallicity, haze production rate, and vertical mixing strength, establishing their generality across sub-Neptune parameter space. Because heterogeneous nucleation is a universal phase-change process, this framework extends naturally to other types of exoplanet atmospheres and potentially any astrophysical environments where condensation onto foreign substrates may occur, including protoplanetary disks and stellar outflows.

\end{abstract}



\section{Introduction} 

Efforts to infer the atmospheric structure and composition of sub-Neptunes have often been hampered by the presence of aerosols that attenuate their spectral features in transmission. Though the nature of these aerosols remains unknown, condensate clouds and photochemical hazes have emerged as the leading hypotheses \citep[e.g.][]{brande_2024,teske_toi776c, wallack_toi836, roy_diversity}. GJ 1214 b stands out as perhaps the most definitive example of this phenomenon. Initially expected to be the most observationally favorable sub-Neptune for atmospheric characterization upon its discovery \citep{charbonneau_gj1214b}, a decade of transit observations from both the ground and space \citep[e.g.][]{fraine_spitzer, bean_2012_gj1214b} revealed no discernible absorption features in its transmission spectrum---precluding characterization of its atmospheric composition. The resulting extremely flat transmission spectrum has been interpreted as strong evidence for the presence of an optically thick aerosol layer in GJ 1214 b's atmosphere \citep{kreidberg_clouds}. 

In recent years, campaigns using JWST have begun to shed new light on the atmosphere of GJ 1214 b. The observation of a thermal emission phase curve by \citet{kempton_miri} revealed a strong indication of an extremely metal-enriched atmosphere ($>1000\times$ solar) and a surprisingly high Bond albedo $\sim$ 0.4 \citep[see also][]{Malsky_2025} suggesting a thick and reflective layer of aerosols. Moreover, the flatness of the infrared transmission spectrum observed using JWST/NIRSpec and JWST/MIRI \citep{gao23, schlawin_24_gj1214b, ohno_24_gj1214b} further strengthens the evidence for abundant aerosols. At the same time, the muted transmission spectra of many other warm sub-Neptunes \citep[e.g.][]{wallack_toi836c, brande_2024, ahrer_escaping_2025} hint at widespread aerosol obscuration.

Explaining these observational traits has motivated significant theoretical work on aerosols in the atmospheres of warm sub-Neptunes. For the specific case of GJ 1214 b, detailed one-dimensional microphysical models of clouds \citep{gao_benneke} found that invoking vigorous vertical mixing with an eddy diffusion coefficient $K_{zz} > 10^{10} \text{cm}^2 \text{s}^{-1}$ (together with metallicities $\geq$ 1000 $\times$ solar) is necessary to loft cloud particles to altitudes high enough to explain the observed transmission spectrum. However, such $K_{zz}$ values are a few orders of magnitude greater than those predicted by General Circulation Models (GCMs) of GJ 1214 b \citep{charnay_mixing, charnay_3dclouds}, casting doubt on the ability of clouds to serve as the lone source of feature attenuation \footnote{Although see \citet{ohno_aggregates} for how fractal aggregate clouds may affect this.}. 

Photochemical hazes, on the other hand, are expected to form at high altitudes due to the efficient photolysis of molecules in the upper atmosphere by incident high-energy radiation. More detailed modeling of this hypothesis, however, suggests that explaining the spectrum of GJ 1214 b with hazes alone requires invoking haze production rates $> 10^{-9} \text{g cm}^2 \text{s}^{-1}$ \citep{ohno_24_gj1214b}---a lower limit that is itself many orders of magnitude greater than predictions made for GJ 1214 b's atmosphere based on photolysis rate calculations \citep[e.g.][]{kawashima_2018, kawashima_ikoma_2019, lavvas_2019}. Furthermore, explaining the high measured Bond albedo with hazes is a challenge, as commonly considered hazes like soots and tholins are thought to be too absorbing \citep{Malsky_2025, kempton_miri, maria_sn_hazes}, although sulfur \citep[e.g.][]{gao_sulfur_haze} and diamond hazes \citep[][]{ohno_diamond} provide intriguing possibilities for reflective hazes. 

The struggles of models that consider clouds and hazes in isolation motivates our focus on scenarios in which clouds and hazes are not only both present, but also microphysically interact. Across the Solar System, cloud formation overwhelmingly proceeds through \textit{heterogeneous nucleation}, in which condensates form on pre-existing seed particles (cloud condensation nuclei, or CCN) before growing through condensation, coagulation, and related processes. Provided the existence of appropriate seeds, heterogeneous nucleation enables abundant cloud formation to proceed at low levels of supersaturation for which clouds would otherwise be unable to form. On Earth, for instance, water clouds nucleate on a broad spectrum of CCN including sea salt, industrial pollutants, volcanic ash, and other aerosols, meaning that cloud formation is tightly regulated by the CCN inventory \citep[e.g.][]{squires_1958,albrecht_report, williamson_npf_ccn, lin_ice_nucleation}. Titan's complex hydrocarbon aerosols (`tholin' hazes) similarly supply a major fraction of the moon's CCN population and strongly influence its albedo and cloud coverage \citep{huygens_condensation_lavvas,xinting_titan_hetnucleation}. Additional examples include possible photochemical-haze-seeded ice clouds on Uranus \citep{irwin_uranus}, sulfuric acid clouds forming on meteoritic dust on Venus \citep{venus_peter, karyu_haze_cosmic_dust}, and Saturnian clouds that may nucleate on ring-derived infalling particles \citep{hsu_ring_rain_saturn}. Together, these cases show that heterogeneous nucleation, and the coupling between condensate cloud formation and other aerosols, is a widespread feature of planetary atmospheres. Moreover, heterogeneous nucleation, as a general phase change process, may also play a key role in astrophysical environments as varied as protoplanetary disks \citep[e.g.][]{ros_nucleation_disks, powell_depletion_of_co}, stellar outflows \citep[e.g.][]{goumans_stardust_silicates}, and the interstellar medium \citep[e.g.][]{seki_hasegawa, bernatowicz_grains_within_grains, tielens_dust_review}. 

The broad prevalence of heterogeneous nucleation suggests that clouds on exoplanets are also likely to form through this pathway. Although predicting the CCN population in a given exoplanet atmosphere remains challenging, we consider the hypothesis that photochemical haze particles may act as CCN that catalyze cloud formation, motivated by studies indicating that such hazes are readily produced in sub-Neptune environments (e.g. \citealt{lavvas_2019, kawashima_ikoma_2019, horst_experiments, chao_he_particle_colors}). Interactions between clouds and hazes are therefore likely to govern the abundance and distribution of aerosols, with direct consequences for interpreting atmospheric observations \citep[e.g.][]{xinting_yu_2021,mang_waterclouds_meteoritic_dust, lavvas2024}. Thus, \textit{a detailed understanding of the impacts of cloud--haze interactions is likely of first-order importance in interpreting sub-Neptune atmospheric observations.} 

Motivated by the need to understand this phenomenon, we have developed a bin-scheme microphysical model of cloud formation on sub-Neptunes that includes interactions between clouds and photochemical hazes. Our model is adapted from the Community Aerosol and Radiation Model for Atmospheres (\texttt{CARMA}) \citep{turcoOneDimensionalModelDescribing1979, toonMultidimensionalModelAerosols1988}. In this work, we use our model to conduct a systematic study of clouds, hazes, and cloud-haze interactions in the atmosphere of GJ 1214 b-like warm sub-Neptunes, which allows us to study trends in aerosol distribution across axes of atmospheric metallicity, haze production rate, and finally, the cloud-haze interaction strength through the microphysical \textit{contact angle} between the two species. In Section \ref{sec:modelling_framework}, we describe our modeling framework. We then apply this model across our grid to compute full aerosol particle size distributions in Section \ref{particle_size_dist_sec}. Using these size distributions, we generate synthetic transmission spectra and study the trends they display in Section \ref{transmission_spectra_sec}. We assess the sensitivity of our results to the strength of vertical mixing in Section \ref{sec:mixing} before discussing our findings in the broader context of the field in Section \ref{sec:discussion} and summarizing our conclusions in Section \ref{sec:conclusions}.

\section{Modeling Framework} \label{sec:modelling_framework}

\begin{figure*}
    \centering
    \includegraphics[width=1.0\linewidth]{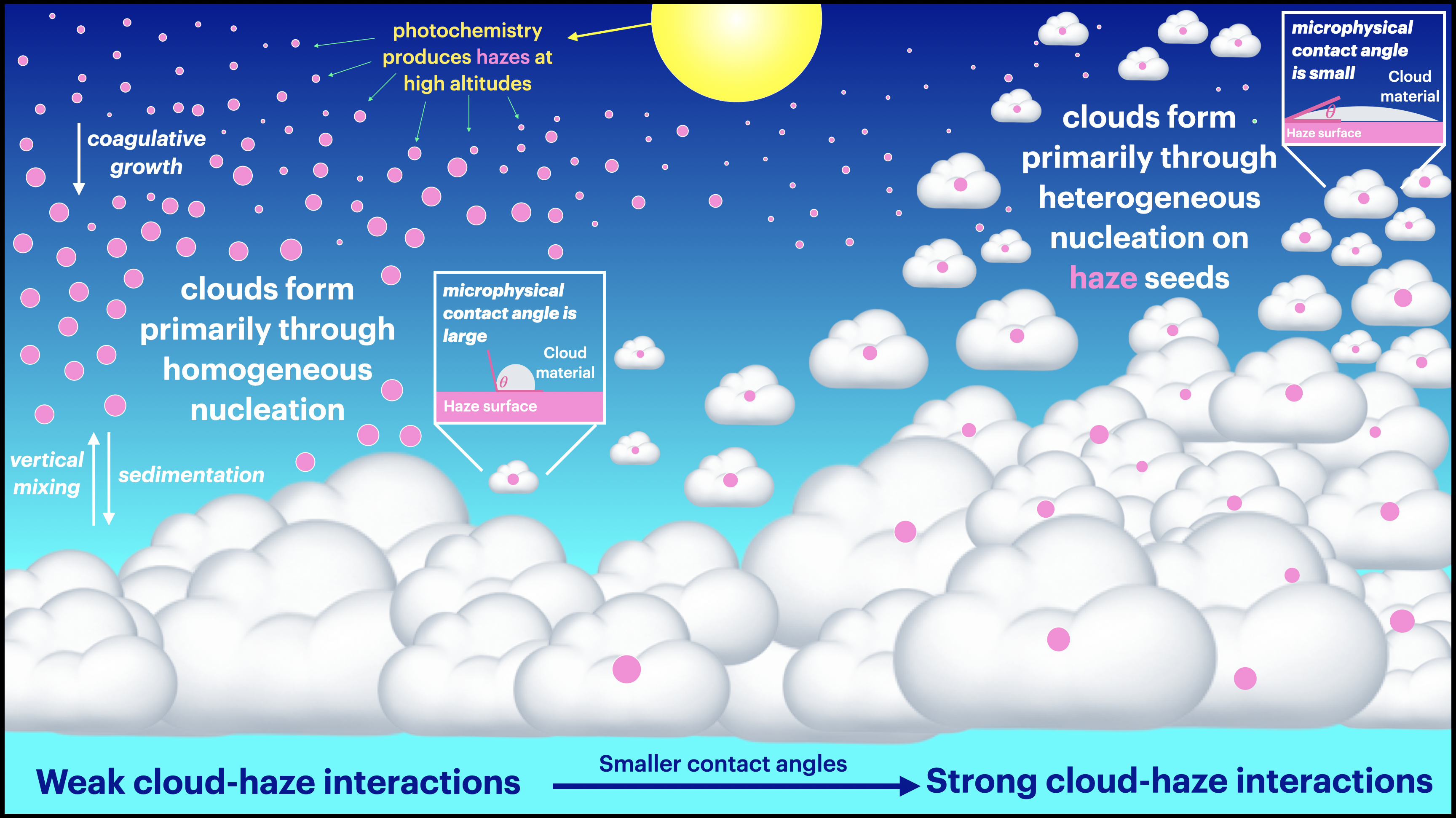}
    \caption{Clouds may form via heterogeneous nucleation on top of `seed' photochemical haze particles. Whether or not this pathway dominates over homogeneous nucleation---the process through which `pure' clouds spontaneously form---depends on the microphysical contact angle between the relevant cloud and haze species. Smaller contact angles correspond to stronger cloud-haze interactions due to the increased affinity between the clouds and hazes. While this contact angle is unknown in the absence of lab data, heterogeneous nucleation is the primary cloud-formation pathway across the Solar System.  }
    \label{fig:schematic}
\end{figure*}

The simulations presented in this work are built on \texttt{CARMA} \citep{turcoOneDimensionalModelDescribing1979, toonMultidimensionalModelAerosols1988}, a bin-scheme aerosol model that explicitly treats the microphysical processes governing cloud formation from first principles. Here we combine the 1D \texttt{ExoCARMA} implementations previously used for microphysical cloud modeling \citep[e.g.][]{diana_silicate, gao_benneke, powellTransitSignaturesInhomogeneous2019} and haze modeling \citep[e.g.][]{gao23} into a unified framework capable of jointly simulating clouds, hazes, and their mutual interactions\footnote{This is similar to what was done by \cite{mang_waterclouds_meteoritic_dust} to simulate water clouds in the atmospheres of cool brown dwarfs and giant planets.}. We apply this framework to study cloud-haze interactions on GJ 1214 b-like planets using models that include three aerosol populations: “pure’’ KCl clouds formed from homogeneous nucleation, photochemical hazes, and “mixed’’ KCl clouds that heterogeneously nucleate and grow on haze particles. Table \ref{planetary_parameters} shows the bulk planetary parameters used in our simulations.\footnote{We derive the planetary radius by averaging the semi-major axis-to-stellar radius ratio and the 3.6 and 4.5 $\mu$m planet-to-star radius ratios from \citet{cloutier_gj1214}, using the updated stellar radius from \citet{gao23}.
} A schematic overview of these aerosol types and cloud-haze interactions is shown in Figure~\ref{fig:schematic}. The remainder of this section describes the modeling details for the background atmosphere, the haze population, the pure KCl clouds, and the mixed cloud particles that form through cloud–haze interactions.

\subsection{Temperature-Pressure \& Vertical Mixing Profiles}

\begin{figure*}
    \centering
    \includegraphics[width=1.0\linewidth]{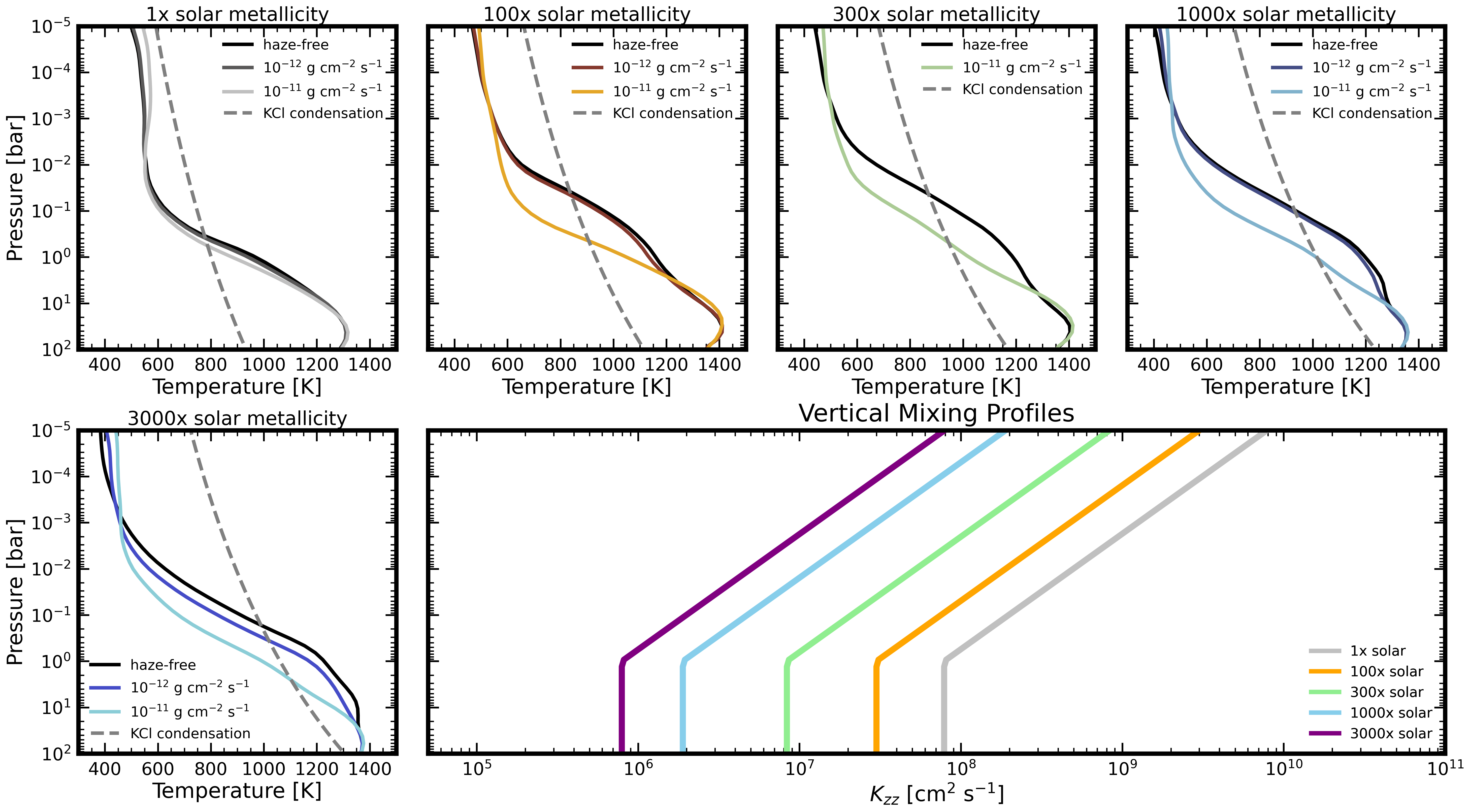}
    \caption{Pressure-Temperature (P-T) and vertical mixing ($K_{zz}$) profiles used for our \texttt{CARMA} simulations. \textit{Square Panels}: P-T profiles used for our simulations, with one panel for each metallicity. These were calculated from the grid of GJ 1214 b GCMs from \citet{maria_sn_hazes}, who also found that for haze production rates $\geq10^{-12} \text{g cm}^2 \text{s}^{-1}$ (colored lines) haze radiative feedback is strong enough to appreciably alter the planet's P-T profile relative to the haze-free expectations (black lines). For haze production rates $\leq 10^{-13} \text{g cm}^2\text{s}^{-1}$, our simulations use haze-free T-P profiles since hazes with such low abundances have a negligible impact on the overall temperature-pressure structure. For higher haze production rates, the radiative effects become important and we thus use T-P profiles with hazes included. On each panel, we also show the KCl condensation curve (grey dashed lines) for the corresponding metallicity. \textit{Bottom-right}: Fiducial vertical mixing ($K_{zz}$) profiles for each metallicity. The vast majority of this study uses these fiducial profiles, except Section \ref{sec:mixing}, in which we explore the effects of changing the mixing strength by factors of 10.}
    \label{fig:pt_structures}
\end{figure*}

To set up our simulations, we calculate 1D pressure-temperature (P-T) profiles from the outputs of the GJ 1214 b General Circulation Models (GCMs) which contain radiatively active hazes from \citet{maria_sn_hazes}. Though technically a grid for one planet, it spans a wide range of atmospheric compositions (from 1 $\times$ to 3000 $\times$ solar metallicity) and column haze production rates ($10^{-12}\text{ g cm}^2 \text{s}^{-1}$ to $10^{-10}\text{ g cm}^2 \text{s}^{-1}$, in addition to a haze-free case) with hazes that are radiatively active\footnote{We note that in addition to the GCMs published in Steinrueck et al. (2025), this study also includes pressure-temperature profiles from a few previously unpublished GCMs simulations with an similar setup, specifically for 1000x solar (hazefree, $10^{-12} \text{g cm}^2 \text{s}^{-1}$ and $10^{-11}\text{g cm}^2 \text{s}^{-1}$) and for solar metallicity ($10^{-11} \text{g cm}^2 \text{s}^{-1}$ with tholin hazes).}. To calculate the input 1D P-T profiles for our \texttt{CARMA} models, we average the GCM output between $\pm 20 \degree$ of the equator and across all longitudes. We show a sub-selection of these for different metallicities, haze types, and haze production rates in Figure \ref{fig:pt_structures}.

\begin{deluxetable}{ll}
\tablecaption{Planetary parameters adopted for this study. These are taken from \citet{cloutier_gj1214} and \citet{gao23}. }
\label{planetary_parameters}
\tablehead{
\colhead{Parameter} & \colhead{Values} 
}
\startdata
Mass & 8.17 $M_{\oplus}$ \\
Radius & 2.628  $R_{\oplus}$ \\ % common value
Orbital Period & 1.5804 days \\ % well measured
Semi-major Axis & 0.01429 AU \\ % from exoplanet.eu
\enddata
\end{deluxetable}

Vertical mixing critically impacts aerosol distributions. Though mixing is a fundamentally three-dimensional process, it can be parameterized in 1D as a diffusional process through the eddy diffusion coefficient $K_{zz}$. Using passive tracers to emulate the three-dimensional transport of clouds, \citet{parmentier2013} demonstrated that scaling laws based on the root-mean-square vertical velocity substantially misrepresent the true mixing. Instead, their results indicated that a power-law profile—fit directly to the tracer distributions in their general circulation model—provides a more accurate representation. Subsequent studies by \citet{charnay_mixing} and \citet{steinrueck_2021} corroborated this conclusion for sub-Neptune atmospheres and haze-like tracers, respectively.

Motivated by these findings, we adopt a power-law vertical mixing profile of the form
\begin{equation} \label{Kzz_equation}
    K_{zz} = K_{zz,0} \left(\frac{P}{\text{1 bar}}\right)^{-0.4} 
\end{equation} as in \cite{gao23}. The exponent of $-0.4$ is taken from \citet{charnay_mixing}. The normalization $K_{zz,0}$ varies across simulations and is scaled relative to the haze-free, 100 $\times$ solar metallicity model according to
\begin{equation}
    K_{zz,0} = 10^7 \left( \frac{H}{H_{100}} \right)^2 \text{cm}^2\text{s}^{-1},
\end{equation} where $H$ and $H_{100}$ denote the local pressure scale heights at 1 bar for the simulation of interest and the 100$\times$ solar haze-free reference case, respectively. This normalization is designed to hold the mixing timescale ($\sim H^2/K_{zz}$) constant and thus ensure comparable vertical mixing between simulations of different metallicities, despite their substantially different physical scale heights. For pressures greater than 1 bar, we hold $K_{zz}$ fixed at $K_{zz,0}$.

\begin{deluxetable}{ll}
\label{model_parameters}
\tablecaption{Model Parameter Space\label{tab:parameter_space}}
\tablehead{
\colhead{Parameter} & \colhead{Values}
}
\startdata
Metallicity [$\times$ solar] &
1, 100, 300, 1000, 3000 \\
Haze Production Rate [$\mathrm{g\ cm^{-2}\ s^{-1}}$] &
$0$, $10^{-19}$, $10^{-18}$, $10^{-17}$, $10^{-16}$, $10^{-15}$, $10^{-14}$, $10^{-13}$, $10^{-12}$, $10^{-11}$ \\
Contact Angle [degrees] &
0.3, 5.1, 9.9, 25.8, 45.6, 60.0, 72.5, 84.3, 154.2, 179.7 \\
$K_{zz}$ [$\times$ fiducial] & 0.1$\times$, 1$\times$, 10$\times$ \\\enddata
\end{deluxetable}

\subsection{Modeling haze formation} \label{haze_modelling}

The haze modeling in this work mostly follows the methodology of \citet{gao23} and we thus only briefly recap the key processes here, referring the interested reader to \citet{gao23} (and references therein). To start, we simulate haze production as a downward flux of spherical particles from the top of the atmosphere. Motivated by previous photolysis-based predictions for the haze production rate on sub-Neptunes like GJ 1214 b \citep[e.g.][]{lavvas_2019, kawashima_ikoma_2019, ohno_24_gj1214b, lavvas2024}, as well as the considerable uncertainties in the physics of haze production, we test haze production rates ranging from $10^{-19}$ to $10^{-11} \text{g cm}^2 \text{s}^{-1}$ in 1 dex increments.  \citet{maria_sn_hazes} showed that haze radiative effects only start altering the global temperature profile at a production rate of $10^{-12} \text{ g cm}^{-2} \text{ s}^{-1}$, and as such we use T-P profiles calculated from haze-free GCM runs for haze production rates $\leq$ $10^{-13}$ g cm$^{-2}$ s$^{-1}$.

The hazes form at the top of the atmosphere and are allowed to vertically mix, coagulate, and settle throughout the atmosphere. For growth through coagulation, we assume a sticking efficiency of unity \citep[as in][]{gao23}, implying perfectly efficient coagulation \citep[although lower values are possible if the hazes are charged, ][]{lavvas_huygens}. Particles generally grow to larger sizes as they settle deeper into the atmosphere. A major process controlling the haze particle size distribution is sedimentation, which is intimately related to the dynamic viscosity of the atmosphere\footnote{As one example, Stokes' fall velocity (which describes the sedimentation velocity for low Reynolds numbers $Re<1$) depends inversely on the dynamic viscosity. We refer the interested reader to \citet{pruppacher_klett} for a more in-depth discussion.}. For atmospheres that are not dominated by a single species, a `mixed' dynamic viscosity profile can be calculated from the viscosities of the individual species constituting the atmosphere. As in \citet{gao23}, we follow the method of \citet{davidson_1993} to compute the mixed viscosity $\eta_{mix}$ which is defined as 

\begin{equation}
    \eta_{mix} = \left( \Sigma_{i,j} \frac{y_i y_j}{\sqrt{\eta_i \eta_j}} E_{i,j}^{3/8} \right) ^{-1}
\end{equation} where $E_{i,j}$ is the efficiency of momentum transfer between species $i$ and $j$. $\eta_i$ is the dynamic viscosity of species $i$, and $y_i$ is its momentum fraction which is given by
\begin{equation}
    y_i = \frac{x_i \sqrt{M_i}}{\Sigma_i x_i \sqrt{M_i}}
\end{equation} where $x_i$ and $M_i$ are the mixing ratio and molecular mass of species $i$ respectively. Thus, in addition to the effect that atmospheric metallicity has on the overall mean molecular weight and P-T profile of the atmosphere, it also has a significant impact on the dynamic viscosity of the atmosphere, which plays a key role in regulating haze mixing and growth. Increasing the atmospheric metallicity typically increases the dynamic viscosity, with the exact structure of the viscosity profile being sensitive to changes in the overall T-P profile and mixing ratios of the atmosphere's constituents. This increase in dynamic viscosity in turn promotes haze growth by lowering the sedimentation velocity. We refer the interested reader to \citet{gao23} (and references contained therein) for a more in-depth discussion of the haze modeling. 

We use the optical properties of Titan-like tholins \citep{Khare_tholins} for this paper. Although our \texttt{CARMA} simulations do not account for the radiative feedback of the aerosols that form, the input T-P structures of \citet{maria_sn_hazes} used to initialize our runs included the effects of haze radiative feedback. We note that the GCMs of \citet{maria_sn_hazes} parameterized aerosol radiative feedback using a single representative haze distribution (100$\times$ solar metallicity with a haze production rate of $10^{-12}$ g cm$^{-2}$ s$^{-1}$). In contrast, our \texttt{CARMA} model grid spans a wide range of atmospheric metallicities, haze production rates, $K_{zz}$ values, and cloud--haze interaction strengths, yielding aerosol populations whose radiative effects would generally differ from those assumed in the GCMs. We therefore adopt the resulting P--T profiles as fixed background atmospheres (with respect to contact angle and vertical mixing) and leave a self-consistent treatment of aerosol radiative feedback (including feedback due to clouds and hazes) to future work.

\subsection{Microphysical Clouds with \texttt{CARMA}}

Though the nature of cloud-forming species in sub-Neptune atmospheres remain uncertain, owing to the lack of features detected from the clouds themselves, previous theoretical modeling has suggested that KCl, ZnS, and Na$_2$S may condense to form clouds at the temperatures and pressures characteristic of GJ 1214 b \citep[e.g. ][]{lodders_alkali, eliza_gj1214_2012}. In this work, we focus on KCl as the singular cloud forming species in our simulations. This is because the analysis by \cite{gao_benneke} shows that clouds of pure ZnS are extremely unlikely to form through homogeneous nucleation, owing to the high surface energy of ZnS\footnote{Though ZnS could theoretically form through heterogeneous nucleation on hazes if the contact angle is small, estimates by \cite{xinting_yu_2021} indicate that large values ($\sim85-88 \degree$) are likely, making this scenario unlikely. }. Clouds of Na$_2$S, if they form, are expected at high temperatures \& pressures deep in the atmosphere. Therefore, Na$_2$S clouds are unlikely to ascend high enough to affect the transmission spectrum of GJ 1214 b. For these reasons, as well as simplicity for the modelling of cloud-haze interactions, we opt to only model KCl clouds in this study.

Cloud formation depends on key microphysical processes such as nucleation, condensation, evaporation, and coagulation. The first step towards forming a cloud is \textit{nucleation}, which refers to the initial transition of a species from its gaseous to solid or liquid phases. Within the framework of classical nucleation theory, cloud formation can proceed via two main pathways: homogeneous and heterogeneous nucleation. In homogeneous nucleation, the cloud-forming phase transition occurs independently within the parent gaseous reservoir. This lies in contrast to heterogeneous nucleation, in which the presence of `seeds' known as cloud condensation nuclei (CCN)  facilitate the formation of cloud particles on their surfaces (see Section \ref{heterogeneous_cloud_modelling}). Previous work \citep[e.g.][]{gao_benneke, diana2D} indicates that pure KCl clouds can form homogeneously, in which case the nucleation rate is given by:

\begin{equation} \label{homogeneous_nucleation}
    J_{hom} = 4\pi a_{c}^2  \Phi Z n  \exp(-F/kT), 
\end{equation} where $n$ is the number density of condensate vapor molecules, T is the temperature, and $a_c$ is the critical radius, which is given by
\begin{equation}
    a_c = \frac{2M \sigma_s}{\rho_p RT \ln S}
\end{equation} where $M$, $\sigma_s$, $\rho_p$, $R$, and $S$ are respectively the molar mass, surface energy, mass density, universal gas constant, and saturation ratio\footnote{The saturation ratio $S$ is defined as the ratio of the partial and saturation pressures. } of the condensible species in question. We use the same values for these quantities as \citealt{gao_2020}. The critical radius marks the size at which a cloud particle can be considered thermodynamically stable. The free energy of formation, $F$, can be related to the critical radius through 
\begin{equation} \label{free_energy}
    F = \frac{4}{3}\pi \sigma_s a_c^{2}
\end{equation} and $\Phi$, the rate of vapor particle diffusion to the forming particle is given by 
\begin{equation} \label{phi}
    \Phi = \frac{p}{\sqrt{2\pi mkT}}
\end{equation} where $p$ is the partial pressure\footnote{As opposed to the atmospheric pressure, as mistakenly stated by \citealt{gao_marley_ackerman}.} of the condensate vapor, $m$ is the mass of the vapor molecule, and $k$ is the Boltzmann constant. Finally, the Zeldovich factor $Z$ accounts for non-equilibrium effects (such as the evaporation of newly formed particles) and is defined as 
\begin{equation} \label{zeldovich}
    Z = \sqrt{\frac{F}{3\pi kT g_m^2}}
\end{equation} where $g_m$ is the number of molecules contained within particles of the critical radius $a_c$. 

To determine the amount of KCl vapor in our simulations, we calculate the equilibrium chemistry prediction for the KCl mixing ratio at the expected cloud base for each simulation. This prediction in turn requires knowing the location of the cloud base, which can be determined by finding the intersection of the KCl condensation curve with the T-P structure of the atmosphere. We calculate the condensation curve by computing the KCl mixing ratio profile throughout the atmosphere using \texttt{FASTCHEM}\footnote{https://github.com/NewStrangeWorlds/FastChem} \citep{fastchem_2018,fastchem2, fastchem_cond} and comparing it to the saturation vapor mixing ratio profile on the same pressure grid. The location where these two profiles intersect corresponds to the predicted cloud base and we set (as a boundary condition) the below cloud abundance of KCl in the \texttt{CARMA} simulations to the mixing ratio at this location. Notably, this approach differs from many previous studies \citep[e.g.,][]{morley_neglected_clouds,morley_gj1214b, gao_benneke}, which assumed that the KCl abundance simply tracks the elemental K abundance rather than calculating the equilibrium KCl abundance directly.

% \begin{equation} \label{KCl_cond_equation}
%     \frac{10^4}{T_{cond}(\text{KCl})} \approx 12.479 - 0.879 \log p_t - 0.879 \text{[Fe/H]},
% \end{equation}where $p_t$ is the total atmospheric pressure and [Fe/H] is the atmospheric metallicity. The cloud-base pressure is defined as the level where this condensation temperature equals the local atmospheric temperature.
% We then use  to compute the equilibrium KCl mixing ratio at this pressure level for each model in our grid. This value is imposed as a fixed lower boundary condition for the KCl vapor in the \texttt{CARMA} simulations. 

\subsection{Cloud-haze interactions: the role of contact angles in heterogeneous nucleation} \label{heterogeneous_cloud_modelling}

\begin{figure}
    \centering
    \includegraphics[width=1.0\linewidth]{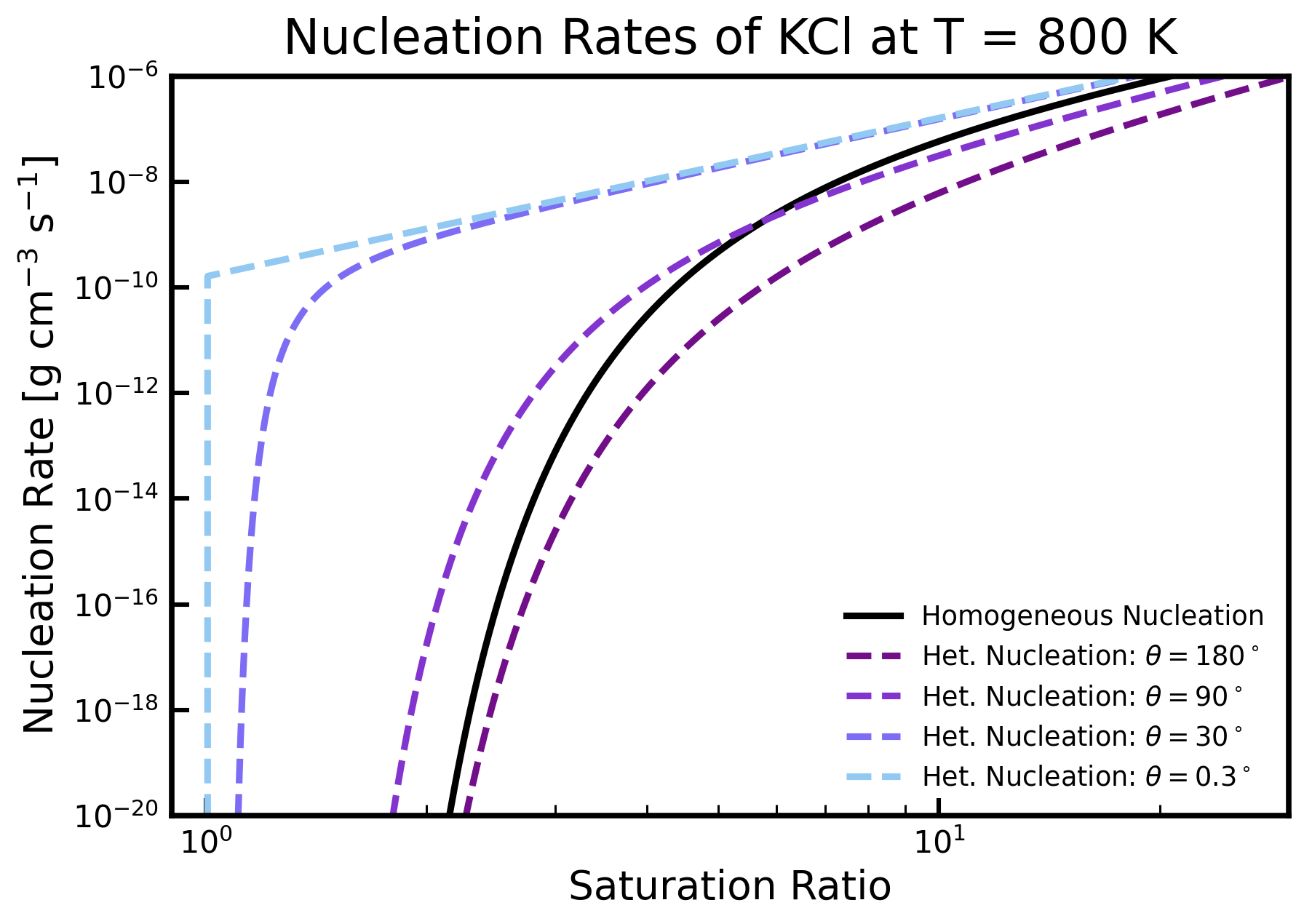}
    \caption{Homogeneous (black line) and heterogeneous (colored dashed lines) nucleation rates for KCl at T = 800 K, the approximate temperature at the predicted cloud-base of GJ 1214 b. The heterogeneous nucleation rates were calculated assuming $r_{\text{CCN}} = 0.2$ $\mu$m (a common haze CCN size in our simulations; see Figure \ref{fig:one_dimensional_psize}) and $F_{des} = 0.5$ eV. Decreasing contact angles $\theta$ are shown using progressively lighter colors. We immediately observe that at low saturation ratios, the heterogeneous nucleation rates for $\theta < 30\degree$ are many orders of magnitude higher than the homogeneous nucleation rate, demonstrating how heterogeneous nucleation is able to drive substantially enhanced cloud formation.}
    \label{fig:nucleation_rate}
\end{figure}

The explicit modeling of cloud-haze interactions through heterogeneous nucleation is the major new addition made by this work. We model this process by varying the microphysical contact angle $\theta$. Analogous to the way in which water droplets on a hydrophilic surface exhibit smaller contact angles with the surface, strong microphysical interactions between two substances can be thought of as having a low equivalent contact angle, and vice-versa (Figure \ref{fig:schematic}). Within the framework of classical nucleation theory, the heterogeneous nucleation rate of a condensate on a CCN can be written as \citep[following][]{pruppacher_klett}:

\begin{equation} \label{heterogeneous_nucleation}
    J_{het} = 4\pi ^2 r_{CN}^2 a_{c}^2 \Phi c_{surf} Z \exp(-Ff/kT), 
    \label{hetnuc}
\end{equation} 
where $r_{CN}$ is the radius of the CCN, $c_{surf}$ is the number density of condensate molecules on the nucleating surface and $f$ is the \textit{shape factor}. $c_{surf}$ in turn also depends on the desorption energy $F_{des}$\footnote{Small desorption energies $\sim$ 0.1 eV are typically associated with weak van der Waals interactions whereas larger energies $\sim 1$ eV are often the result of chemical bond formation between the substrate and adsorbed molecules. We show nucleation rates for an intermediate case with $F_{des} = 0.5$eV in Figure \ref{fig:nucleation_rate}. } and oscillation frequency $\nu$ (both are which are generally uncertain) as
\begin{equation}
    c_{surf} = \frac{\Phi}{\nu} \exp(F_{des}/kT), 
\end{equation}

The contact angle $\theta$ influences the nucleation rate through $f$, which depends on $\mu \equiv \cos(\theta)$ as

\begin{equation} \label{shape_factor}
    f = \frac{1+\left( \frac{1-\mu x}{\phi}\right)^3 + x^3 (2-3f_0+f_0^{3})+3\mu 
    x^2 (f_0-1)}{2}, 
    \label{fparameter}
\end{equation}
where $x={r_{\text{CN}}}/{a_c}$,  $\phi = \sqrt{1-2\mu x + x^2}$, and $f_0 = (x-\mu)/a_c$. Close inspection of Equations \ref{hetnuc} \& \ref{fparameter} shows that the heterogeneous nucleation rate increases as the contact angle decreases, albeit by a magnitude modulated by $x$, the ratio between the radius of the CCN and the critical radius. Figure \ref{fig:nucleation_rate} illustrates how the heterogeneous nucleation rate can substantially exceed the homogeneous nucleation rate at various saturation ratios for a few different contact angles. A consequence of this is that heterogeneous nucleation can drive substantial cloud formation at saturation ratios significantly lower than those required for appreciable homogeneous nucleation to take place. Indeed, the saturation ratio of water vapor in Earth's atmosphere is expected to be too low to produce significant amounts of water clouds through homogeneous nucleation, and the fact that we see substantial cloud formation points towards heterogeneous nucleation being the dominant pathway for cloud formation \citep[e.g.][]{pruppacher_klett}. 

The heterogeneous nucleation rate also depends on the CCN radius through the Kelvin effect. The Kelvin effect relates the saturation vapor pressure over a curved surface $p_{sat,c}$ with radius $r$ to that over a flat surface $p_{sat}$ as follows:
\begin{equation} \label{kelvin_effect}
    \ln\frac{p_{sat, c}}{p_{sat}} = \frac{2\sigma_s V_m}{rRT}, 
\end{equation} where $V_m$ is the molar volume, $T$ is the local temperature, and $\sigma_s$ is once again the surface energy \citep{kelvin_effect_1872}. Smaller particle sizes increase the saturation vapor pressure, making them less efficient CCNs. As a result, we expect that larger haze particles will preferentially serve as CCNs. Similarly, larger surface energies $\sigma_s$ also increase the saturation vapor pressure. The microphysics of the Kelvin Effect therefore helps sculpt the size distributions of the haze particles that serve as CCN, as well as the clouds that form upon them.

Though the cloud--haze contact angle strongly influences heterogeneous nucleation rates, as discussed above, physically realistic values of $\theta$ for exoplanet atmospheres remain poorly constrained. This is because the contact angle is itself set by the balance between the condensate surface energy, the haze surface energy, and the condensate--haze interfacial energy. Their relationship is described by Young's equation,
\begin{equation}
\cos\theta = \frac{\sigma_s - \sigma_{sl}}{\sigma_l},
\end{equation}
where $\sigma_s$, $\sigma_l$, and $\sigma_{sl}$ are respectively the surface energies of the solid haze substrate, condensate material, and their mutual interface. Estimating realistic values of $\theta$ therefore requires knowledge of these material properties, which are generally uncertain for exoplanetary aerosols, and especially so for hazes. Laboratory experiments targeted at measuring these quantities are thus crucial for improving our understanding of cloud-haze interactions. 

Recent experimental studies \citep[e.g.][]{yu_tholin_surface_energy, xinting_yu_2021} have begun to make progress in this area through determining haze surface energies using the Owens--Wendt--Rabel--Kaelble (OWRK) two-liquid method \citep{owens_orwk}, which infers the surface energy of a solid by measuring the contact angles formed by two probe liquids of known surface-energy components. In the OWRK formulation, the total work of adhesion between a solid ($s$) and liquid ($l$) is
\begin{equation}
W_{s,l} = \sigma_l^{\mathrm{tot}}(1 + \cos\theta),
\end{equation}
where $\sigma_l^{\mathrm{tot}}$ is the total surface energy of the liquid and $\theta$ is the measured contact angle. The surface energies of both the liquid and solid are decomposed into dispersive ($d$) and polar ($p$) components, such that
\begin{equation}
\sigma_l^{\mathrm{tot}}(1 + \cos\theta)
=
2\left(
\sqrt{\sigma_s^{d}\sigma_l^{d}}
+
\sqrt{\sigma_s^{p}\sigma_l^{p}}
\right).
\end{equation}
By measuring $\theta$ for two liquids with known $(\sigma_l^d,\sigma_l^p)$, one can solve for the corresponding solid components $(\sigma_s^d,\sigma_s^p)$ and hence the total solid surface energy $\sigma_s$.

Using this framework, \citet{xinting_yu_2021} measured surface energies of laboratory-produced haze analogs and estimated contact angles between KCl condensates and hazes of several compositions. Their inferred values span $\theta \sim 0^\circ$--$77^\circ$, indicating that cloud--haze interactions may range from highly efficient to only weakly enhanced depending on composition (see Section \ref{realistic_contact_angle_section}). Motivated by this experimentally inferred but broad range, we treat $\theta$ as a free parameter and explore its impact on heterogeneous nucleation across a wider domain.\footnote{Although \citet{xinting_yu_2021} reported an upper limit of $77^\circ$ for KCl, we additionally consider larger contact angles up to $179.7^\circ$ in order to map the transition from strongly coupled cloud--haze interactions to the effectively homogeneous nucleation limit.}

\section{Particle size distributions and cloud properties} \label{particle_size_dist_sec}

Our \texttt{CARMA} simulations allow us to predict aerosol particle size distributions as a function of pressure level. These size distributions, in conjunction with the material properties of the aerosols, determine how exoplanet aerosols interact with light and are thus critical for predicting and interpreting atmospheric observations. As such, we devote this section to an in-depth discussion of how cloud-haze interactions (through changes in the microphysical contact angle) affect aerosol particle size distributions (Section \ref{psize_subsec}) across atmospheric metallicity (Section \ref{metallicity_psize_subsubsection}), and haze production rate (Section \ref{hprod_psize_subsubsection}). We then also consider how the cloud mass and optical depth surfaces are affected in Sections \ref{cloud_mass_sec} and \ref{opd_surfaces_section}

\subsection{Particle Size Distributions} \label{psize_subsec}

\begin{figure*}
    \centering
    \includegraphics[width=1.0\linewidth]{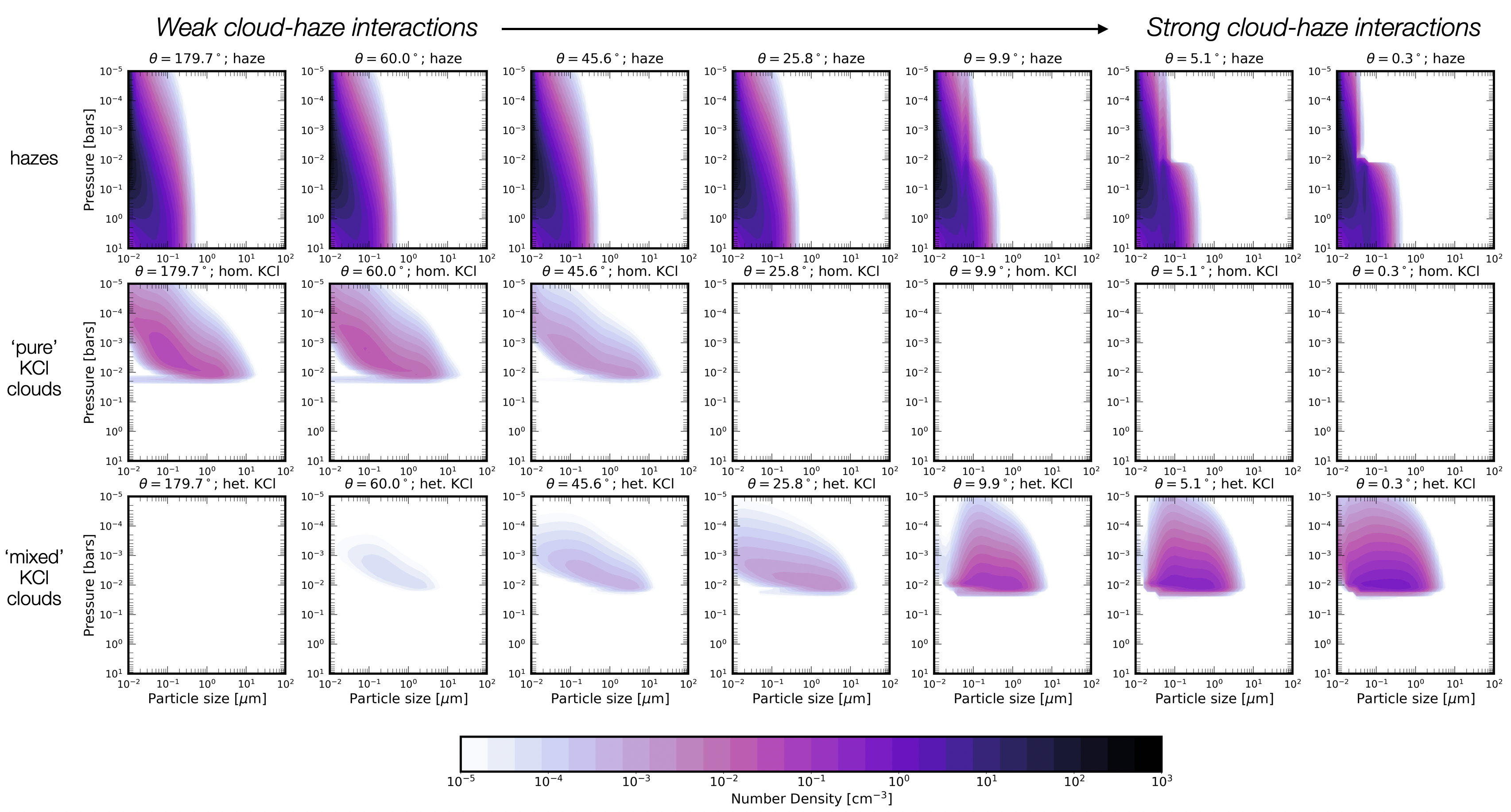}
    \caption{Variation in aerosol particle size distributions across contact angles for our fiducial 100 $\times$ solar case with a $10^{-14}$ g cm$^{-2}$ s$^{-1}$ haze production rate. From top to bottom, the rows show hazes, pure KCl clouds, and `mixed' KCL clouds nucleated upon hazes. The different columns show cases with microphysical contact angle $\theta$ decreasing from $179.7 \degree$ to $0.3 \degree$, corresponding to increasing cloud-haze interaction strength. From left-to-right, the transition from a cloud distribution consisting mainly of pure KCl clouds to a new cloud distribution dominantly composed of mixed clouds is clearly seen, alongside qualitatively different size distributions that result. This effect of cloud nucleation can also be observed in the top row, where the haze distribution is truncated at larger particle sizes starting at the location of the cloud base, which is caused by these haze particles serving as cloud condensation nuclei for the mixed cloud population. }
    \label{fig:mega_grid}
\end{figure*}

\begin{figure*}
    \centering
    \includegraphics[width=1.0\linewidth]{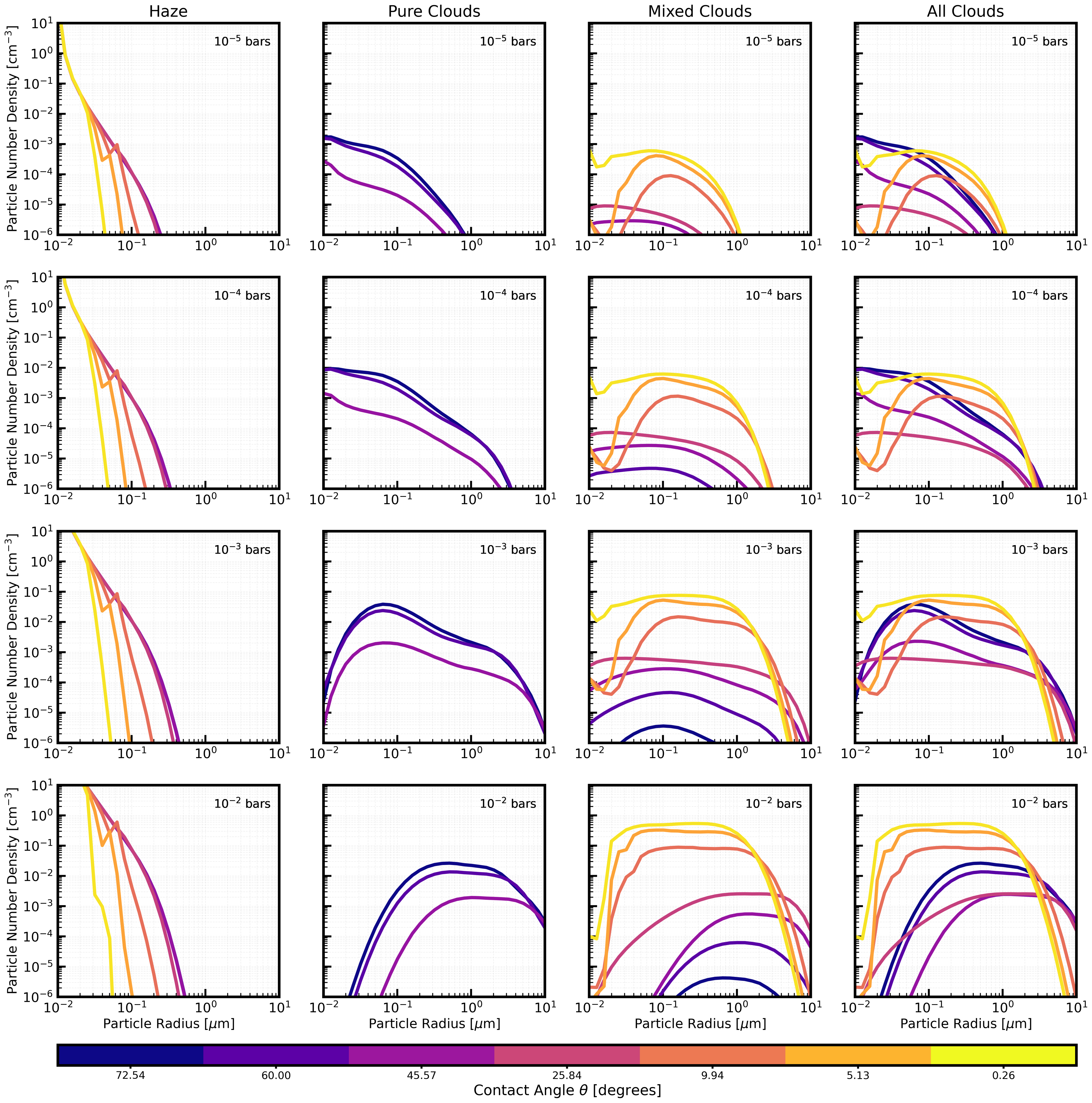}
    \caption{One-dimensional particle size distributions for our fiducial 100 $\times$ solar metallicity, $10^{-14}$ g cm$^{-2}$ s$^{-1}$ haze production rate simulations. From left to right, the columns show the haze, pure cloud, mixed cloud, and total cloud populations. Rows from top to bottom show these distributions at four pressure levels: $10^{-5}$, $10^{-4}$, $10^{-3}$, and $10^{-2}$ bars. }
    \label{fig:one_dimensional_psize}
\end{figure*}

We start by examining how the particle size distribution of each aerosol component changes alongside the microphysical contact angle for a fiducial case with an atmospheric metallicity of 100 $\times$ solar and a low-moderate $10^{-14}$ g cm$^2$ s $^{-1}$ haze production rate (Figure \ref{fig:mega_grid}). To orient this discussion, we again note that decreasing contact angles increases the cloud-haze interaction and the heterogeneous nucleation rate. First, we observe that decreasing the contact angle from $60.0 \degree$ to lower values leads to a steady depletion of pure KCl clouds and a corresponding growth in the mixed cloud population as KCl vapor increasingly condenses upon haze CCN through heterogeneous nucleation. This results in the `wet removal' (or deposition) of haze particles from the upper-atmosphere. Due to the Kelvin effect, clouds preferentially form upon larger haze particles rather than the smaller ones that are much more numerous. Stronger cloud-haze interactions (smaller contact angles) lower the saturation vapor pressure of KCl with respect to the haze, enabling progressively smaller haze particles to serve as CCN. The interplay between the Kelvin Effect and contact angle causes the maximum haze particle size (which is roughly equivalent to the minimum CCN size) to decrease alongside contact angle (Figure \ref{fig:one_dimensional_psize}).

The transformation from homogeneous to heterogeneous nucleation dominated cloud formation is a general feature of cloud-haze interactions when the contact angle is decreased. For our fiducial simulation, the transition between these two regimes occurs gradually, with mixed clouds becoming dominant for $\theta < 25.8 \degree$. The location of this cross-over is sensitive to the haze production rate, metallicity and $K_{zz}$ profiles. Intriguingly, however, the overall cloud number density (ie. the sum of the homogeneous and heterogeneously nucleated population) changes non-monotonically with contact angle. 

Moving from the large contact angles at which cloud-haze interactions are negligible to the moderate contact angles ($72.54\degree > \theta > 25.8\degree$) at which they become significant, further decreases in contact angle substantially reduce the total cloud particle number density across much of the upper atmosphere and a wide range of particle sizes (Figure \ref{fig:one_dimensional_psize}; right column). This reduction in cloud abundance can be understood as follows. In this regime, heterogeneous nucleation enables mixed clouds to form, but only on large haze CCN where the Kelvin effect is overcome. These mixed clouds then undergo condensational growth, leading to sedimentation and evaporation at depth. Simultaneously, the resulting decrease in gas saturation ratio suppresses homogeneous nucleation (Figure \ref{fig:nucleation_rate}). Together, these two mechanisms facilitate the wet removal of preferentially large haze particles from the upper atmosphere --- analogous to `nucleation scavenging' on Earth \citep[e.g.][]{jenson_charlson_nucleation_scavenging, svenningson_nucleation_scavenging} --- while also inhibiting cloud formation via homogeneous nucleation. \textit{Cloud-haze interactions in the moderate-interaction regime therefore produce clearer atmospheres.}

However, strong cloud-haze interactions qualitatively alter this picture, driving substantially enhanced cloud formation at high altitudes. For contact angles $\theta < 25.8\degree$, much smaller haze particles can become CCN (Figure \ref{fig:one_dimensional_psize}; left column). Since small haze particles are far more numerous at high altitudes, KCl gas can now nucleate on an abundant population of seeds in the upper atmosphere, producing a large number of new mixed cloud particles. Figures \ref{fig:mega_grid} and \ref{fig:one_dimensional_psize} illustrate this effect, and also show the accompanying progressive depletion of haze as a larger fraction of its size distribution serves as CCN. The mixed cloud population produced in this way is enhanced in $0.1$--$1\mu$m particles relative to the homogeneously nucleated cloud population expected in the absence of cloud-haze interactions (Figure \ref{fig:one_dimensional_psize}; right column). \textit{Strong cloud-haze interactions therefore substantially enhance the population of high-altitude cloud particles}, with important implications for the total cloud mass and transmission spectra, which we discuss further in Sections \ref{cloud_mass_sec} and \ref{transmission_spectra_sec}.

\subsubsection{Variations with Metallicity} \label{metallicity_psize_subsubsection}

\begin{figure}
    \centering
    \includegraphics[width=1.0\linewidth]{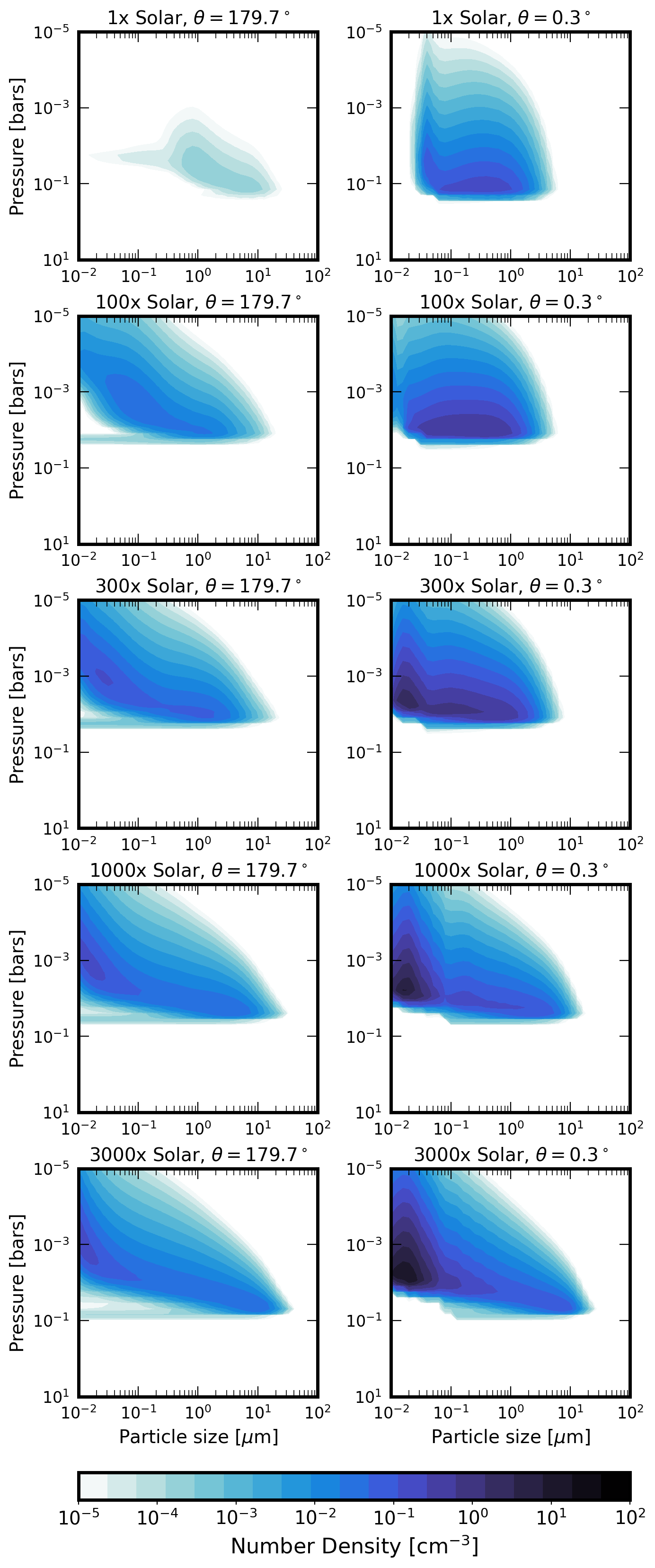}
    \caption{Evolution in total cloud particle size distributions between two extremes in cloud-haze interaction strength, $\theta=179.7 \degree$ (left column) and $\theta=0.3\degree$ (right column), for a haze production rate of $10^{-14} \text{g cm}^2 \text{s}^{-1}$ and each metallicity in our grid (increasing from top to bottom). We see that the overall cloud particle number density substantially increases due to cloud-haze interactions in all cases, and that the lower metallicity cases exhibit the most dramatic changes in particle size distribution morphology. }
    \label{fig:clouds_vs_metallicity}
\end{figure}

Atmospheric metallicity plays a key role in determining aerosol properties, as it strongly influences the availability of condensable KCl vapor and alters aerosol dynamics through changes in the mean molecular weight\footnote{ Metallicity may also affect the haze production rate, although theoretical \citep[e.g.][]{lavvas_2019} and experimental studies \citep[e.g.][]{horst_experiments} show that this relationship is complex and perhaps even non-monotonic. As such, we treat haze production rate as a free parameter in this work.}. Both of these effects are illustrated by Figure \ref{fig:clouds_vs_metallicity}, in which we show the two end-member cases of negligible and strong cloud-haze interactions ($\theta = 179.7 \degree$ and $0.3 \degree$ respectively) for a set of simulations spanning the entire metallicity range of our grid, keeping the haze production rate fixed to $10^{-14} \text{ g cm}^2\text{s}^{-1}$. First, we observe an overall increase in cloudiness with metallicity for the negligible cloud-haze interaction case, in accordance with the increasing abundance of KCl at higher metallicities. Furthermore, we find that  strong cloud-haze interactions result in significantly more abundant clouds at higher altitudes (lower pressure levels) in all of these cases. This effect is most stark for the 1 $\times$ solar metallicity case, for which there is negligible cloud formation through homogeneous nucleation, but substantial cloud formation through heterogeneous nucleation in the scenario with strong cloud-haze interactions. This suggests that cloud-haze interactions can have significant impacts on the distribution of aerosols (and atmospheric observables) for even low-metallicity atmospheres, which may be especially applicable to planets larger than the sub-Neptunes we focus on in this work, as discussed further in Section \ref{sec:discussion}. 

\begin{figure*}
    \centering
    \includegraphics[width=1.0\linewidth]{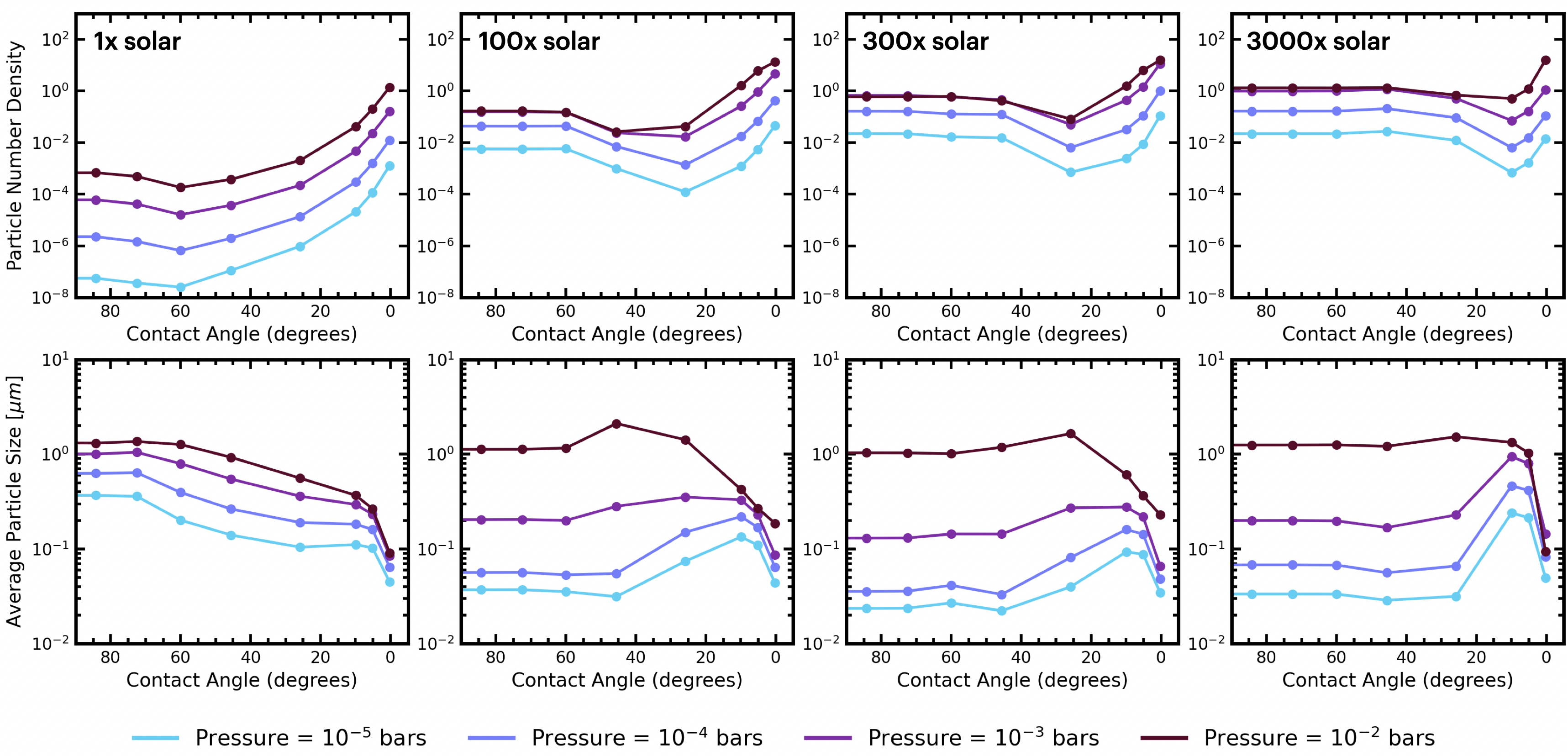}
    \caption{Total cloud particle number density [cm$^{-3}$] (integrated over particle size; top row) and average particle size (bottom row) at four pressure levels (indicated by the different colors) for the different metallicity cases at a haze production rate of $10^{-14} \text{g cm}^2\text{s}^{-1}$. }
    \label{fig:pnumdens_pavgsize_vs_metallicity_contactangle}
\end{figure*}

Moving to the cases with moderate 100 $\times$ and 300 $\times$ solar metallicities \citep[as have been inferred for many sub-Neptunes to date; e.g.][]{wallack_toi836,hu_k218b, roy_diversity, wallack_hd15337}, we observe that the principal effect of strong cloud-haze interactions is once again a substantial increase in cloud particle number density at high altitudes. In addition, strong cloud-haze interactions lead average particle sizes to decrease at depth but increase at high-altitudes. Both these effects are visible in Figure \ref{fig:pnumdens_pavgsize_vs_metallicity_contactangle}, which shows the average particle number density (top panels) and particle size (bottom panels) at select pressure levels as a function of contact angle for four of the considered metallicities. Closer examination of these trends further reveals that the general shift to lower average particle sizes occurs gradually and often non-monotonically alongside decreases in the contact angle (as discussed in Section \ref{psize_subsec}). For example, all metallicities above 1~$\times$ solar exhibit a rise in the average high-altitude ($10^{-4}$ to $10^{-5}$ bar) cloud particle size between $\sim 40\degree$ and $\sim 10\degree$. Finally, very high metallicity cases such as 3000~$\times$ solar exhibit qualitatively similar behavior, although the relative magnitude of the changes in particle number densities is somewhat reduced.  

\subsubsection{Variations with Haze Production Rate} \label{hprod_psize_subsubsection}

\begin{figure}
    \centering
    \includegraphics[width=1.0\linewidth]{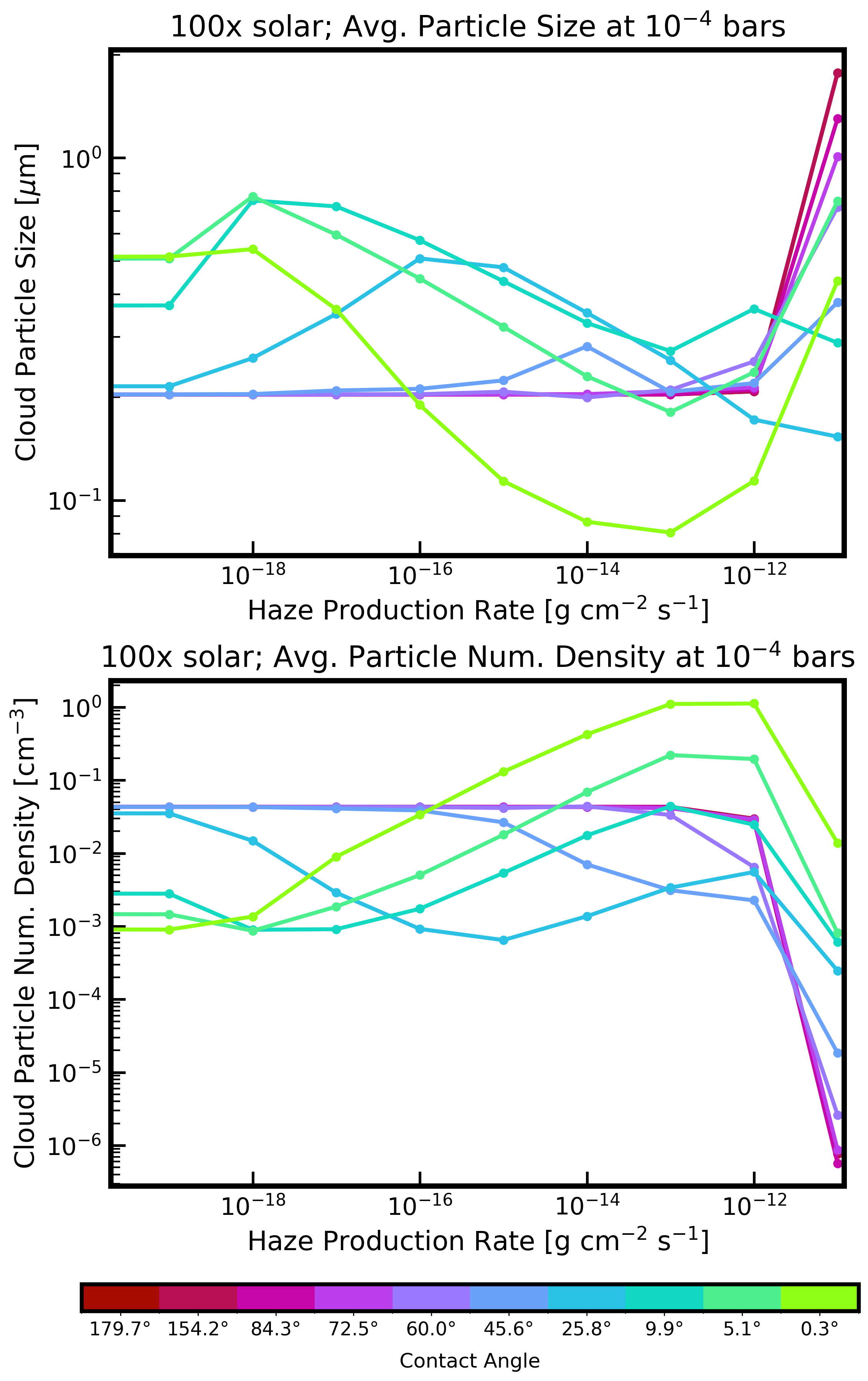}
    \caption{Average cloud particle size (top) and number density (bottom) at $10^{-4}$ bars as a function of haze production rate and contact angle (color) for 100 $\times$ solar metallicity atmospheres.  }
    \label{fig:avgpsize_numdens_vs_hprod_contactangle}
\end{figure}

The haze production rates (Table \ref{tab:parameter_space}) we test in this study allow us to examine how cloud-haze interactions manifest across a wide range in atmospheric haziness. To start, in Figure \ref{fig:avgpsize_numdens_vs_hprod_contactangle} we examine how the average cloud particle size (top panel) and number density (bottom panel) at 10$^{-4}$ bar evolve with haze production rate and microphysical contact angle for our 100 $\times$ solar metallicity simulations. Immediately, we observe that cloud-haze interactions trigger appreciable changes in cloud properties \textit{even in atmospheres with extremely low haze production} that would not typically be considered `hazy'. For example, at a haze production rate of $10^{-18} \text{ g cm}^2\text{ s}^{-1}$, varying the contact angle can nearly quadruple the average particle size at 0.1 mbar from 0.2 to $\sim 0.8$ $\mu$m while simultaneously decreasing the average particle number density at this pressure-level by more than an order of magnitude. In short, stronger cloud-haze interactions in this haze-poor regime substantially increase the average cloud particle size while decreasing their number density. 

The increase in cloud particle size (and decrease in number density) due to cloud-haze interactions at low haze production rates can be understood as arising due to a `seed-limited' regime of heterogeneous nucleation. When the number density of available CCN is low, fewer cloud particles form. However, cloud particles in this scenario are also able to reach larger sizes as growth processes efficiently consolidate condensate mass on fewer particles. This persists over the wide haze production rate range of $10^{-19}$ to $10^{-14} \text{ g cm}^2\text{ s}^{-1}$ for contact angles $< 70\degree$, with the magnitude of the effect peaking at smaller haze production rates as the contact angle decreases. Overall, this demonstrates, somewhat counter-intuitively, that\textit{even a small amount of photochemical haze is enough for cloud-haze interactions to significantly transform the cloud distribution in an exoplanet's atmosphere}. 

At much larger haze production rates of $10^{-13}$ and $10^{-12}\text{ g cm}^2\text{ s}^{-1}$, however, cloud-haze interactions have an opposite effect on the average particle size. Here, the case with the strongest cloud-haze interactions has an average particle size that \textit{decreases} by $\sim2$x to 0.1 $\mu$m relative to the $0.2$ $\mu$m seen in the case with negligible cloud-haze interactions. This occurs because the higher numbers of haze CCN allows the formation of greater numbers of mixed cloud particles. The available condensible vapor is then split between these particles, leading to smaller average particle sizes. In accordance with this picture, the reduction in particle size is accompanied by a drastic ($\sim$$25$ x) increase in particle number density, demonstrating that the net-effect of strong cloud-haze interactions in this regime is an \textit{enhancement of cloud particles at high altitudes}. This, in particular, may affect the albedo of the atmosphere, since it is especially sensitive to the abundance and optical properties of high-altitude aerosols \citep[e.g.][]{jordan_albedo}. 

Finally, at the highest haze production rate of $10^{-11}\text{ g cm}^2\text{ s}^{-1}$, the average cloud particle size begins to rise once again. This behavior is driven by two effects. First, more efficient coagulative growth at such high haze production rates increases the average haze particle size, which in turn raises the average cloud particle size because the size of the haze CCN sets the minimum possible size of a heterogeneously nucleated cloud particle. Second, at a haze production rate of $10^{-11}\text{g cm}^2\text{s}^{-1}$, radiative cooling of the deep atmosphere due to hazes shifts the KCl cloud base to higher pressures (see Figure \ref{fig:pt_structures}). Clouds forming in this higher density environment are able to more efficiently grow to large particle sizes, sequestering condensate material in fewer particles.  

\subsection{Cloud Mass} \label{cloud_mass_sec}
 
\begin{figure*}
    \centering
    \includegraphics[width=1.0\linewidth]{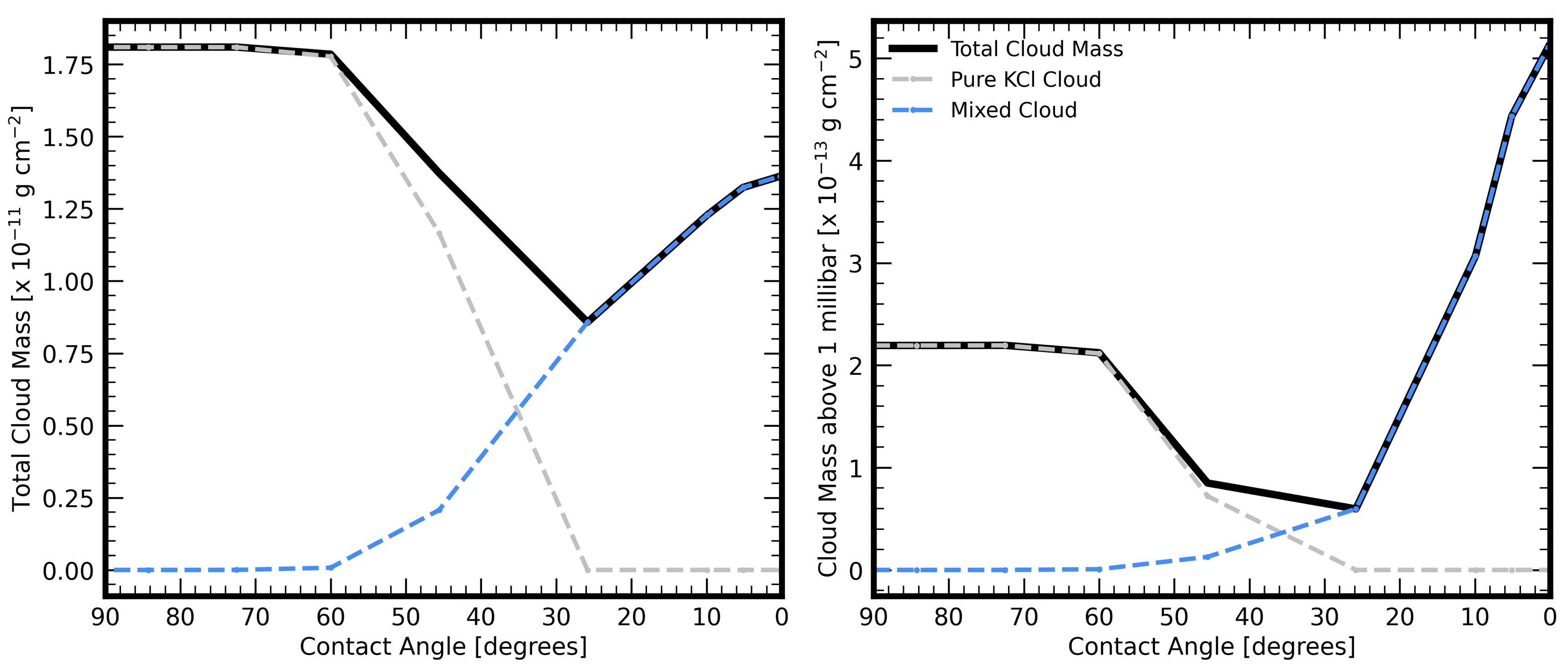}
    \caption{Variation in total (integrated across the entire atmosphere; left panel) and upper-atmosphere ($<1$ mbar; right panel) cloud mass (black solid lines) due to cloud-haze interactions for a 100 $\times$ solar metallicity atmosphere with $10^{-14} \text{g cm}^2\text{s}^{-1}$ haze production rate. The contributions of the homogeneously nucleated `pure' KCl and heterogeneously nucleated `mixed' KCl clouds are shown using grey and blue dashed lines respectively. We observe that cloud-haze interactions first (moderate contact angles) cause a decrease in total and upper-atmosphere cloud mass reflecting the `wet removal' regime before significantly increasing the upper atmosphere cloud mass at small contact angles ($\lesssim 20 \degree$).}
    \label{fig:total_cloud_mass}
\end{figure*}

\begin{figure*}
    \centering
    \includegraphics[width=1.0\linewidth]{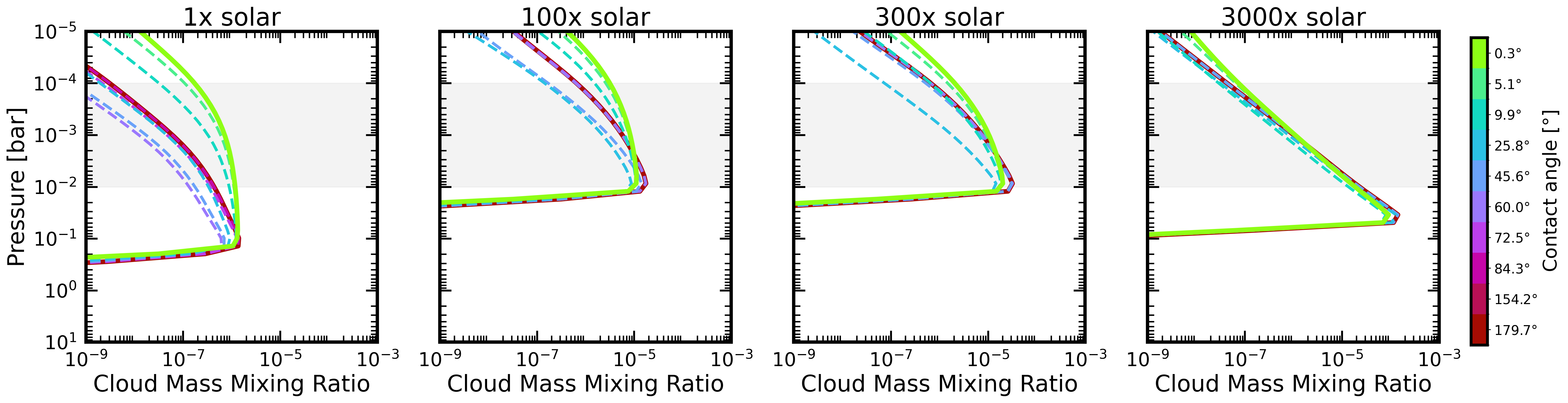}
    \caption{Variation in cloud mass for a haze production rate of $10^{-14}$ g cm$^2$ s$^{-1}$ and different metallicities due to cloud-haze interactions. Different panels show the different metallicity cases, with red to lime colors showing progressively smaller contact angles in dashed lines, except for the end-member cases which are highlighted in solid lines. The gray shaded regions highlight the pressure-range (10$^{-2}$ to 10$^{-4}$ bars) typically probed by transmission spectroscopy, although we note that the pressure range probed depends on atmospheric composition and aerosol abundance.}
    \label{fig:cloud_mass_multipanel}
\end{figure*}

Cloud mass provides a key diagnostic of how cloud–haze interactions reshape both the vertical distribution and overall abundance of condensates in exoplanet atmospheres. In Figure \ref{fig:total_cloud_mass} we show how the total cloud mass and cloud mass at pressure levels $< 10^{-3}$ bar (representative of the region typically probed by transmission spectroscopy) vary with contact angle for our fiducial 100~$\times$~solar metallicity simulation with a $10^{-14}$ g cm$^{-2}$ s$^{-1}$ haze production rate. Starting from contact angles $\sim 70\degree$, the total cloud mass decreases until $\theta=25 \degree$, before rising back up as $\theta$ approaches 0\degree\ and thus the strong cloud-haze interaction regime. Interestingly, the total cloud mass actually decreases by 25\% between the two endpoints of cloud-haze interactions. This is because, when cloud–haze interactions are weak, clouds form primarily through homogeneous nucleation near the cloud base, where high-pressure conditions enable efficient condensational growth to larger particle sizes and thus increase the total cloud mass. For strong cloud-haze interactions, however, heterogeneous nucleation upon haze CCN at this haze production rate leads to a greater abundance of cloud particles with smaller average radii. Since these mixed clouds are preferentially located at lower pressures than the homogeneous population, they are unable to grow as efficiently through condensation, resulting in a lower total cloud mass. Crucially, however, the cloud mass located at the high altitudes probed by transmission spectroscopy exhibits the opposite trend. In this case, proceeding once again from large to small contact angles, the total cloud-mass first decreases until $\theta=25 \degree$ before rising up to nearly double the value in the case with no cloud-haze interactions---reflecting the 'wet removal' and cloud enhancement regimes seen in the particle size distributions and discussed in Sections \ref{psize_subsec} and \ref{overview_subsection}. This significant increase in the cloud-mass at high altitudes has important effects on the transmission spectra, as shown in Section \ref{transmission_spectra_sec}.

Expanding our attention beyond this fiducial case, we show the cloud mass mixing ratio as  a function of pressure level and atmospheric metallicity while maintaining a $10^{-14}$ g cm$^{-2}$ s$^{-1}$ haze production rate in Figure \ref{fig:cloud_mass_multipanel}. Immediately, we observe that the total cloud mass increases significantly alongside metallicity, owing to the greater abundances of KCl vapor in these cases. Furthermore, we see that the high altitude cloud mass increase driven by strong cloud-haze interactions occurs at all metallicities, though the magnitude of this effect decreases with increasing metallicity. Indeed, looking at an example pressure-level of 10$^{-4}$ bars, we observe  that the cloud-mass mixing ratio between minimal and strong cloud-haze interacting cases increases by $>$ 2 orders of magnitude for a 1 $\times$ solar metallicity atmosphere, and $\sim2$x for 3000x solar. As in Figure \ref{fig:total_cloud_mass}, for all of these metallicity cases the cloud mass does not trend monotonically with contact angle, but displays a minimum at moderate contact angles.

\subsection{Optical Depth Surfaces} \label{opd_surfaces_section}

Changes in cloud and haze distributions due to variations in the strength of cloud-haze interactions directly translate to changes in aerosol optical depths. Figure \ref{fig:opd_grid} shows the layer-by-layer optical depth contributions of all types of aerosols in our models for our fiducial $10^{-14}$ g cm$^{-2}$ s$^{-1}$ haze production rate across two axes in our grid: atmospheric metallicity and contact angle. On each panel, we indicate the surfaces at which the nadir optical depth equals 0.01, 0.1, and 1 (if such a surface exist) through solid, dashed, and dotted white lines. Echoing the results of the previous two sub-sections, we find that these surfaces first drop to lower altitudes (corresponding to the wet removal regime) before rising to higher altitudes as the contact angle decreases (corresponding to the cloud- enhancement regime). The cloud-enhancement regime also shows an overall increase in the nadir optical depth $\tau_n$ of the atmosphere driven by the aerosols. Correspondingly, the slant optical depth $\tau_s$ also rises since this quantity can be approximated using \begin{equation} \label{taus_taun}
    \frac{\tau_s}{\tau_n} = \sqrt{\frac{2\pi R_p}{H}}
\end{equation} where $H$ is the atmospheric pressure scale height and $R_p$ is the planetary radius \citep{fortney_2005}. From this relation, we can compute the nadir optical depth and pressure level at which $\tau_s = 1$, which will help us understand and interpret transmission spectra since these are observed in slant geometry. As an example, for our fiducial 100 $\times$ solar metallicity case with a $10^{-14} \text{g cm}^2 \text{s}^{-1}$ haze production rate, $\tau_s = 1$ when $\tau_n = 0.04$. From Figure \ref{fig:opd_grid}, we can infer that the $\tau_s = 1$ level for this simulation rises from $\sim 10^{-2.8}$ bars in the case with negligible cloud-haze interactions ($\theta = 179.7 \degree$) to $\sim 10^{-3.5}$ bars in the $\theta=0.3\degree$ case with strong cloud-haze interactions. At the lowest metallicity, this pressure level similarly shifts up from $\sim 10^{-1.5}$ bars to $\sim 10^{-2.5}$ bars. On the contrary, the highest metallicity case shows a negligible change along the contact angle axis, which has implications for how transmission spectra for this case might vary due to cloud-haze interactions, as will be further discussed in Section \ref{transmission_spectra_sec}. 

\subsection{Overview of the impact of cloud-haze interactions} \label{overview_subsection}

The previous sub-sections describe in detail how cloud-haze interactions affect particle size distributions, cloud mass, and aerosol optical depth across our grid. Here we step back to summarize the major trends.
Holding bulk properties such as metallicity and haze production rate fixed, we find that cloud-haze interactions at weak to moderate strengths (contact angles $\sim 72.5 - 25.8\degree$) typically clear the upper atmosphere of aerosols. This occurs because high-altitude haze particles can act as CCN for KCl, enabling heterogeneously nucleated KCl clouds to form. Through condensation, these mixed KCl cloud particles grow to significantly larger sizes than the haze CCN, causing them to settle into deeper, warmer layers where they evaporate. At the same time, the decrease in the KCl saturation vapor ratio reduces the homogeneous nucleation rate. The net effect is the preferential wet removal of the largest haze particles in the upper atmosphere via the Kelvin Effect and decrease in total condensate mass, reducing aerosol opacity there. At even smaller contact angles, smaller haze particles — which are more abundant, particularly at high altitudes — are activated as CCN as well. This accelerates cloud formation and drives a cloud-enhancement regime.

\begin{figure*}[t]
    \centering
    \includegraphics[width=1.0\textwidth]{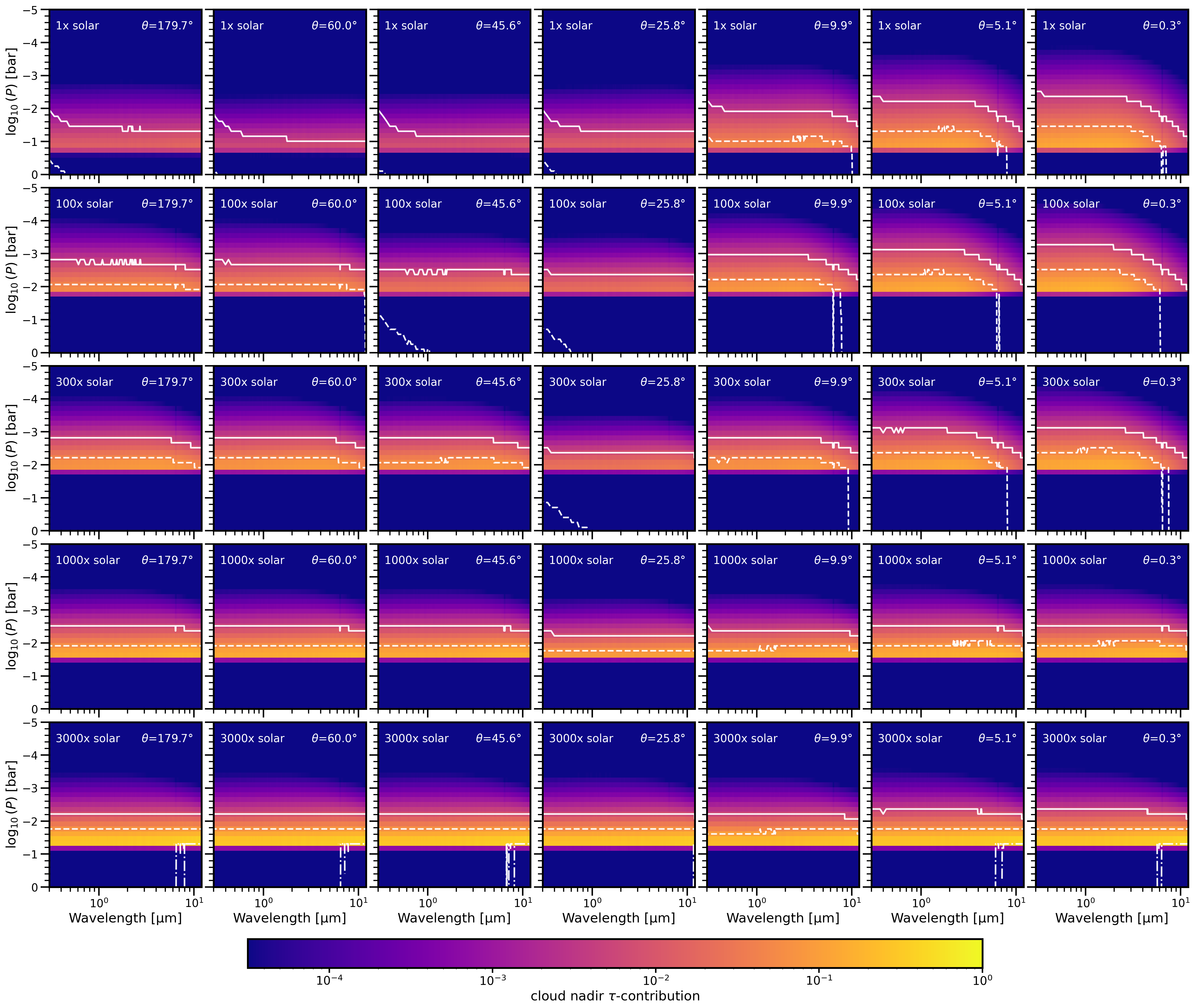}
    \caption{Layer-by-layer nadir aerosol optical depth contributions across metallicities (top to bottom) and contact angles (left to right) for our fiducial $10^{-14}$ g cm$^{-2}$ s$^{-1}$ haze production rate. On each panel, the solid, dashed, and dash-dotted white lines indicate the pressure level at which the nadir optical depth $\tau_n$ equals 0.01, 0.1, and 1 respectively. We observe that the principal effect of strong cloud-haze interactions is to raise the opacity contributed by aerosols high in the atmosphere, which follows directly from the behavior exhibited by their particle size distributions, as shown by Figure \ref{particle_size_dist_sec}. }
    \label{fig:opd_grid}
\end{figure*}

These regimes are modulated by both metallicity and haze production rate. The effects of cloud-haze interactions on particle size distributions are strongest at low metallicities and weaken as metallicity rises (Figure \ref{fig:clouds_vs_metallicity}). At low haze production rates, cloud formation is seed-limited: interactions typically increase the average cloud particle size in the upper atmosphere while decreasing number density. Because fewer haze particles are able to act as CCN at larger contact angles, the seed-limited regime extends to higher haze production rates in those cases. At sufficiently high haze production rates, cloud formation is no longer seed-limited and can enter the cloud-enhancement regime depending on the contact angle, dramatically raising cloud particle number densities in the upper atmosphere.
Examining the optical depth as a function of contact angle illustrates how wet removal and cloud enhancement shape aerosol opacity across the grid (Figure \ref{fig:opd_grid}), with the relative effects most pronounced at low to moderate metallicities (1–100 $\times$
solar). Taken together, these trends produce systematic changes in aerosol opacity with direct consequences for transmission spectra---a topic we turn to now.

% However, stronger vertical mixing enables cloud-haze interactions to significantly alter transmission spectra even in the 3000 $\times$ solar metallicity case, as we show in Section \ref{sec:mixing}. 

\section{Transmission Spectra} \label{transmission_spectra_sec}

We use \texttt{PICASO 3.0} \citep{batalha_picaso_2019, mukherjee_picaso} combined with the particle size distributions from our \texttt{CARMA} models to generate synthetic transmission spectra that include gas-phase and aerosol opacity. Our spectral calculations assume a reference pressure of 1 mbar for the planetary radius. Since our simulations treat the clouds \& hazes as spheres, we model this aerosol opacity contribution with Mie-scattering theory using the publicly-available code \texttt{pyMieScat} \citep{pymiescat}. Using \texttt{PICASO 3.0} with a custom aerosol specification requires providing the cumulative extinction coefficient $\beta_{e, i}$, single-scattering albedo $\omega_{i}$, and asymmetry parameter $g_{i}$ in each atmospheric pressure-level $i$. Following the method of \citet{petty_textbook}, we calculate these quantities as follows:

\begin{equation}
    \beta_{e, i} = \int_{0}^{\infty} n_{i}(r) Q_{e} (r) \pi r^2 dr, 
\end{equation}
\begin{equation}
    \omega \equiv \beta_{s, i}/\beta_{e, i} = \frac{\int_{0}^{\infty} n_{i}(r) Q_{s, i} (r) \pi r^2 dr}{\int_{0}^{\infty} n_{i}(r) Q_{e, i} (r) \pi r^2 dr}
\end{equation}
\begin{equation}
    g_{i} = \frac{1}{\beta_{s, i}}\int_{0}^{\infty} n_{i}(r) Q_{s} (r) \pi r^2 g(r) dr, 
\end{equation} where $n_{i}(r)$ is the number density of particles with size $r$, and $\beta_{s, i}$ is the total scattering coefficient. $Q_s$ and $Q_e$ are the Mie efficiencies for scattering and extinction respectively. All of these quantities are also wavelength dependent. For the hazes, we use the optical properties from \citet{Khare_tholins} for Titan-like tholins and for the KCl clouds we use the optical properties from \citet{wakeford_sing, palik_kcl} accessed via the \texttt{POSEIDON} aerosol database \citep{poseidon_joss,mullens_2024}. We then compute equilibrium chemistry abundances with \texttt{FASTCHEM} before using the previously calculated aerosol extinction coefficients and scattering properties to generate transmission spectra for each of our simulations that account for both gas phase and aerosol opacity. 

\begin{figure*}
    \centering
    \includegraphics[width=1.0\linewidth]{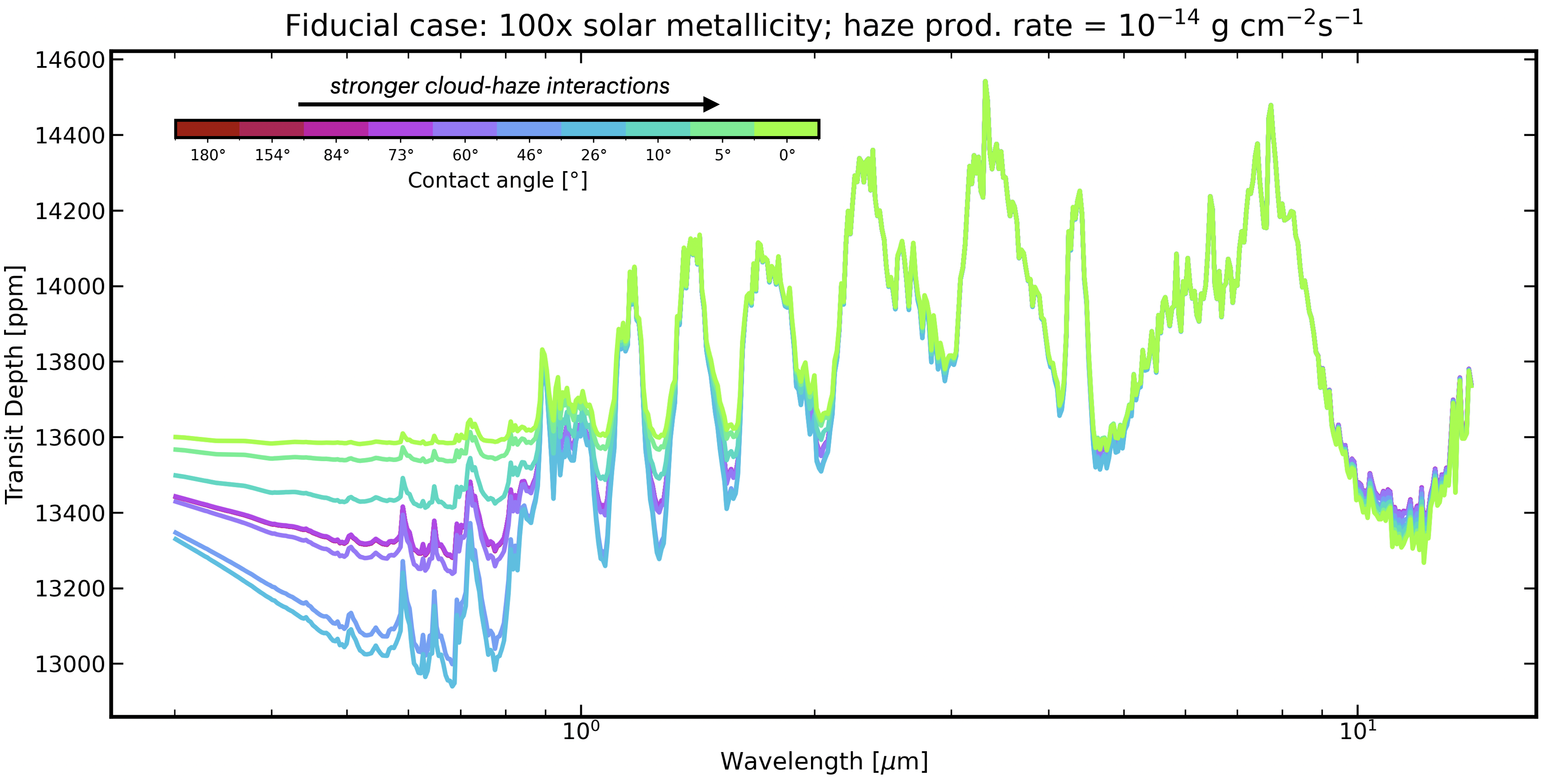}
    \caption{Variations in transmission spectra caused by changing the microphysical contact angle for our fiducial 100 $\times$ solar metallicity atmosphere with a $10^{-14}$ g cm$^{-2}$ s$^{-1}$ haze production rate. We observe that, while wavelengths longer than $\sim 2.5 \mu$m are largely insensitive to cloud-haze interactions, shorter wavelengths show significant flattening along the transition to strong cloud-haze interactions. }
    \label{fig:fiducial_transmission_spectra}
\end{figure*}

We start by examining the contact-angle dependence of predicted transmission spectra for our fiducial case with 100 $\times$ solar metallicity and a haze production rate of $10^{-14}$ g cm$^{-2}$ s$^{-1}$ in Figure \ref{fig:fiducial_transmission_spectra}. As can be seen, the changes in aerosol particle size distributions due to cloud-haze interactions exert a significant impact on the transmission spectrum at optical and near-infrared wavelengths $\lessapprox2.5$ $\mu\text{m}$ whereas longer wavelengths are barely affected. This is driven by the fact that the cloud and haze particles in the high-altitude (low-pressure) regions probed in transmission rarely reach large sizes in large enough abundances to affect these longer wavelengths significantly\footnote{This is not strictly true for the very highest haze production rate: $10^{-11}$ g cm$^{-2}$ s$^{-1}$. In this case, the haze distribution can extend to large enough sizes to affect spectra at longer wavelengths. However, the effect is fairly minor (sub $\sim$ scale-height) and relatively constant between contact angles.}. As a result, we focus the following discussion on the shorter wavelengths most affected by cloud-haze interactions. 

% \subsection{The case of metallicities below 300x solar}

\begin{figure*}
    \centering
    \includegraphics[width=1.0\linewidth]{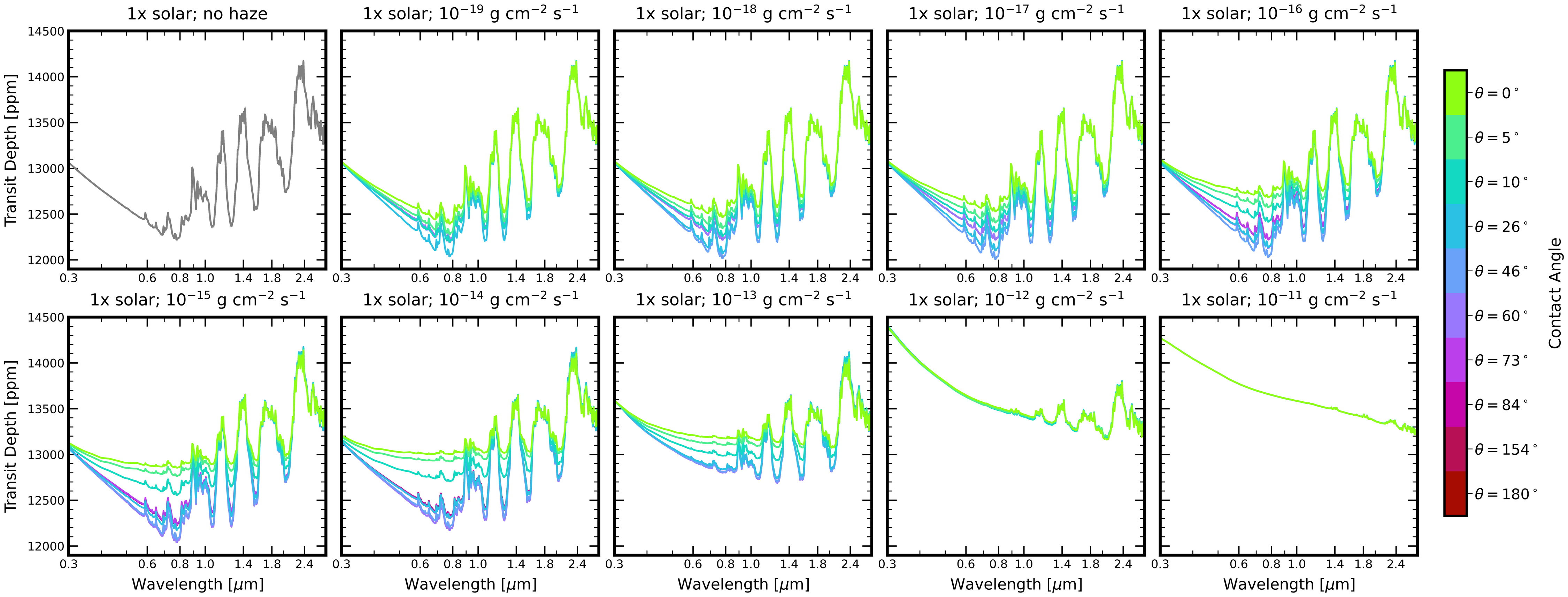}
    \caption{Predicted transmission spectra for 1 $\times$ solar metallicity atmospheres as a function of contact angle and (tholin) haze production rate. Each panel shows a different haze production rate, ranging from a haze-free case in the the top left panel to $10^{-11}$ g cm$^{-2}$ s$^{-1}$ for the bottom right panel. Colors on each panel indicate the contact angle, with lower values (corresponding to stronger cloud-haze interactions) shown in progressively lighter colors.  }
    \label{fig:1xsolar_contactangle_spectra}
\end{figure*}

In Figure \ref{fig:1xsolar_contactangle_spectra} we show spectra at short wavelengths for the 1 $\times$ solar metallicity subset of our simulations and all haze production rates considered. We find that the range of potential spectra created by cloud-haze interactions steadily widens as we increase the haze production rate from zero to $10^{-15} \text{ g cm}^2 \text{ s}^{-1}$ before reaching a maximum at our fiducial haze production rate of $10^{-14} \text{ g cm}^2 \text{ s}^{-1}$. For a planet with GJ 1214 b's parameters specifically, this envelope represents changes exceeding hundreds of ppm in the optical and near-infrared, which translates to 2-3 scale height variations in both the continuum and absorption features. \textit{Cloud-haze microphysics is thus capable of driving substantial spectral variation for warm sub-Neptunes}. Specifically, atmospheres with the smallest contact angles show the flattest transmission spectra while moderate contact angles can lead to clearer spectra. This is a consequence of the cloud enhancement and wet removal regimes discussed previously. Further increasing the haze production rate narrows this range until no discernible differences between contact angles can be observed for production rates $ > 10^{-12} \text{cm}^2 \text{s}^{-1}$, when most absorption features give way to a strong scattering slope. This is due to haze opacity dominating over cloud opacity in its contribution to the transmission spectra at these high production rates. At these high haze production rates, cloud-haze interactions produce only modest changes to the haze particle size distribution while continuing to substantially reshape the cloud population. Because haze opacity dominates the transmission spectrum, these large cloud changes contribute little to the observable spectra. Consequently, transmission spectra become nearly independent of contact angle.

\begin{figure*}
    \centering
    \includegraphics[width=1.0\linewidth]{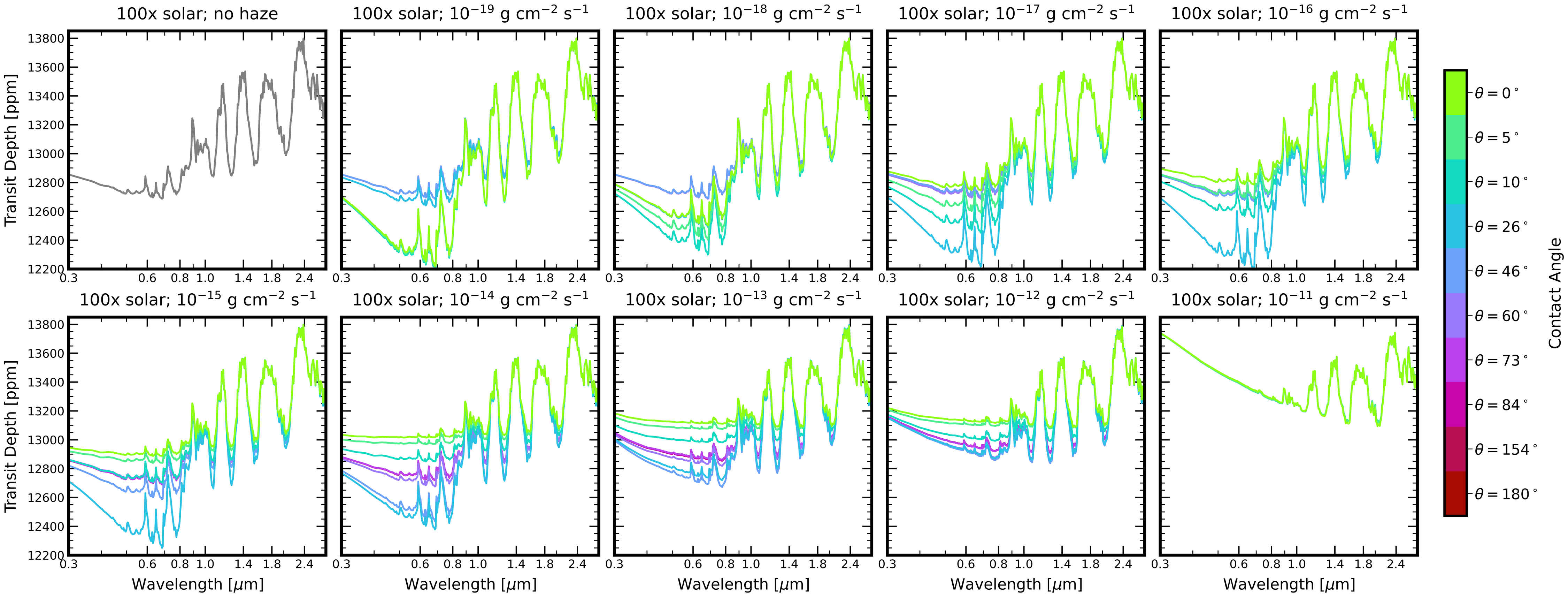}
    \caption{Analogous to Figure \ref{fig:1xsolar_contactangle_spectra} but for a metallicity of 100 $\times$ solar.}
    \label{fig:100xsolar_contactangle_spectra}
\end{figure*}

The trends observed for the 1 $\times$ solar metallicity case shown in Figure \ref{fig:1xsolar_contactangle_spectra} are largely echoed by the results for higher metallicities more characteristic of many sub-Neptunes characterized to-date, albeit accompanied by an overall suppression in absolute feature sizes due to the increased mean molecular weight. In Figure \ref{fig:100xsolar_contactangle_spectra}, we show how transmission spectra evolve with contact angle and haze production rate for  100 $\times$ solar metallicity simulations. Once again, we observe that haze production rates between $10^{-15}$ and $10^{-13} \text{ g cm}^2 \text{ s}^{-1}$ correspond to the cases with the largest spread in possible transmission spectra due to cloud-haze interactions, spanning 2-5 scale heights. The fact that this effect persists for more than two orders of magnitude in metallicity implies that using feature sizes in this wavelength region to infer population-level trends in the atmospheric metallicities and aerosol abundances of sub-Neptunes may suffer from significant degeneracies with the unknown microphysics of cloud-haze interactions, a topic we quantify and discuss further in Section \ref{water_feature_section}. The behavior for 300 $\times$ solar metallicity closely resembles that for 100 $\times$ solar metallicity, as shown in the Appendix.

% While decreasing the contact angle initially leads to a clearing of the atmosphere, an example of so-called `rainout', cases with \textit{the lowest contact angle} ($\theta=0.3 \degree$) \textit{routinely show the flattest transmission spectra}---a consequence of the shift to larger particle sizes and number densities seen in the particle size distributions in Section \ref{particle_size_dist_sec}. Increasing the haze production rate decreases the magnitude of spectral variation with contact angle, due to the increasing dominance of the haze opacity, culminating in virtually identical spectra across contact angles for the case with a haze production rate of $10^{-11}$ g cm$^{-2}$ s$^{-1}$. Overall, we observe that cloud-haze interactions can alter the predicted transmission spectra by hundreds of ppm at wavelengths shorter than $\sim$ 2.5 microns, while leaving longer wavelengths relatively unaffected.

\begin{figure*}
    \centering
    \includegraphics[width=1.0\linewidth]{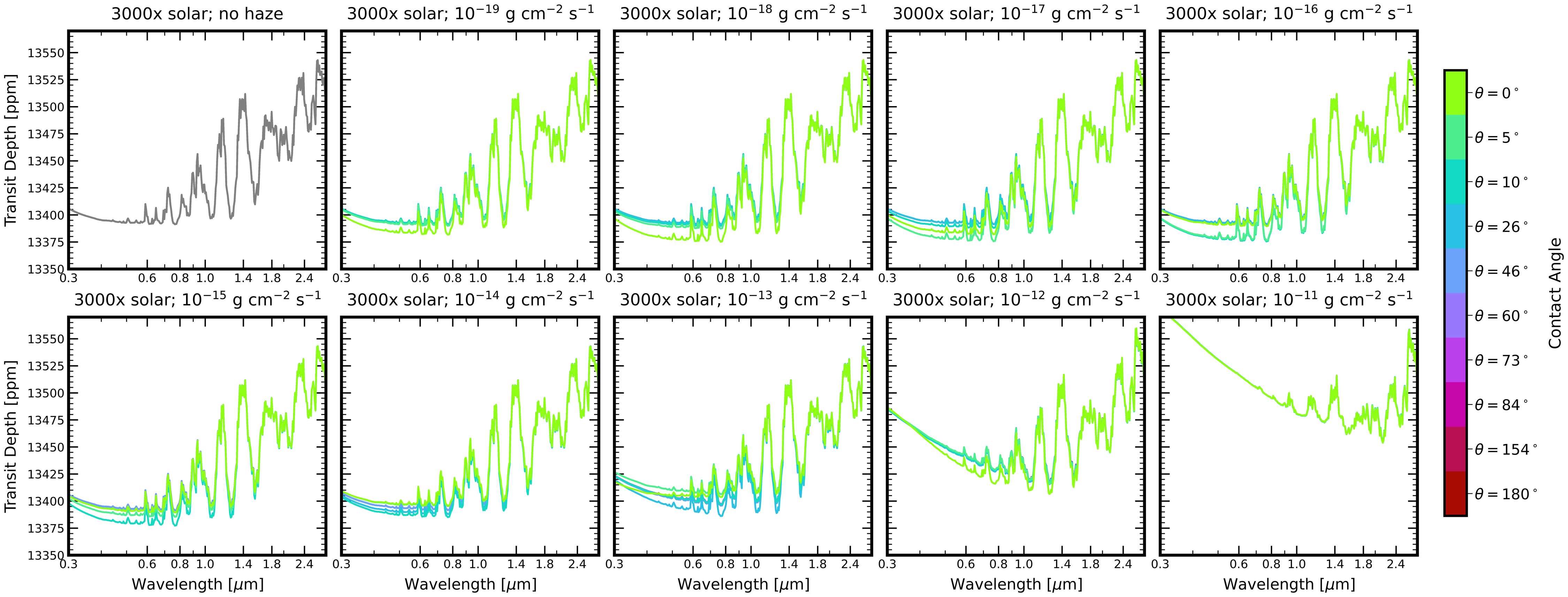}
    \caption{Analogous to Figure \ref{fig:1xsolar_contactangle_spectra} but for a metallicity of 3000 $\times$ solar.}
    \label{fig:3000xsolar_contactangle_spectra}
\end{figure*}

% \subsection{The case of metallicities above 1000x solar}\label{3000xsolar_variation}

In contrast to the lower metallicity cases, cloud-haze interactions have a much weaker effect on predicted transmission spectra for atmospheres with metallicities of 1000 (shown in the Appendix) and 3000 $\times$ solar (Figure \ref{fig:3000xsolar_contactangle_spectra}). This weaker effect reflects the trend with optical depth contributions seen in Figure \ref{fig:opd_grid} and persists across the entire range of haze production rates tested in this study. One explanation for this is that the super-saturation at the KCl cloud base is greater for these higher metallicity cases, as can be inferred by looking at Figure \ref{fig:pt_structures}. This supersaturation increases the efficiency of particle growth at the cloud-base, leading to the preferential formation of large particles at depth, even in the presence of significant cloud-haze interactions. As a result, the population of cloud particles at the high-altitudes probed in transmission does not change substantially with different contact angles. This conclusion, however, is sensitive to the strength of vertical mixing assumed, as we will discuss in Section \ref{sec:mixing}. %we will discuss how this is no longer the case for stronger vertical mixing with 10x larger $K_{zz}$ than our fiducial profiles. 

\subsection{When do clouds/hazes dominate the aerosol opacity?}

\begin{figure*}
    \centering
    \includegraphics[width=1.0\linewidth]{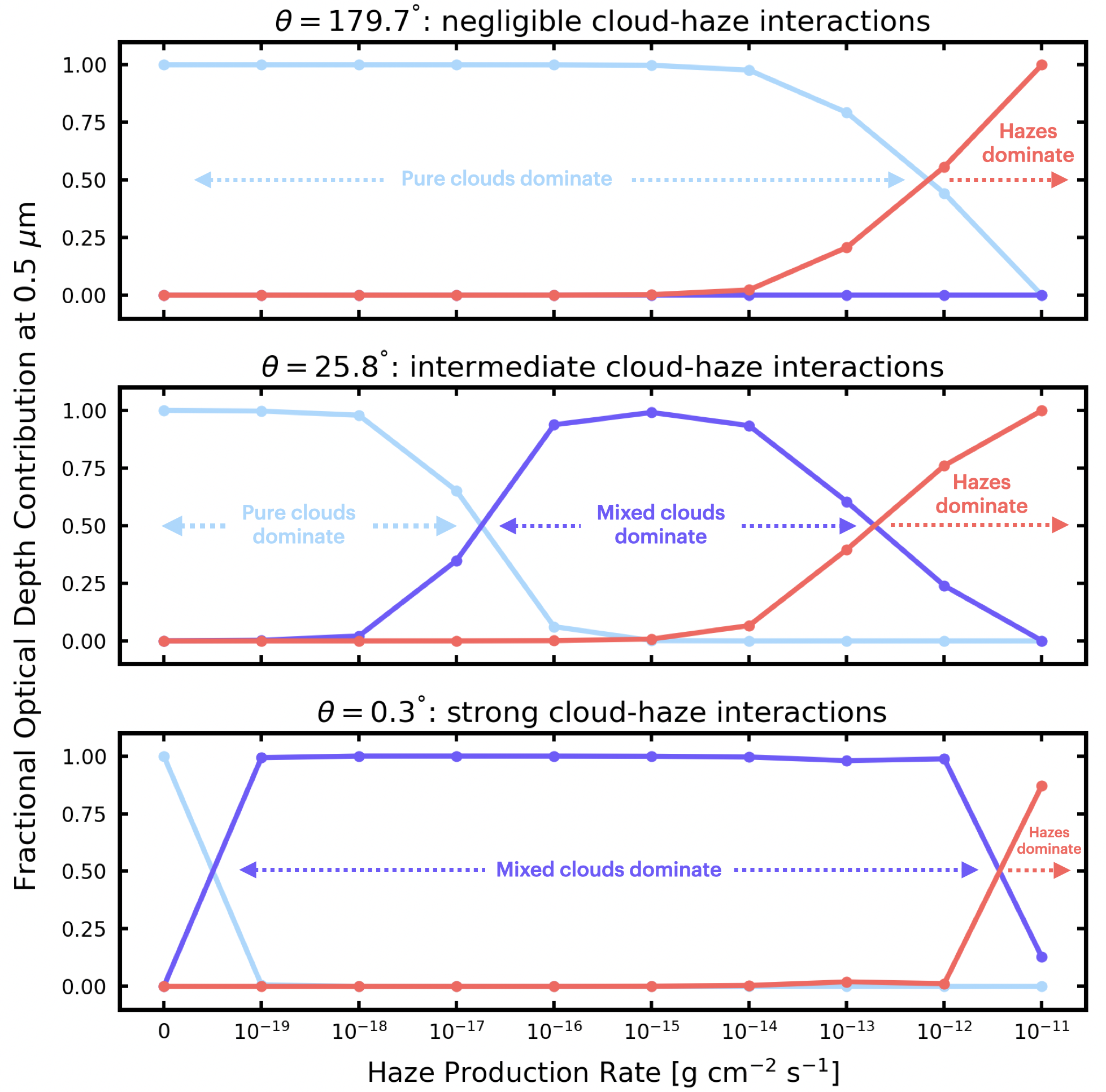}
    \caption{Fractional optical depth contributions at $0.5$ $\mu$m for pure KCl clouds (light blue), hazes (red), and mixed clouds (blue) as a function of haze production rate for 100 $\times$ solar metallicity simulations. We show cases with negligible ($\theta=179.7\degree$; top), moderate ($\theta=25.8\degree$; medium), and strong ($\theta=0.3\degree$; bottom) cloud-haze interactions. Aerosols contained between 0.1 and 10 mbar are included for this calculation. }
    \label{fig:fractional_opd}
\end{figure*}

Given the wide range of uncertainties in haze production rate and cloud properties, it is helpful to determine under what conditions cloud opacity dominates over haze opacity in transmission and vice versa. For our fiducial 100 $\times$ solar metallicity simulations, we quantify this by calculating the fractional upper atmosphere ($10^{-2}$ to $10^{-4}$ bar) optical depth contributed by each aerosol type as a function of haze production rate (Figure \ref{fig:fractional_opd}). When cloud-haze interactions are minimal ($\theta=179.7\degree$), pure KCl clouds dominate the opacity until a haze production rate of $10^{-12}$ g cm$^{-2}$ s$^{-1}$, above which hazes dominate. In a case of moderate cloud-haze interactions such as $\theta=25.8 \degree$, pure KCl clouds dominate for low haze production rates $< 10^{-17}$ g cm$^{-2}$ s$^{-1}$. Above this value, the increased abundance of haze CCN facilitates a transition to a new mixed cloud dominated regime from $10^{-16}$ to $10^{-13}$ g cm$^{-2}$ s$^{-1}$. For higher haze production rates, the haze population grows too abundant for KCl clouds to entirely condense upon. As a result, hazes once again become the dominant aerosol in transmission above $10^{-12}$ g cm$^{-2}$ s$^{-1}$. 

If the contact angle is further decreased to $\theta=0.3\degree$ such that cloud-haze interactions become very strong, we observe that mixed clouds entirely dominate the aerosol opacity from $10^{-19}$ g cm$^{-2}$ s$^{-1}$ to  $10^{-12}$ g cm$^{-2}$ s$^{-1}$. Unlike the previous cases, in which the fractional opacity due to hazes rises steadily starting around $10^{-14}$ g cm$^{-2}$ s$^{-1}$ before becoming dominant at $10^{-12}$ g cm$^{-2}$ s$^{-1}$, the haze contribution in this  remains completely negligible up till $10^{-12}$ g cm$^{-2}$ s$^{-1}$ in this scenario. It is only at the highest haze production rate of $10^{-11}$ g cm$^{-2}$ s$^{-1}$ that haze opacity begins to exceed the opacity due to mixed clouds, This demonstrates how changes in the contact angle can qualitatively alter the dominant aerosol contributor, depending on the haze production rate. The sensitivity of this behavior to vertical mixing (and metallicity) is discussed in Section \ref{sec:mixing}.  

\subsection{Trends in the 1.4 $\mu\text{m}$ water-feature} \label{water_feature_section}

\begin{figure*}
    \centering
    \includegraphics[width=1.0\linewidth]{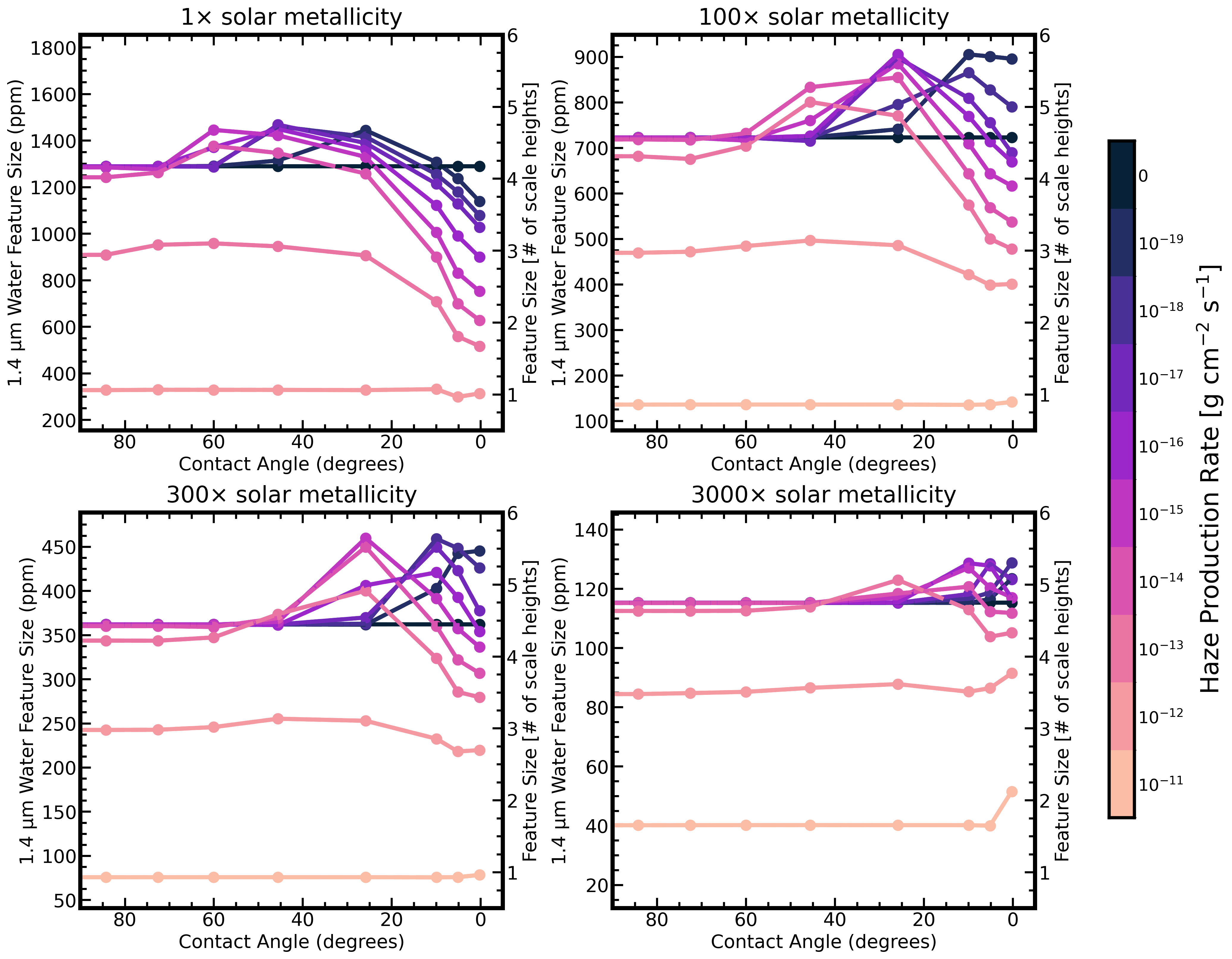}
    \caption{Variation in amplitude of the $1.4 \mu$m water absorption feature due to cloud-haze interactions. Different panels show the different metallicity cases with different colors representing different haze production rates. We show the feature size in both transit depth units and normalized to the true mean-molecular weight-corrected atmospheric scale heights. We observe that changes in the microphysical contact angle are capable of driving multiple scale height changes in the feature amplitude. }
    \label{fig:waterfeature}
\end{figure*}

Attempts to chart population-level trends in exoplanet atmosphere properties have used the amplitude of the 1.4 $\mu$m water absorption feature in transmission as a diagnostic that can be compared between planets \citep[e.g., ][]{crossfield_kreidberg, brande_2024, stevenson_clouds}. Motivated by these studies, in this section, we take a closer look at our predicted transmission spectra to examine the variation in the amplitude of the 1.4 $\mu$m water-feature that can arise solely from the microphysics of cloud-haze interactions.

Figure \ref{fig:waterfeature} shows this feature amplitude as a function of atmospheric metallicity, microphysical contact angle, and haze production rate. For all metallicities, we observe that smaller contact angles $<10 \degree$ often significantly suppress the 1.4 micron water feature size in absolute terms for all but the smallest haze production rates, reflecting the trends discussed in Sections \ref{particle_size_dist_sec} and \ref{transmission_spectra_sec}. Since the atmospheric mean-molecular-weight is anti-correlated with feature size (through the pressure scale height $H = k_bT/\mu g$) we also express the feature size in scale heights on each panel as a normalizing factor between these cases. Overall, we see that changes in the contact angle $\theta$ can directly lead to $2-4$ scale height variations in the $1.4 \mu m$ water-feature size for low to moderate metallicities (1 to 300 $\times$ solar) and weaker variations for the higher metallicity 3000 $\times$ solar case. We also find that the dependence of the 1.4 micron feature amplitude on cloud-haze interactions is strongest at moderate haze production rates, with both higher and lower rates leading to decreased dependence. Our findings thus indicate that the 1.4 micron feature amplitude can be substantially altered by the microphysics of cloud-haze interactions even when bulk atmospheric properties like temperature, composition, and vertical mixing are held fixed.

\section{Sensitivity to Vertical Mixing} \label{sec:mixing}

Vertical mixing critically modulates atmospheric chemistry and aerosol distributions, yet stands out as one of the most poorly constrained phenomena in our current understanding of planetary atmospheres. Though GCMs can provide insights into how mixing may occur, calculating one-dimensional estimates of mixing such as $K_{zz}$ from GCMs is non-trivial and subject to large typical uncertainties exceeding an order of magnitude \citep[e.g. ][]{moses_2011, parmentier2013,xi_zhang_global_mean_2}. Given how crucial vertical mixing is for cloud formation \citep[e.g. ][]{ackerman_marley,gao_marley_ackerman,diana_silicate}, with more vigorous mixing associated with increased cloud vertical extent, understanding how variations in K$_{zz}$ impact cloud-haze interactions is key for grounding our understanding and necessary to bound observational predictions. As such, we devote this section to assessing the sensitivity of our results to choice of vertical mixing strength. We do this by running our entire grid of simulations using $K_{zz}$ values that are 0.1 and 10 $\times$ our nominal value. These profiles are shown for each metallicity in Figure \ref{fig:pt_structures}. 

\begin{figure*}
    \centering
    \includegraphics[width=1.0\linewidth]{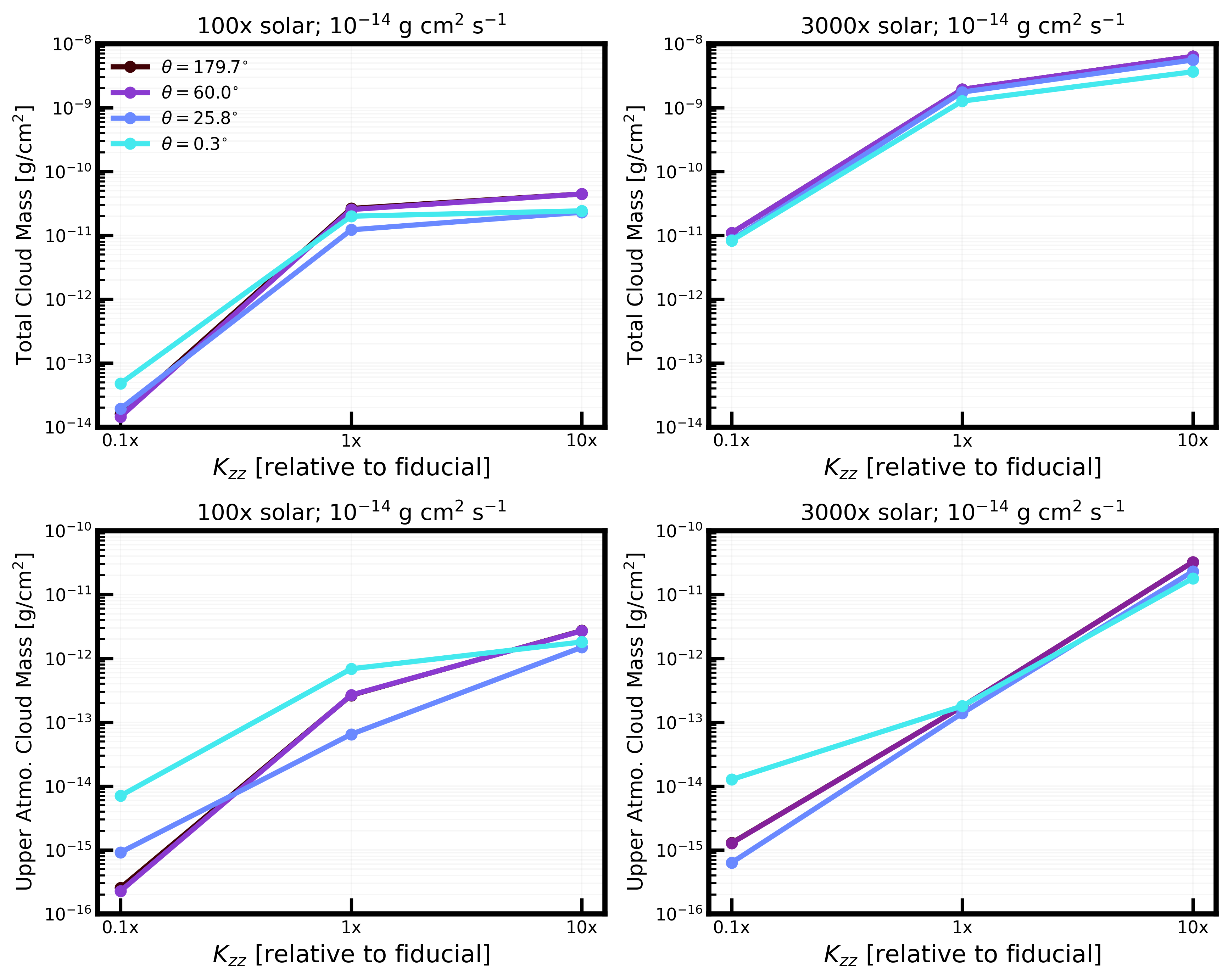}
    \caption{Sensitivity of total (top panels) and upper atmosphere (P$<10^{-4}$ bars) cloud mass (bottom panels) for 100 (left) and 3000 $\times$ solar (right) metallicities with a $10^{-14}$ g cm$^2$ s$^{-1}$ haze production rate. Different contact angles are indicated using different colors. We see that stronger vertical mixing increases the total and upper atmospheric cloud mass in all cases, with the 3000 $\times$ solar simulation showing the greatest increase in upper atmospheric cloud mass. }
    \label{fig:kzz_cloudmass}
\end{figure*}

We begin by exploring how the vertical mixing strength impacts the total and upper-atmosphere cloud mass. Figure \ref{fig:kzz_cloudmass} shows these quantities for 100 $\times$ and 3000 $\times$ solar metallicity simulations with a $10^{-14} \text{ g cm}^2 \text{ s}^{-1}$ haze production rate, with the left panels clearly illustrating how a 10x increase in $K_{zz}$ leads to significantly increased cloud formation in the upper atmosphere. This behavior holds true from the largest ($179.7\degree$) to smallest ($0.3\degree$) contact angles we test. Conversely, decreasing $K_{zz}$ by a similar magnitude reduces the total cloud mass by greater than two orders of magnitude for both metallicities. The upper-atmosphere (P $<10^{-4}$ bars) cloud mass exhibits similar dependence on $K_{zz}$---albeit with a magnitude dependent on metallicity. For 100 $\times$ solar metallicity case, the upper-atmospheric cloud mass density rises and drops by $\sim1-2$ orders of magnitude between our fiducial, 0.1 $\times$ and $10\times$ $K_{zz}$ (Figure \ref{fig:kzz_cloudmass}; bottom left). 

\begin{figure*}
    \centering
    \includegraphics[width=1.0\linewidth]{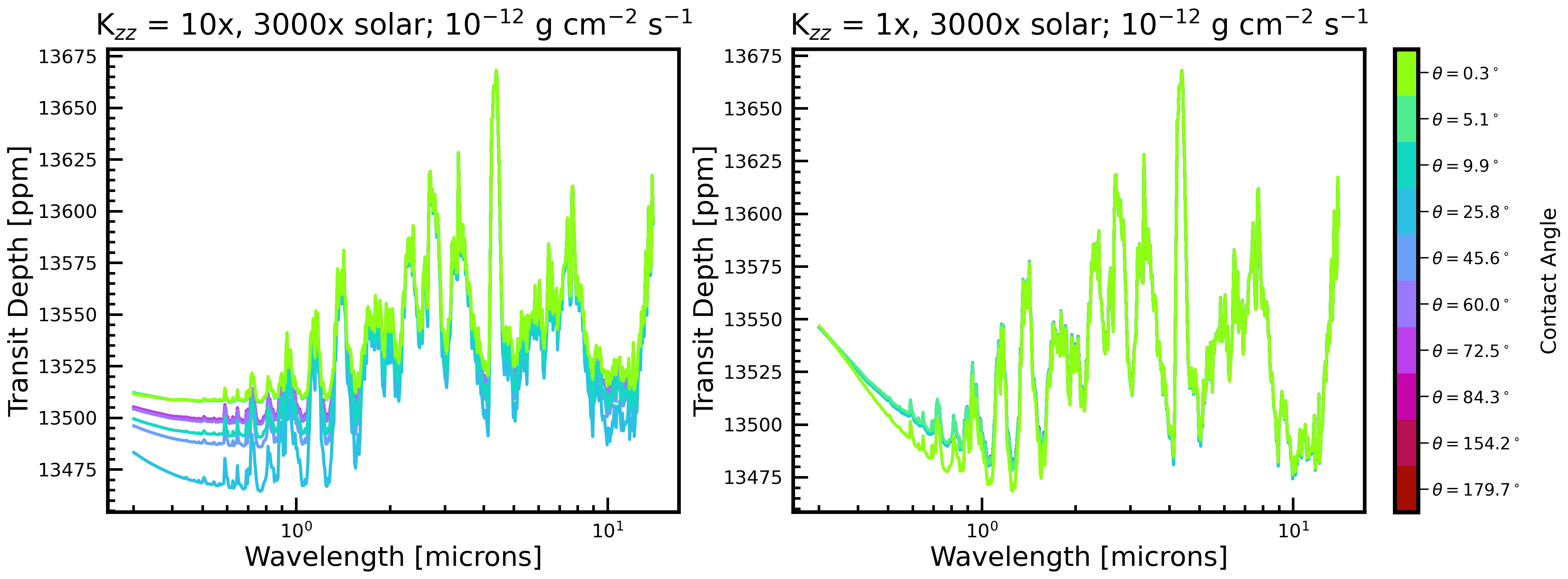}
    \caption{Variations in predicted transmission spectra due to cloud-haze interactions for a 3000 $\times$ solar metallicity atmosphere with a haze production rate of $10^{-12}$ g cm$^2$ s$^{-1}$ for our fiducial (left) and 10 $\times$ $K_{zz}$ (right) profiles. We observe that strong cloud-haze interactions dramatically change the predicted spectrum, in contrast to the lack of variation seen with the fiducial $K_{zz}$ profiles. This demonstrates how the effects of cloud-haze interactions manifest more prominently under stronger vertical mixing. }
    \label{fig:highkzz_spectra}
\end{figure*}

The effects of higher $K_{zz}$ become especially pronounced for the $3000\times$ solar metallicity case, which experiences a $> 3$ order of magnitude increase in the upper-atmosphere cloud mass density (Figure \ref{fig:kzz_cloudmass}; bottom right). This is because higher $K_{zz}$ tends to lead to more vertically extended (rather than compact) cloud layers while also promoting more vigorous cloud formation by enhancing the supply of condensate vapor to growing particles. Enhanced vertical mixing ($K_{zz}$) in high-metallicity atmospheres increases the prevalence of high-altitude clouds, thereby amplifying the role of aerosol opacity in shaping the transmission spectrum relative to fiducial $K_{zz}$ profiles. Illustrating this effect, Figure \ref{fig:highkzz_spectra} shows predicted transmission spectra for 10 $\times$ $K_{zz}$ simulations with 3000 $\times$ solar metallicity and a haze production rate of $10^{-12}$ g cm$^{-2}$ s$^{-1}$. These spectra show significant variations with contact angle, in contrast to the lack of variation exhibited at this metallicity and haze production rate under the fiducial $K_{zz}$ profiles. This additionally underscores how cloud-haze interactions and their impact on transmission spectra are sensitive to the vertical mixing strength. 

\begin{figure*}
    \centering
    \includegraphics[width=1.0\linewidth]{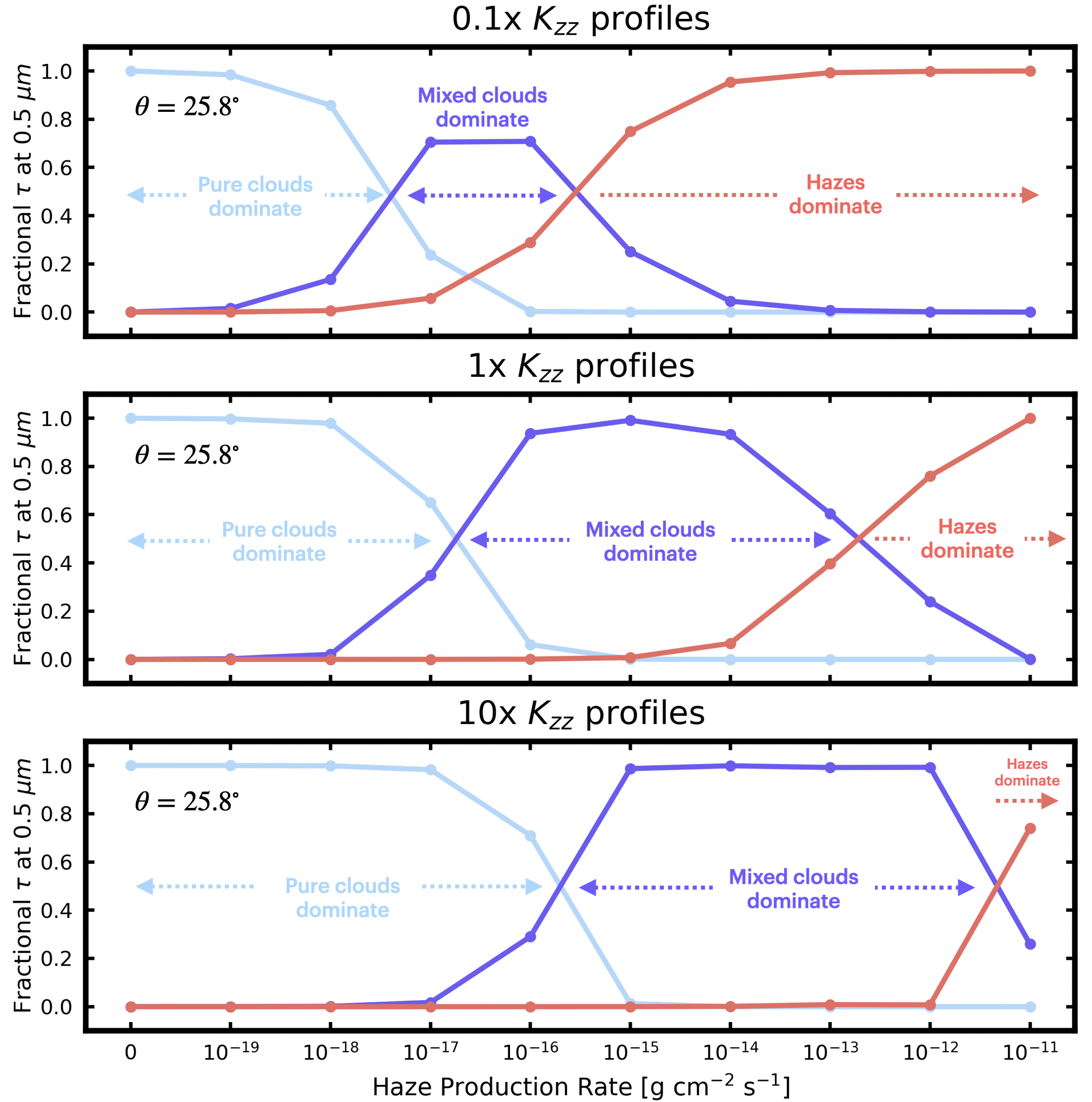}
    \caption{Aerosol fractional optical depth contributions at $0.5\mu$m for a fixed contact angle ($\theta=25.8 \degree$) and metallicity (100 $\times$ solar) for varying levels of vertical mixing. The top, middle, and bottom panels show the cases using our 0.1$\times$, 1$\times$, and 10$\times$ fiducial K$_{zz}$ profiles respectively. Cloud opacity becomes increasingly important in the presence of stronger vertical mixing, with the opposite trend exhibited by haze. }
    \label{fig:kzz_fractional_opd}
\end{figure*}

Variations in the strength of vertical mixing also change which aerosol dominates the opacity contributed in transmission. Holding the metallicity (100$\times$ solar) and contact angle ($\theta=25.8\degree$) fixed, we show how the choice of  $K_{zz}$ shifts the regimes in which pure clouds, mixed clouds, and hazes dominate the upper-atmosphere's opacity at $0.5 \mu$m in Figure \ref{fig:kzz_fractional_opd}. Since cloud formation, growth, and transport to the upper atmosphere all increase with stronger mixing, larger $K_{zz}$ profiles increase the opacity contributed by clouds in the upper atmosphere. Concurrently, more efficient vertical mixing impedes haze coagulative growth by transporting particles to greater depth before they have time to grow, decreasing the opacity contributed by haze in the optical \& near-infrared \citep[e.g.][]{lavvas_koskinen, kawashima_2018,ohno_kawashima_super_rayleigh, gao_zhang_superpuff}. Thus, the overall effect of higher $K_{zz}$ is to enable clouds to dominate over hazes up to much larger magnitudes of haze production, and vice-versa. Providing one illustrative example of this general phenomenon, we see from Figure \ref{fig:kzz_fractional_opd} that the transition between cloud-opacity and haze-opacity dominated regimes occurs at $10^{-15}$ g cm$^{-2}$ s$^{-1}$ for the 0.1 $\times$ $K_{zz}$ case but only at $10^{-11}$ g cm$^{-2}$ s$^{-1}$ for the 10$\times$ $K_{zz}$ case.

\section{Discussion} \label{sec:discussion}

\subsection{What is a realistic contact angle?} \label{realistic_contact_angle_section}

Motivated by the significant uncertainties that remain in our understanding of exoplanet hazes, condensate clouds, and their material properties, we surveyed contact angles ranging from $\sim 0^{\circ}$ to $\sim 180^{\circ}$ and found that different values drive qualitatively distinct transformations in aerosol size distributions and observables. A natural follow-up question is therefore what contact angles are physically realistic for clouds and hazes in sub-Neptunian atmospheres.

As discussed above, \citet{xinting_yu_2021} used surface energy measurements to constrain the contact angle of KCl on lab-synthesized hazes \citep{chao_he_haze_formation} to between $0^{\circ}$ and $77^{\circ}$ — a range that spans most of the behavior mapped out in this work, from haze rainout to cloud-dominated profiles. Terrestrial analogues offer additional guidance: surface energy measurements of inorganic and organic CCN candidates on Earth (e.g.\ NaCl, $\mathrm{(NH_4)_2SO_4}$, $\mathrm{C_5H_9NO_4}$) yield contact angles near $0^{\circ}$ with water \citep{raymond_cloud_activation}. Surface-active substances common in Earth's atmosphere, such as sea-salt spray, can further reduce surface energies and enhance CCN efficiency \citep[e.g.][]{kleinheins_seasalt_ccn}; analogous processes in exoplanet atmospheres may similarly drive contact angles toward lower values.

Beyond Earth, \citet{yu_tholin_surface_energy} measured surface energies of Titan-like tholins and predicted contact angles with respect to methane ($\mathrm{CH_4}$) and ethane ($\mathrm{C_2H_6}$) — both condensates on Titan. In both cases the predicted contact angles were small: estimates for ethane range from $14.7^{\circ}$ to $28.8^{\circ}$, while methane yields near-perfect nucleation efficiency with $\theta \sim 0^{\circ}$. These low values are qualitatively consistent with other Titan-focused laboratory studies \citep[e.g.][]{curtis_2008_lab}.

Taken together, Solar System evidence suggests that virtually the full range of contact angles explored in this work are a priori plausible on exoplanets. This motivates dedicated laboratory experiments to characterize haze formation, material properties, and cloud-haze contact angles across the range of compositions relevant to warm sub-Neptune atmospheres.

\subsection{Cloud-Haze Interactions and Heterogeneous Nucleation in Different Regimes}

While we framed this paper to focus on the specific case of KCl clouds nucleating upon hazes in warm sub-Neptunes such as GJ 1214 b, the general microphysics of clouds nucleating on haze is applicable to a much broader range of planets. For example, cloud-haze interactions may play an important role in mediating cloud formation on Hot Jupiters, in which silicate clouds may nucleate upon hazes \citep{arfaux_cloudhaze, gao_aerosol_composition} whereas CH$_4$ nucleation on haze at much cooler temperatures has been suggested as a key process setting the structure of Uranus and Neptune's clouds \citep[e.g.][]{carlson_giant_planet_microphysics,irwin_hazy_blue_worlds}. 

Furthermore, the physical pathway we explore is also more generally applicable to any scenario in which particles raining down from the top of an atmosphere serve as CCN. Examples include the infall of dust---which has been suggested as a CCN for Earth, Venus, and exoplanets \citep[e.g.][]{petty_textbook, arras_dust_accretion,mang_waterclouds_meteoritic_dust, mang_2024_parameterizations}---as well as that from a planetary ring, as hypothesized for Saturn \citep{hsu_ring_rain_saturn}. Given how our results indicate that cloud-haze interactions cause larger relative changes for the lower-metallicity atmospheres typical of giant planets \citep[e.g.][]{thorngren_mass_metallicity,welbanks_trends, chachan_2025}, follow up studies dedicated to understanding cloud-haze interactions and their observational implications for giant planets across a wide temperature range are well motivated.

\subsection{Effects of Haze Optical Properties and Shapes}

In the modeling in this paper, we assumed perfectly spherical aerosol particles. Real exoplanet aerosols, however, likely span a wide range of morphologies. Fractal aggregate particles, for example, are common across the Solar System \citep[e.g.][]{west_smith_1991_aggregates, xi_aerosol_influence, gao_pluto_aggregate, fan_pluto_bimodal}. They have additionally attracted substantial interest in an exoplanet context due to the fact that aggregates can grow to larger particle sizes before sedimentation, potentially allowing them help explain highly muted transmission spectra \citep[e.g.][]{adams_aggregate_hazes, ohno_aggregates, lodge_2024a,lodge_2024b,moran_virga,vahidinia_moran_aggregates, lodge_virga}. How cloud-haze interactions manifest in scenarios where either the clouds or hazes (or both) are aggregates remains, as of now, unclear. Lab experiments suggest that surface roughness may significantly decrease the contact angle relative to cases with spherical particles with smooth surfaces \citep{swain_cassie_wenzel,razavifar_roughness_contactangle}. If this is true, it may imply that fractal aggregate haze particles may enable increased heterogeneous nucleation, improving their ability to serve as CCN through a potential `inverse Kelvin Effect' \citep[e.g. ][]{SJOGREN2007157,ice_pores} that occurs when the CCN surface is concave rather than convex. In this scenario, the equilibrium vapor pressure drops below the flat-surface value, which may enable condensation to occur even at sub-saturation ($S<1$). While a more detailed exploration of this possibility is out of this study's scope, investigating the microphysics of cloud-haze interactions with aggregate and porous particles is a promising avenue for future work. 

Though we assumed Titan-like tholin optical properties in this work, the composition and diversity of sub-Neptunian hazes remains poorly understood. Beyond tholins, viable candidates include hydrocarbon soots, graphite \citep{li_he_graphite}, diamonds formed through chemical vapor deposition \citep{ohno_diamond}, and sulfur-based hazes \citep{gao_sulfur_haze, chao_he_sulfur, ohno_24_gj1214b, lavvas2024}. Future photochemical modeling — aimed at identifying dominant haze formation pathways and quantifying production rates across a diverse range of warm sub-Neptune atmospheres — will help clarify the relative importance of clouds and hazes in these environments. Complementary lab experiments measuring the optical properties of varied haze samples \citep[e.g.][]{corrales_2023_hazes_c/o_ratio,huseby_hazes_uv, pesciotta_hazes} will be essential for predicting and interpreting exoplanet spectra; measurements of surface energies will further allow us to constrain contact angles and assess how cloud-haze interactions may vary across the sub-Neptunian population.

\subsection{Moving Beyond One-Dimensional Models}

Though one-dimensional models serve as a good starting point for understanding the aerosol distributions in an atmosphere, they cannot capture the effects of key processes such as horizontal advection resulting from planetary-scale heat transport. This process has been shown to have significant implications for the cloud distributions on hot Jupiters by \cite{diana2D}, who find that the total cloud mass in the upper atmosphere for these planets increases when considering the effects of horizontal advection on microphysical cloud formation. Sub-Neptunes may also show similar effects when we account for transport via horizontal advection, which could alter the expected particle size distributions of aerosols under the cloud-haze interactions scenario.  Furthermore, since photochemical hazes are expected to predominantly form on a planet's dayside, modeling their formation, transport, and growth in two and three-dimensional settings \citep[e.g. ][]{steinrueck_2021, mak_2025_3d} while explicitly accounting for the microphysics of these processes will be key for understanding the degree to which longitudinal transport affects the abundance and distribution of haze cover on sub-Neptunes.

\subsection{Observational Strategies and Implications}

A key open question is how clouds, photochemical hazes, and cloud-haze interactions manifest across the broader sub-Neptunian population. Despite the emerging diversity of this population in bulk atmospheric properties \citep[e.g.][]{roy_diversity, caroline_steam, benneke_toi270d}, detailed microphysical aerosol modeling has largely been carried out in the specific context of GJ 1214 b. As JWST delivers data for sub-Neptunes across a wider range of temperatures and gravities, there is a clear need to expand our aerosol modeling accordingly. For example, GCMs run by \cite{charnay_k218b} for K2-18b suggest that the abundance and distribution of water clouds on this temperate sub-Neptune is highly sensitive to CCN number density. Given the importance of water cloud condensation for modulating the energy balance and upper-atmosphere chemistry of temperate sub-Neptunes — and the immense astrobiological interest in this planet class — understanding cloud-haze interactions in this regime is especially pressing.
A population-level understanding of sub-Neptunian aerosols will also be essential for interpreting observations across the board, from featureless spectra to those with discernible features. As subtle spectral properties in panchromatic data are increasingly used to infer atmospheric and bulk planetary properties, disentangling the effects of aerosols \citep[e.g.,][]{liu_rouhan_k218b} will be key for placing such inferences on firmer footing.

Finally, whether cloud-haze interactions produce a unique observational signature remains an open and interesting question. Recent theoretical work has suggested that heterogeneous aerosols may have distinct optical properties and  features in transmission \citep[e.g][]{kiefer_heterogeneous, ohno_diamond}, which could present one avenue to search for observational signatures of cloud-haze interactions. However, given the complexities involved in identifying dominant CCN populations even on Earth, smoking-gun observational diagnostics are probable to remain difficult in the near future.  On longer timescales, transmission and reflected-light observations by future missions such as the Extremely Large Telescope (ELT) and planned Habitable Worlds Observatory \citep[][HWO]{HWO_pathways}, including polarimetry \citep{hamill_polarimetry}, may offer the most discriminating power between different types of aerosols.

\section{Summary \& Conclusions} \label{sec:conclusions}

%% Please use the acknowledgment and contribution environments. This will 
%% be anonomyized when the "anonymous" style option is used.

In this work, we have presented a new microphysical bin-scheme model for clouds, hazes, and their interactions in exoplanet atmospheres. Based on the \texttt{CARMA} framework for simulating aerosols in planetary atmospheres, our approach considers the formation of clouds atop haze `seeds' through the microphysical pathway of heterogeneous nucleation. Known to regulate cloud formation on Earth and across the Solar System, heterogeneous nucleation of this kind is a general process that should likewise exert a first-order influence on aerosols across the exoplanet population. 

We focused on the case of warm sub-Neptunes such as GJ 1214 b in this paper, motivated by the prevalence of aerosols in this population. Our results indicate that interactions between clouds and photochemical hazes substantially transform aerosol distributions on warm sub-Neptunes, with the strength of the effect governed by the microphysical contact angle $\theta$, the relative abundances of hazes and condensible vapor, and the strength of vertical mixing. Our main conclusions are as follows:

\begin{enumerate}
    \item Cloud-haze interactions through heterogeneous nucleation can dramatically transform the size distributions and vertical extent of clouds in exoplanet atmospheres across many orders of magnitude in bulk atmospheric properties such as metallicity and haze production rate. While we focus on KCl clouds here, this process likely affects other condensate species in colder and warmer atmospheres as well. The strength of this effect is strongly dependent on the microphysical contact angle between cloud and haze particles—a quantity that remains largely unconstrained for KCl and most other likely condensates, motivating future laboratory measurements.

    \item  Cloud-haze interactions become apparent at contact angles $\lessapprox 70\degree$. For moderate values ($25\degree \lessapprox \theta \lessapprox 70\degree$), the largest haze particles act as cloud condensation nuclei (CCN), seeding clouds that grow by condensation and gravitationally settle to depth. This `wet removal' of hazes --- analogous to the scenario posited by \citet{xinting_yu_2021} --- clears the upper atmosphere of aerosol opacity.

    \item At contact angles $\lessapprox 25\degree$, smaller and more abundant haze particles are activated as CCN, facilitating accelerated cloud formation through heterogeneous nucleation. This produces an abundant population of `mixed' clouds with larger average particle sizes at high altitudes than clouds formed through homogeneous nucleation. Strong cloud-haze interactions thus lead to a cloud enhancement regime in which the total aerosol opacity is typically enhanced. 

    \item The effects of cloud-haze interactions are more prominent for low to moderate (1-100 $\times$ solar) metallicities and become less important at higher metallicities (1000-3000 $\times$ solar).

    \item Intermediate haze production rates ($10^{-16}$ to $10^{-12}$ g cm$^{-2}$ s$^{-1}$) typically see the maximal effects of cloud-haze interactions. At lower production rates, the haze abundance is too low to significantly alter cloud formation  and clouds dominate. At higher rates, hazes dominate regardless. Stronger vertical mixing shifts this sweet spot to higher haze production rates, whereas weaker mixing shifts it to lower rates. 

    \item Transmission spectra are strongly affected at wavelengths shorter than $\sim 2.5$ microns, with variations in contact angle alone producing up to $\sim 4$ scale height changes in near-infrared features such as the 1.4 $\mu$m water absorption feature. This holds key implications for optical-to-near-infrared observations by instruments such as JWST NIRISS/SOSS and optical ground-based facilities, and we advocate caution when using such short-wavelength features as population-level diagnostics for bulk atmospheric properties.

    \item Wavelengths longer than $\sim 3$ $\mu$m are less affected by cloud-haze interactions, since upper-atmosphere aerosols in our models rarely reach the large particle sizes needed to significantly affect longer wavelengths. While strong vertical mixing can occasionally promote sufficient aerosol formation to affect these wavelengths, this result reinforces the idea that observations of sub-Neptunes at $\gtrsim 3\ \mu$m are more sensitive to mean molecular weight than to aerosols.

    \item The impact of cloud-haze interactions is highly sensitive to the strength of vertical mixing. Illustrating this, factor of 10 changes in $K_{zz}$ can shift the regimes in which pure clouds, mixed clouds, and hazes dominate the aerosol opacity by multiple orders of magnitude in haze production rate. 
    
\end{enumerate}

Our study establishes cloud-haze interactions as a microphysical pathway that induces dramatic changes in the clouds of warm sub-Neptunes, with significant wavelength-dependent consequences for observables such as transmission spectra. This is just one example demonstrating how the microphysics of aerosols can strongly influence key macro-scale properties of an atmosphere. As the study of exoplanet atmospheres progresses deeper into the era of JWST and we look towards upcoming facilities such as Ariel, the ELTs, and the planned Habitable Worlds Observatory, accounting for the detailed effects of aerosols and their microphysics is likely to become critical for interpreting  observations and advancing our knowledge of exoplanets and their atmospheres. 

\begin{acknowledgments}

\end{acknowledgments}

This work benefited from the 2025 Exoplanet Summer Program in the Other Worlds Laboratory (OWL) at the University of California, Santa Cruz, a program funded by the Heising-Simons Foundation and NASA. 

VN acknowledges support from the National Science Foundation Graduate Research Fellowship under Grant No. DGE 2140001. MS was supported through a 51 Pegasi b fellowship by the Heising-Simons Foundation. GCM simulations were performed at the Max Planck Computing and Data Facility. We would like to thank Kazumasa Ohno, Xi Zhang, Panayotis Lavvas, Ruth Murray-Clay, Xinting Yu, and members of the UChicago Exoplanet Atmospheres group for useful exchanges.  

\begin{contribution}

VN led the writing of this paper and the development of the coupled cloud-haze microphysical model. MS, DP, and PG co-led the development of the model and supervised the writing of this paper. MS also ran the GCM grid whose output was used to initialize our \texttt{CARMA} models. DS and WC contributed to model development, spectral post-processing, and helped edit this paper.
%%This section gives authors the space to recognize author contributions. The text inside this environment is NOT counted towards the total word quanta. At a minimum, manuscripts are expected to include this text:

%% But authors are expected to provide more specific details, e.g. 
%%
%%SC was responsible for writing and submitting the manuscript.
%%WWM came up with the initial research concept and edited the manuscript.
%%OTS obtained the funding and edited the manuscript.
%%EBF provided the formal analysis and validation. He also edited the manuscript.
%%GEH Supervised the undergraduates, wrote the software and administers the project github and Zenodo repositories.
%%
%% Authors can use the Contributor Role Taxonomy (CRediT) at
%% for ideas on how write a good statement tailored to their needs.

\end{contribution}


% \facilities{HST(STIS), Swift(XRT and UVOT), AAVSO, CTIO:1.3m, CTIO:1.5m, CXO}

\software{\texttt{astropy} \citep{astropy2013, astropy2022}, \texttt{numpy} \citep{numpy2020}, \texttt{scipt} \citep{scipy2020}, \texttt{CARMA} \citep{turcoOneDimensionalModelDescribing1979, toonMultidimensionalModelAerosols1988, gao_benneke, diana_silicate}, \texttt{PICASO} \citep{batalha_picaso_2019, mukherjee_picaso, mang_picaso4}, \texttt{pyMieScat} \citep{pymiescat}
          }


\appendix \label{appendix}

\begin{figure}
    \centering
    \includegraphics[width=1.0\linewidth]{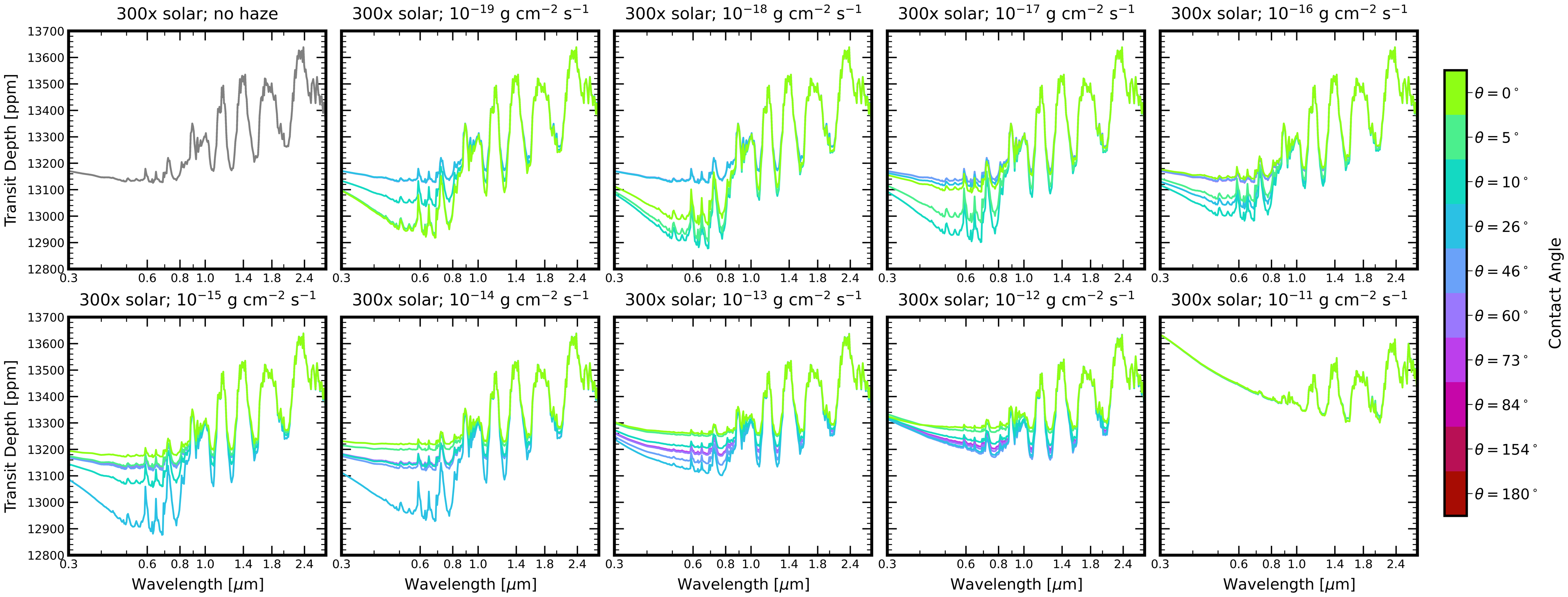}
    \caption{Analogous to Figure \ref{fig:1xsolar_contactangle_spectra} but for a metallicity of 300 $\times$ solar. }
    \label{fig:300xsolar_contactangle_spectra}
    \end{figure}

\begin{figure}
    \centering
    \includegraphics[width=1.0\linewidth]{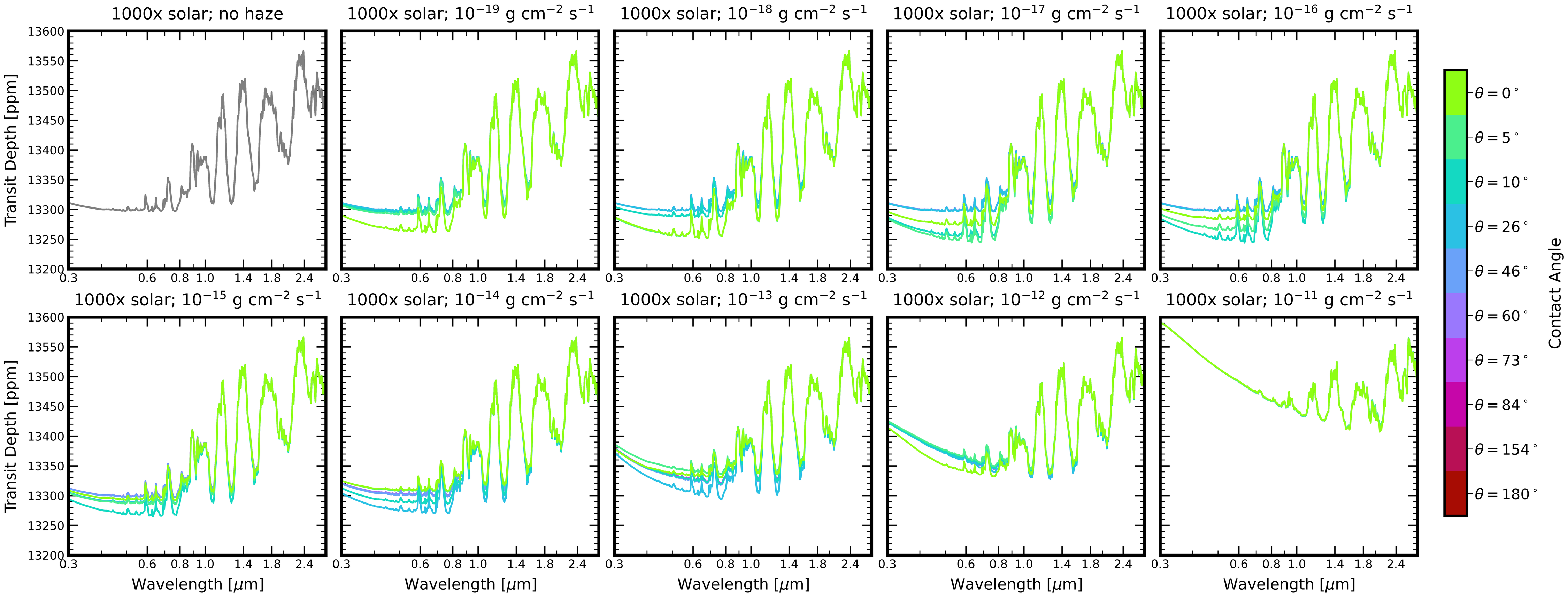}
    \caption{Analogous to Figure \ref{fig:1xsolar_contactangle_spectra} but for a metallicity of 1000 $\times$ solar. }
    \label{fig:1000xsolar_contactangle_spectra}
    \end{figure}

In this Appendix, we show the transmission spectra for the 300 and 1000$\times$solar metallicity simulations in Figures \ref{fig:300xsolar_contactangle_spectra} and \ref{fig:1000xsolar_contactangle_spectra} respectively.

% \begin{figure*}
%     \centering
%     \includegraphics[width=1.0\linewidth]{figures/noninteracting_clouds_vs_haze_spectra.png}
%     \caption{Transmission spectra for a subset of haze production rates and 100 $\times$ solar metallicity highlighting the relative contributions of pure KCl clouds (light blue) and hazes (orange) for cases with minimal cloud-haze interactions. Pure KCl clouds dominate the aerosol opacity until a haze production rate of $\sim 10^{-13}$ g cm$^2$ s$^{-1}$, above which hazes take over.}
%     \label{fig:noninteracting_cloud_haze_spectra}
% \end{figure*}

% \begin{figure*}
%     \centering
%     \includegraphics[width=1.0\linewidth]{figures/stronginteracting_clouds_vs_haze_spectra.png}
%     \caption{Analogous to Figure \ref{fig:noninteracting_cloud_haze_spectra} but for the strongly interacting case with $\theta = 0.3\degree$, with the contribution of mixed KCl clouds highlighted instead of pure KCl clouds. We see that strong cloud-haze interactions enable cloud opacity to remain dominant until the haze production rate reaches $\sim 10^{-11}$ g cm$^2$ s$^{-1}$.}
%     \label{fig:stronginteracting_cloud_haze_spectra}
% \end{figure*}

%% For this sample we use BibTeX plus aasjournalv7.bst to generate the

\bibliography{sample7}{}

@ARTICLE{xinting_yu_2021,
       author = {{Yu}, Xinting and {He}, Chao and {Zhang}, Xi and {H{\"o}rst}, Sarah M. and {Dymont}, Austin H. and {McGuiggan}, Patricia and {Moses}, Julianne I. and {Lewis}, Nikole K. and {Fortney}, Jonathan J. and {Gao}, Peter and {Kempton}, Eliza M. -R. and {Moran}, Sarah E. and {Morley}, Caroline V. and {Powell}, Diana and {Valenti}, Jeff A. and {Vuitton}, V{\'e}ronique},
        title = "{Haze evolution in temperate exoplanet atmospheres through surface energy measurements}",
      journal = {Nature Astronomy},
         year = 2021,
        month = jul,
       volume = {5},
        pages = {822-831},
          doi = {10.1038/s41550-021-01375-3},
archivePrefix = {arXiv},
       eprint = {2107.07069},
 primaryClass = {astro-ph.EP},
       adsurl = {https://ui.adsabs.harvard.edu/abs/2021NatAs...5..822Y}
}

@ARTICLE{hu_k218b,
       author = {{Hu}, Renyu and {Bello-Arufe}, Aaron and {Tokadjian}, Armen and {Yang}, Jeehyun and {Damiano}, Mario and {Roy}, Pierre-Alexis and {Coulombe}, Louis-Philippe and {Madhusudhan}, Nikku and {Constantinou}, Savvas and {Benneke}, Bj{\"o}rn},
        title = "{A water-rich interior in the temperate sub-Neptune K2-18 b revealed by JWST}",
      journal = {arXiv e-prints},
         year = 2025,
        month = jul,
          eid = {arXiv:2507.12622},
        pages = {arXiv:2507.12622},
          doi = {10.48550/arXiv.2507.12622},
archivePrefix = {arXiv},
       eprint = {2507.12622},
 primaryClass = {astro-ph.EP},
       adsurl = {https://ui.adsabs.harvard.edu/abs/2025arXiv250712622H}
}

@ARTICLE{chao_he_haze_formation,
       author = {{He}, Chao and {H{\"o}rst}, Sarah M. and {Lewis}, Nikole K. and {Yu}, Xinting and {Moses}, Julianne I. and {Kempton}, Eliza M.-R. and {Marley}, Mark S. and {McGuiggan}, Patricia and {Morley}, Caroline V. and {Valenti}, Jeff A. and {Vuitton}, V{\'e}ronique},
        title = "{Photochemical Haze Formation in the Atmospheres of Super-Earths and Mini-Neptunes}",
      journal = {\aj},
         year = 2018,
        month = jul,
       volume = {156},
       number = {1},
          eid = {38},
        pages = {38},
          doi = {10.3847/1538-3881/aac883},
archivePrefix = {arXiv},
       eprint = {1805.10488},
 primaryClass = {astro-ph.EP},
       adsurl = {https://ui.adsabs.harvard.edu/abs/2018AJ....156...38H}
}

@ARTICLE{williamson_npf_ccn,
       author = {{Williamson}, Christina J. and {Kupc}, Agnieszka and {Axisa}, Duncan and {Bilsback}, Kelsey R. and {Bui}, ThaoPaul and {Campuzano-Jost}, Pedro and {Dollner}, Maximilian and {Froyd}, Karl D. and {Hodshire}, Anna L. and {Jimenez}, Jose L. and {Kodros}, John K. and {Luo}, Gan and {Murphy}, Daniel M. and {Nault}, Benjamin A. and {Ray}, Eric A. and {Weinzierl}, Bernadett and {Wilson}, James C. and {Yu}, Fangqun and {Yu}, Pengfei and {Pierce}, Jeffrey R. and {Brock}, Charles A.},
        title = "{A large source of cloud condensation nuclei from new particle formation in the tropics}",
      journal = {\nat},
         year = 2019,
        month = oct,
       volume = {574},
       number = {7778},
        pages = {399-403},
          doi = {10.1038/s41586-019-1638-9},
       adsurl = {https://ui.adsabs.harvard.edu/abs/2019Natur.574..399W}
}

@ARTICLE{pruppacher_klett,
       author = {{Pruppacher}, Hans R. and {Klett}, James D.},
        title = "{Microphysics of Clouds and Precipitation}",
      journal = {\nat},
         year = 1980,
        month = mar,
       volume = {284},
       number = {5751},
        pages = {88},
          doi = {10.1038/284088b0},
       adsurl = {https://ui.adsabs.harvard.edu/abs/1980Natur.284...88P}
}

@ARTICLE{caroline_steam,
       author = {{Piaulet-Ghorayeb}, Caroline and {Benneke}, Bj{\"o}rn and {Radica}, Michael and {Raul}, Eshan and {Coulombe}, Louis-Philippe and {Ahrer}, Eva-Maria and {Kubyshkina}, Daria and {Howard}, Ward S. and {Krissansen-Totton}, Joshua and {MacDonald}, Ryan J. and {Roy}, Pierre-Alexis and {Louca}, Amy and {Christie}, Duncan and {Fournier-Tondreau}, Marylou and {Allart}, Romain and {Miguel}, Yamila and {Schlichting}, Hilke E. and {Welbanks}, Luis and {Cadieux}, Charles and {Dorn}, Caroline and {Evans-Soma}, Thomas M. and {Fortney}, Jonathan J. and {Pierrehumbert}, Raymond and {Lafreni{\`e}re}, David and {Acu{\~n}a}, Lorena and {Komacek}, Thaddeus and {Innes}, Hamish and {Beatty}, Thomas G. and {Cloutier}, Ryan and {Doyon}, Ren{\'e} and {Gagnebin}, Anna and {Gapp}, Cyril and {Knutson}, Heather A.},
        title = "{JWST/NIRISS Reveals the Water-rich ``Steam World'' Atmosphere of GJ 9827 d}",
      journal = {\apjl},
         year = 2024,
        month = oct,
       volume = {974},
       number = {1},
          eid = {L10},
        pages = {L10},
          doi = {10.3847/2041-8213/ad6f00},
archivePrefix = {arXiv},
       eprint = {2410.03527},
 primaryClass = {astro-ph.EP},
       adsurl = {https://ui.adsabs.harvard.edu/abs/2024ApJ...974L..10P}
}

@ARTICLE{mak_2025_3d,
       author = {{Mak}, Mei Ting and {Sergeev}, Denis E. and {Mayne}, Nathan J. and {Zamyatina}, Maria and {Steinrueck}, Maria E. and {Manners}, James and {H{\'e}brard}, {\'E}ric and {Sing}, David K. and {Kohary}, Krisztian},
        title = "{The impact of different haze types on the atmospheres and observations of hot Jupiters: 3D simulations of HD 189733b, HD 209458b, and WASP-39b}",
      journal = {\mnras},
         year = 2025,
        month = sep,
       volume = {542},
       number = {3},
        pages = {1873-1900},
          doi = {10.1093/mnras/staf1250},
archivePrefix = {arXiv},
       eprint = {2507.20366},
 primaryClass = {astro-ph.EP},
       adsurl = {https://ui.adsabs.harvard.edu/abs/2025MNRAS.542.1873M}
}

@article{owens_orwk,
author = {Owens, D. K. and Wendt, R. C.},
title = {Estimation of the surface free energy of polymers},
journal = {Journal of Applied Polymer Science},
volume = {13},
number = {8},
pages = {1741-1747},
doi = {https://doi.org/10.1002/app.1969.070130815},
url = {https://onlinelibrary.wiley.com/doi/abs/10.1002/app.1969.070130815},
eprint = {https://onlinelibrary.wiley.com/doi/pdf/10.1002/app.1969.070130815},
year = {1969}
}

@ARTICLE{swain_cassie_wenzel,
       author = {{Swain}, P.~S. and {Lipowsky}, Reinhard},
        title = "{Contact angles on heterogeneous surfaces; a new look at Cassie's and Wenzel's laws}",
      journal = {arXiv e-prints},
         year = 1998,
        month = sep,
          eid = {cond-mat/9809089},
        pages = {cond-mat/9809089},
          doi = {10.48550/arXiv.cond-mat/9809089},
archivePrefix = {arXiv},
       eprint = {cond-mat/9809089},
 primaryClass = {cond-mat.soft},
       adsurl = {https://ui.adsabs.harvard.edu/abs/1998cond.mat..9089S}
}

@ARTICLE{fortney_2005,
       author = {{Fortney}, Jonathan J.},
        title = "{The effect of condensates on the characterization of transiting planet atmospheres with transmission spectroscopy}",
      journal = {\mnras},
         year = 2005,
        month = dec,
       volume = {364},
       number = {2},
        pages = {649-653},
          doi = {10.1111/j.1365-2966.2005.09587.x},
archivePrefix = {arXiv},
       eprint = {astro-ph/0509292},
 primaryClass = {astro-ph},
       adsurl = {https://ui.adsabs.harvard.edu/abs/2005MNRAS.364..649F}
}

@ARTICLE{wallack_toi836,
       author = {{Wallack}, Nicole L. and {Batalha}, Natasha E. and {Alderson}, Lili and {Scarsdale}, Nicholas and {Adams Redai}, Jea I. and {Aguichine}, Artyom and {Alam}, Munazza K. and {Gao}, Peter and {Wolfgang}, Angie and {Batalha}, Natalie M. and {Kirk}, James and {L{\'o}pez-Morales}, Mercedes and {Moran}, Sarah E. and {Teske}, Johanna and {Wakeford}, Hannah R. and {Wogan}, Nicholas F.},
        title = "{JWST COMPASS: A NIRSpec/G395H Transmission Spectrum of the Sub-Neptune TOI-836c}",
      journal = {\aj},
         year = 2024,
        month = aug,
       volume = {168},
       number = {2},
          eid = {77},
        pages = {77},
          doi = {10.3847/1538-3881/ad3917},
archivePrefix = {arXiv},
       eprint = {2404.01264},
 primaryClass = {astro-ph.EP},
       adsurl = {https://ui.adsabs.harvard.edu/abs/2024AJ....168...77W}
}

@ARTICLE{teske_toi776c,
       author = {{Teske}, Johanna and {Batalha}, Natasha E. and {Wallack}, Nicole L. and {Kirk}, James and {Wogan}, Nicholas F. and {Gordon}, Tyler A. and {Alam}, Munazza K. and {Aguichine}, Artyom and {Wolfgang}, Angie and {Wakeford}, Hannah R. and {Scarsdale}, Nicholas and {Adams Redai}, Jea and {Moran}, Sarah E. and {L{\'o}pez-Morales}, Mercedes and {Meech}, Annabella and {Gao}, Peter and {Batalha}, Natalie M. and {Alderson}, Lili and {Gagnebin}, Anna},
        title = "{JWST COMPASS: NIRSpec/G395H Transmission Observations of TOI-776 c, a 2 R$_{{\ensuremath{\oplus}}}$ M Dwarf Planet}",
      journal = {\aj},
         year = 2025,
        month = may,
       volume = {169},
       number = {5},
          eid = {249},
        pages = {249},
          doi = {10.3847/1538-3881/adb975},
archivePrefix = {arXiv},
       eprint = {2502.20501},
 primaryClass = {astro-ph.EP},
       adsurl = {https://ui.adsabs.harvard.edu/abs/2025AJ....169..249T}
}

@ARTICLE{horst_experiments,
       author = {{H{\"o}rst}, Sarah M. and {He}, Chao and {Lewis}, Nikole K. and {Kempton}, Eliza M.-R. and {Marley}, Mark S. and {Morley}, Caroline V. and {Moses}, Julianne I. and {Valenti}, Jeff A. and {Vuitton}, V{\'e}ronique},
        title = "{Haze production rates in super-Earth and mini-Neptune atmosphere experiments}",
      journal = {Nature Astronomy},
         year = 2018,
        month = mar,
       volume = {2},
        pages = {303-306},
          doi = {10.1038/s41550-018-0397-0},
archivePrefix = {arXiv},
       eprint = {1801.06512},
 primaryClass = {astro-ph.EP},
       adsurl = {https://ui.adsabs.harvard.edu/abs/2018NatAs...2..303H}
}

@ARTICLE{squires_1958,
       author = {{Squires}, P.},
        title = "{The Microstructure and Colloidal Stability of Warm Clouds}",
      journal = {Tellus},
         year = 1958,
        month = may,
       volume = {10},
       number = {2},
        pages = {256-261},
          doi = {10.1111/j.2153-3490.1958.tb02011.x10.3402/tellusa.v10i2.9229},
       adsurl = {https://ui.adsabs.harvard.edu/abs/1958Tell...10..256S}
}

@ARTICLE{lin_ice_nucleation,
       author = {{Lin}, Lin and {Liu}, Xiaohong and {Zhao}, Xi and {Shan}, Yunpeng and {Ke}, Ziming and {Lyu}, Kai and {Bowman}, Kenneth P.},
        title = "{Ice nucleation by volcanic ash greatly alters cirrus cloud properties}",
      journal = {Science Advances},
         year = 2025,
        month = may,
       volume = {11},
       number = {19},
          eid = {eads0572},
        pages = {eads0572},
          doi = {10.1126/sciadv.ads0572},
       adsurl = {https://ui.adsabs.harvard.edu/abs/2025SciA...11S.572L}
}

@ARTICLE{albrecht_report,
       author = {{Albrecht}, Bruce A.},
        title = "{Aerosols, Cloud Microphysics, and Fractional Cloudiness}",
      journal = {Science},
         year = 1989,
        month = sep,
       volume = {245},
       number = {4923},
        pages = {1227-1230},
          doi = {10.1126/science.245.4923.1227},
       adsurl = {https://ui.adsabs.harvard.edu/abs/1989Sci...245.1227A}
}

@ARTICLE{chao_he_particle_colors,
       author = {{He}, Chao and {H{\"o}rst}, Sarah M. and {Lewis}, Nikole K. and {Yu}, Xinting and {Moses}, Julianne I. and {Kempton}, Eliza M.-R. and {McGuiggan}, Patricia and {Morley}, Caroline V. and {Valenti}, Jeff A. and {Vuitton}, V{\'e}ronique},
        title = "{Laboratory Simulations of Haze Formation in the Atmospheres of Super-Earths and Mini-Neptunes: Particle Color and Size Distribution}",
      journal = {\apjl},
         year = 2018,
        month = mar,
       volume = {856},
       number = {1},
          eid = {L3},
        pages = {L3},
          doi = {10.3847/2041-8213/aab42b},
archivePrefix = {arXiv},
       eprint = {1803.01706},
 primaryClass = {astro-ph.EP},
       adsurl = {https://ui.adsabs.harvard.edu/abs/2018ApJ...856L...3H}
}

@ARTICLE{venus_peter,
       author = {{Gao}, Peter and {Zhang}, Xi and {Crisp}, David and {Bardeen}, Charles G. and {Yung}, Yuk L.},
        title = "{Bimodal distribution of sulfuric acid aerosols in the upper haze of Venus}",
      journal = {\icarus},
         year = 2014,
        month = mar,
       volume = {231},
        pages = {83-98},
          doi = {10.1016/j.icarus.2013.10.013},
archivePrefix = {arXiv},
       eprint = {1312.3750},
 primaryClass = {astro-ph.EP},
       adsurl = {https://ui.adsabs.harvard.edu/abs/2014Icar..231...83G}
}

@ARTICLE{charnay_3dclouds,
       author = {{Charnay}, B. and {Meadows}, V. and {Misra}, A. and {Leconte}, J. and {Arney}, G.},
        title = "{3D Modeling of GJ1214b{\textquoteright}s Atmosphere: Formation of Inhomogeneous High Clouds and Observational Implications}",
      journal = {\apjl},
         year = 2015,
        month = nov,
       volume = {813},
       number = {1},
          eid = {L1},
        pages = {L1},
          doi = {10.1088/2041-8205/813/1/L1},
archivePrefix = {arXiv},
       eprint = {1510.01706},
 primaryClass = {astro-ph.EP},
       adsurl = {https://ui.adsabs.harvard.edu/abs/2015ApJ...813L...1C}
}

@ARTICLE{benneke_toi270d,
       author = {{Benneke}, Bj{\"o}rn and {Roy}, Pierre-Alexis and {Coulombe}, Louis-Philippe and {Radica}, Michael and {Piaulet}, Caroline and {Ahrer}, Eva-Maria and {Pierrehumbert}, Raymond and {Krissansen-Totton}, Joshua and {Schlichting}, Hilke E. and {Hu}, Renyu and {Yang}, Jeehyun and {Christie}, Duncan and {Thorngren}, Daniel and {Young}, Edward D. and {Pelletier}, Stefan and {Knutson}, Heather A. and {Miguel}, Yamila and {Evans-Soma}, Thomas M. and {Dorn}, Caroline and {Gagnebin}, Anna and {Fortney}, Jonathan J. and {Komacek}, Thaddeus and {MacDonald}, Ryan and {Raul}, Eshan and {Cloutier}, Ryan and {Acuna}, Lorena and {Lafreni{\`e}re}, David and {Cadieux}, Charles and {Doyon}, Ren{\'e} and {Welbanks}, Luis and {Allart}, Romain},
        title = "{JWST Reveals CH$_4$, CO$_2$, and H$_2$O in a Metal-rich Miscible Atmosphere on a Two-Earth-Radius Exoplanet}",
      journal = {arXiv e-prints},
         year = 2024,
        month = mar,
          eid = {arXiv:2403.03325},
        pages = {arXiv:2403.03325},
          doi = {10.48550/arXiv.2403.03325},
archivePrefix = {arXiv},
       eprint = {2403.03325},
 primaryClass = {astro-ph.EP},
       adsurl = {https://ui.adsabs.harvard.edu/abs/2024arXiv240303325B}
}

@ARTICLE{ackerman_marley,
       author = {{Ackerman}, Andrew S. and {Marley}, Mark S.},
        title = "{Precipitating Condensation Clouds in Substellar Atmospheres}",
      journal = {\apj},
         year = 2001,
        month = aug,
       volume = {556},
       number = {2},
        pages = {872-884},
          doi = {10.1086/321540},
archivePrefix = {arXiv},
       eprint = {astro-ph/0103423},
 primaryClass = {astro-ph},
       adsurl = {https://ui.adsabs.harvard.edu/abs/2001ApJ...556..872A}
}

@ARTICLE{huygens_condensation_lavvas,
       author = {{Lavvas}, P. and {Griffith}, C.~A. and {Yelle}, R.~V.},
        title = "{Condensation in Titan's atmosphere at the Huygens landing site}",
      journal = {\icarus},
         year = 2011,
        month = oct,
       volume = {215},
       number = {2},
        pages = {732-750},
          doi = {10.1016/j.icarus.2011.06.040},
       adsurl = {https://ui.adsabs.harvard.edu/abs/2011Icar..215..732L}
}

@ARTICLE{eliza_gj1214_2012,
       author = {{Miller-Ricci Kempton}, Eliza and {Zahnle}, Kevin and {Fortney}, Jonathan J.},
        title = "{The Atmospheric Chemistry of GJ 1214b: Photochemistry and Clouds}",
      journal = {\apj},
         year = 2012,
        month = jan,
       volume = {745},
       number = {1},
          eid = {3},
        pages = {3},
          doi = {10.1088/0004-637X/745/1/3},
archivePrefix = {arXiv},
       eprint = {1104.5477},
 primaryClass = {astro-ph.EP},
       adsurl = {https://ui.adsabs.harvard.edu/abs/2012ApJ...745....3M}
}

@article{kelvin_effect_1872, title={4. On the Equilibrium of Vapour at a Curved Surface of Liquid}, volume={7}, DOI={10.1017/S0370164600041729}, journal={Proceedings of the Royal Society of Edinburgh}, author={Thomson, William}, year={1872}, pages={63–68}}

@ARTICLE{lodders_alkali, 
       author = {{Lodders}, Katharina},
        title = "{Alkali Element Chemistry in Cool Dwarf Atmospheres}",
      journal = {\apj},
         year = 1999,
        month = jul,
       volume = {519},
       number = {2},
        pages = {793-801},
          doi = {10.1086/307387},
       adsurl = {https://ui.adsabs.harvard.edu/abs/1999ApJ...519..793L}
}

@book{davidson_1993,
  author       = {Davidson, T A},
  title        = {A simple and accurate method for calculating viscosity of gaseous mixtures},
  url          = {https://www.osti.gov/biblio/6129940},
  place        = {United States},
  publisher    = {Pittsburgh, PA (United States); Bureau of Mines},
  year         = {1993},
  month        = {01}}

@ARTICLE{morley_gj1214b,
       author = {{Morley}, Caroline V. and {Fortney}, Jonathan J. and {Kempton}, Eliza M.-R. and {Marley}, Mark S. and {Visscher}, Channon and {Zahnle}, Kevin},
        title = "{Quantitatively Assessing the Role of Clouds in the Transmission Spectrum of GJ 1214b}",
      journal = {\apj},
         year = 2013,
        month = sep,
       volume = {775},
       number = {1},
          eid = {33},
        pages = {33},
          doi = {10.1088/0004-637X/775/1/33},
archivePrefix = {arXiv},
       eprint = {1305.4124},
 primaryClass = {astro-ph.EP},
       adsurl = {https://ui.adsabs.harvard.edu/abs/2013ApJ...775...33M}
}

@ARTICLE{roy_diversity,
       author = {{Roy}, Pierre-Alexis and {Benneke}, Bj{\"o}rn and {Fournier-Tondreau}, Marylou and {Coulombe}, Louis-Philippe and {Piaulet-Ghorayeb}, Caroline and {Lafreni{\`e}re}, David and {Allart}, Romain and {Cowan}, Nicolas B. and {Dang}, Lisa and {Johnstone}, Doug and {Langeveld}, Adam B. and {Pelletier}, Stefan and {Radica}, Michael and {Taylor}, Jake and {Albert}, Lo{\"\i}c and {Doyon}, Ren{\'e} and {Flagg}, Laura and {Jayawardhana}, Ray and {MacDonald}, Ryan J. and {Turner}, Jake D.},
        title = "{Diversity in the haziness and chemistry of temperate sub-Neptunes}",
      journal = {Nature Astronomy},
         year = 2025,
        month = dec,
          doi = {10.1038/s41550-025-02723-3},
archivePrefix = {arXiv},
       eprint = {2512.10876},
 primaryClass = {astro-ph.EP},
       adsurl = {https://ui.adsabs.harvard.edu/abs/2025NatAs.tmp..256R}
}

@ARTICLE{charnay_k218b,
       author = {{Charnay}, B. and {Blain}, D. and {B{\'e}zard}, B. and {Leconte}, J. and {Turbet}, M. and {Falco}, A.},
        title = "{Formation and dynamics of water clouds on temperate sub-Neptunes: the example of K2-18b}",
      journal = {\aap},
         year = 2021,
        month = feb,
       volume = {646},
          eid = {A171},
        pages = {A171},
          doi = {10.1051/0004-6361/202039525},
archivePrefix = {arXiv},
       eprint = {2011.11553},
 primaryClass = {astro-ph.EP},
       adsurl = {https://ui.adsabs.harvard.edu/abs/2021A&A...646A.171C}
}

@ARTICLE{fastchem2,
       author = {{Stock}, Joachim W. and {Kitzmann}, Daniel and {Patzer}, A. Beate C.},
        title = "{FASTCHEM 2 : an improved computer program to determine the gas-phase chemical equilibrium composition for arbitrary element distributions}",
      journal = {\mnras},
         year = 2022,
        month = dec,
       volume = {517},
       number = {3},
        pages = {4070-4080},
          doi = {10.1093/mnras/stac2623},
archivePrefix = {arXiv},
       eprint = {2206.08247},
 primaryClass = {astro-ph.EP},
       adsurl = {https://ui.adsabs.harvard.edu/abs/2022MNRAS.517.4070S}
}

@ARTICLE{fastchem_2018,
       author = {{Stock}, Joachim W. and {Kitzmann}, Daniel and {Patzer}, A. Beate C. and {Sedlmayr}, Erwin},
        title = "{FastChem: A computer program for efficient complex chemical equilibrium calculations in the neutral/ionized gas phase with applications to stellar and planetary atmospheres}",
      journal = {\mnras},
         year = 2018,
        month = sep,
       volume = {479},
       number = {1},
        pages = {865-874},
          doi = {10.1093/mnras/sty1531},
archivePrefix = {arXiv},
       eprint = {1804.05010},
 primaryClass = {astro-ph.EP},
       adsurl = {https://ui.adsabs.harvard.edu/abs/2018MNRAS.479..865S}
}

@ARTICLE{fastchem_cond,
       author = {{Kitzmann}, Daniel and {Stock}, Joachim W. and {Patzer}, A. Beate C.},
        title = "{FASTCHEM COND: equilibrium chemistry with condensation and rainout for cool planetary and stellar environments}",
      journal = {\mnras},
         year = 2024,
        month = jan,
       volume = {527},
       number = {3},
        pages = {7263-7283},
          doi = {10.1093/mnras/stad3515},
archivePrefix = {arXiv},
       eprint = {2309.02337},
 primaryClass = {astro-ph.EP},
       adsurl = {https://ui.adsabs.harvard.edu/abs/2024MNRAS.527.7263K}
}

@ARTICLE{morley_neglected_clouds,
       author = {{Morley}, Caroline V. and {Fortney}, Jonathan J. and {Marley}, Mark S. and {Visscher}, Channon and {Saumon}, Didier and {Leggett}, S.~K.},
        title = "{Neglected Clouds in T and Y Dwarf Atmospheres}",
      journal = {\apj},
         year = 2012,
        month = sep,
       volume = {756},
       number = {2},
          eid = {172},
        pages = {172},
          doi = {10.1088/0004-637X/756/2/172},
archivePrefix = {arXiv},
       eprint = {1206.4313},
 primaryClass = {astro-ph.SR},
       adsurl = {https://ui.adsabs.harvard.edu/abs/2012ApJ...756..172M}
}

@ARTICLE{gao_pluto_aggregate,
       author = {{Gao}, Peter and {Fan}, Siteng and {Wong}, Michael L. and {Liang}, Mao-Chang and {Shia}, Run-Lie and {Kammer}, Joshua A. and {Yung}, Yuk L. and {Summers}, Michael E. and {Gladstone}, G. Randall and {Young}, Leslie A. and {Olkin}, Catherine B. and {Ennico}, Kimberly and {Weaver}, Harold A. and {Stern}, S. Alan and {New Horizons Science Team}},
        title = "{Constraints on the microphysics of Pluto's photochemical haze from New Horizons observations}",
      journal = {\icarus},
         year = 2017,
        month = may,
       volume = {287},
        pages = {116-123},
          doi = {10.1016/j.icarus.2016.09.030},
archivePrefix = {arXiv},
       eprint = {1610.01679},
 primaryClass = {astro-ph.EP},
       adsurl = {https://ui.adsabs.harvard.edu/abs/2017Icar..287..116G}
}

@ARTICLE{liu_rouhan_k218b,
       author = {{Liu}, Ruohan and {Lavvas}, Panayotis and {Tinetti}, Giovanna and {Maldonado}, Jesus and {Ma}, Sushuang and {Saba}, Arianna},
        title = "{Hydrocarbon Hazes on Temperate sub-Neptune K2-18b supported by data from the James Webb Space Telescope}",
      journal = {arXiv e-prints},
         year = 2025,
        month = sep,
          eid = {arXiv:2509.10947},
        pages = {arXiv:2509.10947},
          doi = {10.48550/arXiv.2509.10947},
archivePrefix = {arXiv},
       eprint = {2509.10947},
 primaryClass = {astro-ph.EP},
       adsurl = {https://ui.adsabs.harvard.edu/abs/2025arXiv250910947L}
}

@article{turcoOneDimensionalModelDescribing1979,
  title = {A {{One-Dimensional Model Describing Aerosol Formation}} and {{Evolution}} in the {{Stratosphere}}: {{I}}. {{Physical Processes}} and {{Mathematical Analogs}}},
  shorttitle = {A {{One-Dimensional Model Describing Aerosol Formation}} and {{Evolution}} in the {{Stratosphere}}},
  author = {Turco, R. P. and Hamill, P. and Toon, O. B. and Whitten, R. C. and Kiang, C. S.},
  year = 1979,
  month = apr,
  journal = {Journal of the Atmospheric Sciences},
  volume = {36},
  number = {4},
  pages = {699--717},
  issn = {0022-4928, 1520-0469},
  doi = {10.1175/1520-0469(1979)036<0699:AODMDA>2.0.CO;2},
  urldate = {2025-11-18},
  langid = {english}
}

@article{toonMultidimensionalModelAerosols1988,
  title = {A {{Multidimensional Model}} for {{Aerosols}}: {{Description}} of {{Computational Analogs}}},
  shorttitle = {A {{Multidimensional Model}} for {{Aerosols}}},
  author = {Toon, O. B. and Turco, R. P. and Westphal, D. and Malone, R. and Liu, M.},
  year = 1988,
  month = aug,
  journal = {Journal of the Atmospheric Sciences},
  volume = {45},
  number = {15},
  pages = {2123--2144},
  issn = {0022-4928, 1520-0469},
  doi = {10.1175/1520-0469(1988)045<2123:AMMFAD>2.0.CO;2},
  urldate = {2025-08-15},
  langid = {english}
}

@ARTICLE{hsu_ring_rain_saturn,
       author = {{Hsu}, Hsiang-Wen and {Schmidt}, J{\"u}rgen and {Kempf}, Sascha and {Postberg}, Frank and {Moragas-Klostermeyer}, Georg and {Sei{\ss}}, Martin and {Hoffmann}, Holger and {Burton}, Marcia and {Ye}, ShengYi and {Kurth}, William S. and {Hor{\'a}nyi}, Mih{\'a}ly and {Khawaja}, Nozair and {Spahn}, Frank and {Schirdewahn}, Daniel and {O'Donoghue}, James and {Moore}, Luke and {Cuzzi}, Jeff and {Jones}, Geraint H. and {Srama}, Ralf},
        title = "{In situ collection of dust grains falling from Saturn's rings into its atmosphere}",
      journal = {Science},
         year = 2018,
        month = oct,
       volume = {362},
       number = {6410},
          eid = {aat3185},
        pages = {aat3185},
          doi = {10.1126/science.aat3185},
       adsurl = {https://ui.adsabs.harvard.edu/abs/2018Sci...362.3185H}
}

@ARTICLE{bean_2012_gj1214b,
       author = {{Bean}, Jacob L. and {D{\'e}sert}, Jean-Michel and {Kabath}, Petr and {Stalder}, Brian and {Seager}, Sara and {Miller-Ricci Kempton}, Eliza and {Berta}, Zachory K. and {Homeier}, Derek and {Walsh}, Shane and {Seifahrt}, Andreas},
        title = "{The Optical and Near-infrared Transmission Spectrum of the Super-Earth GJ 1214b: Further Evidence for a Metal-rich Atmosphere}",
      journal = {\apj},
         year = 2011,
        month = dec,
       volume = {743},
       number = {1},
          eid = {92},
        pages = {92},
          doi = {10.1088/0004-637X/743/1/92},
archivePrefix = {arXiv},
       eprint = {1109.0582},
 primaryClass = {astro-ph.EP},
       adsurl = {https://ui.adsabs.harvard.edu/abs/2011ApJ...743...92B}
}

@ARTICLE{fraine_spitzer,
       author = {{Fraine}, Jonathan D. and {Deming}, Drake and {Gillon}, Micha{\"e}l and {Jehin}, Emmanu{\"e}l and {Demory}, Brice-Olivier and {Benneke}, Bjoern and {Seager}, Sara and {Lewis}, Nikole K. and {Knutson}, Heather and {D{\'e}sert}, Jean-Michel},
        title = "{Spitzer Transits of the Super-Earth GJ1214b and Implications for its Atmosphere}",
      journal = {\apj},
         year = 2013,
        month = mar,
       volume = {765},
       number = {2},
          eid = {127},
        pages = {127},
          doi = {10.1088/0004-637X/765/2/127},
archivePrefix = {arXiv},
       eprint = {1301.6763},
 primaryClass = {astro-ph.EP},
       adsurl = {https://ui.adsabs.harvard.edu/abs/2013ApJ...765..127F}
}

@ARTICLE{mang_2024_parameterizations,
       author = {{Mang}, James and {Morley}, Caroline V. and {Robinson}, Tyler D. and {Gao}, Peter},
        title = "{Microphysical Prescriptions for Parameterized Water Cloud Formation on Ultra-cool Substellar Objects}",
      journal = {\apj},
         year = 2024,
        month = oct,
       volume = {974},
       number = {2},
          eid = {190},
        pages = {190},
          doi = {10.3847/1538-4357/ad6c4c},
archivePrefix = {arXiv},
       eprint = {2408.08958},
 primaryClass = {astro-ph.EP},
       adsurl = {https://ui.adsabs.harvard.edu/abs/2024ApJ...974..190M}
}

@ARTICLE{thorngren_mass_metallicity,
       author = {{Thorngren}, Daniel P. and {Fortney}, Jonathan J. and {Murray-Clay}, Ruth A. and {Lopez}, Eric D.},
        title = "{The Mass-Metallicity Relation for Giant Planets}",
      journal = {\apj},
         year = 2016,
        month = nov,
       volume = {831},
       number = {1},
          eid = {64},
        pages = {64},
          doi = {10.3847/0004-637X/831/1/64},
archivePrefix = {arXiv},
       eprint = {1511.07854},
 primaryClass = {astro-ph.EP},
       adsurl = {https://ui.adsabs.harvard.edu/abs/2016ApJ...831...64T}
}

@ARTICLE{welbanks_trends,
       author = {{Welbanks}, Luis and {Madhusudhan}, Nikku and {Allard}, Nicole F. and {Hubeny}, Ivan and {Spiegelman}, Fernand and {Leininger}, Thierry},
        title = "{Mass-Metallicity Trends in Transiting Exoplanets from Atmospheric Abundances of H$_{2}$O, Na, and K}",
      journal = {\apjl},
         year = 2019,
        month = dec,
       volume = {887},
       number = {1},
          eid = {L20},
        pages = {L20},
          doi = {10.3847/2041-8213/ab5a89},
archivePrefix = {arXiv},
       eprint = {1912.04904},
 primaryClass = {astro-ph.EP},
       adsurl = {https://ui.adsabs.harvard.edu/abs/2019ApJ...887L..20W}
}

@ARTICLE{chachan_2025,
       author = {{Chachan}, Yayaati and {Fortney}, Jonathan J. and {Ohno}, Kazumasa and {Thorngren}, Daniel and {Murray-Clay}, Ruth},
        title = "{Revising the Giant Planet Mass─Metallicity Relation: Deciphering the Formation Sequence of Giant Planets}",
      journal = {\apj},
         year = 2025,
        month = nov,
       volume = {994},
       number = {1},
          eid = {43},
        pages = {43},
          doi = {10.3847/1538-4357/ae0cbf},
archivePrefix = {arXiv},
       eprint = {2509.20428},
 primaryClass = {astro-ph.EP},
       adsurl = {https://ui.adsabs.harvard.edu/abs/2025ApJ...994...43C}
}

@ARTICLE{charbonneau_gj1214b,
       author = {{Charbonneau}, David and {Berta}, Zachory K. and {Irwin}, Jonathan and {Burke}, Christopher J. and {Nutzman}, Philip and {Buchhave}, Lars A. and {Lovis}, Christophe and {Bonfils}, Xavier and {Latham}, David W. and {Udry}, St{\'e}phane and {Murray-Clay}, Ruth A. and {Holman}, Matthew J. and {Falco}, Emilio E. and {Winn}, Joshua N. and {Queloz}, Didier and {Pepe}, Francesco and {Mayor}, Michel and {Delfosse}, Xavier and {Forveille}, Thierry},
        title = "{A super-Earth transiting a nearby low-mass star}",
      journal = {\nat},
         year = 2009,
        month = dec,
       volume = {462},
       number = {7275},
        pages = {891-894},
          doi = {10.1038/nature08679},
archivePrefix = {arXiv},
       eprint = {0912.3229},
 primaryClass = {astro-ph.EP},
       adsurl = {https://ui.adsabs.harvard.edu/abs/2009Natur.462..891C}
}

@ARTICLE{batalha_picaso_2019,
       author = {{Batalha}, Natasha E. and {Marley}, Mark S. and {Lewis}, Nikole K. and {Fortney}, Jonathan J.},
        title = "{Exoplanet Reflected-light Spectroscopy with PICASO}",
      journal = {\apj},
         year = 2019,
        month = jun,
       volume = {878},
       number = {1},
          eid = {70},
        pages = {70},
          doi = {10.3847/1538-4357/ab1b51},
archivePrefix = {arXiv},
       eprint = {1904.09355},
 primaryClass = {astro-ph.EP},
       adsurl = {https://ui.adsabs.harvard.edu/abs/2019ApJ...878...70B}
}

@ARTICLE{Khare_tholins,
       author = {{Khare}, B.~N. and {Sagan}, C. and {Arakawa}, E.~T. and {Suits}, F. and {Callcott}, T.~A. and {Williams}, M.~W.},
        title = "{Optical constants of organic tholins produced in a simulated Titanian atmosphere: From soft x-ray to microwave frequencies}",
      journal = {\icarus},
         year = 1984,
        month = oct,
       volume = {60},
       number = {1},
        pages = {127-137},
          doi = {10.1016/0019-1035(84)90142-8},
       adsurl = {https://ui.adsabs.harvard.edu/abs/1984Icar...60..127K}
}

@ARTICLE{wakeford_sing,
       author = {{Wakeford}, H.~R. and {Sing}, D.~K.},
        title = "{Transmission spectral properties of clouds for hot Jupiter exoplanets}",
      journal = {\aap},
         year = 2015,
        month = jan,
       volume = {573},
          eid = {A122},
        pages = {A122},
          doi = {10.1051/0004-6361/201424207},
archivePrefix = {arXiv},
       eprint = {1409.7594},
 primaryClass = {astro-ph.EP},
       adsurl = {https://ui.adsabs.harvard.edu/abs/2015A&A...573A.122W}
}

@ARTICLE{mullens_2024,
       author = {{Mullens}, Elijah and {Lewis}, Nikole K. and {MacDonald}, Ryan J.},
        title = "{Implementation of Aerosol Mie Scattering in POSEIDON with Application to the Hot Jupiter HD 189733 b's Transmission, Emission, and Reflected Light Spectrum}",
      journal = {\apj},
         year = 2024,
        month = dec,
       volume = {977},
       number = {1},
          eid = {105},
        pages = {105},
          doi = {10.3847/1538-4357/ad8575},
archivePrefix = {arXiv},
       eprint = {2410.19253},
 primaryClass = {astro-ph.EP},
       adsurl = {https://ui.adsabs.harvard.edu/abs/2024ApJ...977..105M}
}

@ARTICLE{mang_waterclouds_meteoritic_dust,
       author = {{Mang}, James and {Gao}, Peter and {Hood}, Callie E. and {Fortney}, Jonathan J. and {Batalha}, Natasha and {Yu}, Xinting and {de Pater}, Imke},
        title = "{Microphysics of Water Clouds in the Atmospheres of Y Dwarfs and Temperate Giant Planets}",
      journal = {\apj},
         year = 2022,
        month = mar,
       volume = {927},
       number = {2},
          eid = {184},
        pages = {184},
          doi = {10.3847/1538-4357/ac51d3},
archivePrefix = {arXiv},
       eprint = {2202.01355},
 primaryClass = {astro-ph.EP},
       adsurl = {https://ui.adsabs.harvard.edu/abs/2022ApJ...927..184M}
}

@book{petty_textbook,
  title = {A First Course in Atmospheric Radiation},
  author = {Petty, Grant W.},
  year = 2006,
  edition = {2. ed},
  publisher = {Sundog Publ},
  address = {Madison, Wisc},
  isbn = {978-0-9729033-1-8},
  langid = {english}
}

@BOOK{palik_kcl,
       author = {{Palik}, Edward D.},
        title = "{Handbook of optical constants of solids}",
         year = 1985,
       adsurl = {https://ui.adsabs.harvard.edu/abs/1985hocs.book.....P}
}

@ARTICLE{irwin_uranus,
       author = {{Irwin}, P.~G.~J. and {Teanby}, N.~A. and {Fletcher}, L.~N. and {Toledo}, D. and {Orton}, G.~S. and {Wong}, M.~H. and {Roman}, M.~T. and {P{\'e}rez-Hoyos}, S. and {James}, A. and {Dobinson}, J.},
        title = "{Hazy Blue Worlds: A Holistic Aerosol Model for Uranus and Neptune, Including Dark Spots}",
      journal = {Journal of Geophysical Research (Planets)},
         year = 2022,
        month = jun,
       volume = {127},
       number = {6},
          eid = {e07189},
        pages = {e07189},
          doi = {10.1029/2022JE007189},
archivePrefix = {arXiv},
       eprint = {2201.04516},
 primaryClass = {astro-ph.EP},
       adsurl = {https://ui.adsabs.harvard.edu/abs/2022JGRE..12707189I}
}

@article{razavifar_roughness_contactangle,
  title = {Quantifying the Impact of Surface Roughness on Contact Angle Dynamics under Varying Conditions},
  author = {Razavifar, Mehdi and Abdi, Arastoo and Nikooee, Ehsan and Aghili, Omidreza and Riazi, Masoud},
  year = 2025,
  month = may,
  journal = {Scientific Reports},
  volume = {15},
  number = {1},
  pages = {16611},
  issn = {2045-2322},
  doi = {10.1038/s41598-025-01127-7}
}

@ARTICLE{ohno_aggregates,
       author = {{Ohno}, Kazumasa and {Okuzumi}, Satoshi and {Tazaki}, Ryo},
        title = "{Clouds of Fluffy Aggregates: How They Form in Exoplanetary Atmospheres and Influence Transmission Spectra}",
      journal = {\apj},
         year = 2020,
        month = mar,
       volume = {891},
       number = {2},
          eid = {131},
        pages = {131},
          doi = {10.3847/1538-4357/ab44bd},
archivePrefix = {arXiv},
       eprint = {1908.02201},
 primaryClass = {astro-ph.EP},
       adsurl = {https://ui.adsabs.harvard.edu/abs/2020ApJ...891..131O}
}

@ARTICLE{vahidinia_moran_aggregates,
       author = {{Vahidinia}, Sanaz and {Moran}, Sarah E. and {Marley}, Mark S. and {Cuzzi}, Jeffrey N.},
        title = "{Aggregate Cloud Particle Effects in Exoplanet Atmospheres}",
      journal = {\pasp},
         year = 2024,
        month = aug,
       volume = {136},
       number = {8},
          eid = {084404},
        pages = {084404},
          doi = {10.1088/1538-3873/ad6cf2},
archivePrefix = {arXiv},
       eprint = {2408.11215},
 primaryClass = {astro-ph.EP},
       adsurl = {https://ui.adsabs.harvard.edu/abs/2024PASP..136h4404V}
}

@ARTICLE{kreidberg_clouds,
       author = {{Kreidberg}, Laura and {Bean}, Jacob L. and {D{\'e}sert}, Jean-Michel and {Benneke}, Bj{\"o}rn and {Deming}, Drake and {Stevenson}, Kevin B. and {Seager}, Sara and {Berta-Thompson}, Zachory and {Seifahrt}, Andreas and {Homeier}, Derek},
        title = "{Clouds in the atmosphere of the super-Earth exoplanet GJ1214b}",
      journal = {\nat},
         year = 2014,
        month = jan,
       volume = {505},
       number = {7481},
        pages = {69-72},
          doi = {10.1038/nature12888},
archivePrefix = {arXiv},
       eprint = {1401.0022},
 primaryClass = {astro-ph.EP},
       adsurl = {https://ui.adsabs.harvard.edu/abs/2014Natur.505...69K}
}

@ARTICLE{schlawin_24_gj1214b,
       author = {{Schlawin}, Everett and {Ohno}, Kazumasa and {Bell}, Taylor J. and {Murphy}, Matthew M. and {Welbanks}, Luis and {Beatty}, Thomas G. and {Greene}, Thomas P. and {Fortney}, Jonathan J. and {Parmentier}, Vivien and {Edelman}, Isaac R. and {Gill}, Samuel and {Anderson}, David R. and {Wheatley}, Peter J. and {Henry}, Gregory W. and {Mehta}, Nishil and {Kreidberg}, Laura and {Rieke}, Marcia J.},
        title = "{Possible Carbon Dioxide above the Thick Aerosols of GJ 1214 b}",
      journal = {\apjl},
         year = 2024,
        month = oct,
       volume = {974},
       number = {2},
          eid = {L33},
        pages = {L33},
          doi = {10.3847/2041-8213/ad7fef},
archivePrefix = {arXiv},
       eprint = {2410.10183},
 primaryClass = {astro-ph.EP},
       adsurl = {https://ui.adsabs.harvard.edu/abs/2024ApJ...974L..33S}
}

@ARTICLE{stevenson_clouds,
       author = {{Stevenson}, Kevin B.},
        title = "{Quantifying and Predicting the Presence of Clouds in Exoplanet Atmospheres}",
      journal = {\apjl},
         year = 2016,
        month = feb,
       volume = {817},
       number = {2},
          eid = {L16},
        pages = {L16},
          doi = {10.3847/2041-8205/817/2/L16},
archivePrefix = {arXiv},
       eprint = {1601.03492},
 primaryClass = {astro-ph.EP},
       adsurl = {https://ui.adsabs.harvard.edu/abs/2016ApJ...817L..16S}
}

@ARTICLE{adams_aggregate_hazes,
       author = {{Adams}, Danica and {Gao}, Peter and {de Pater}, Imke and {Morley}, Caroline V.},
        title = "{Aggregate Hazes in Exoplanet Atmospheres}",
      journal = {\apj},
         year = 2019,
        month = mar,
       volume = {874},
       number = {1},
          eid = {61},
        pages = {61},
          doi = {10.3847/1538-4357/ab074c},
archivePrefix = {arXiv},
       eprint = {1902.05231},
 primaryClass = {astro-ph.EP},
       adsurl = {https://ui.adsabs.harvard.edu/abs/2019ApJ...874...61A}
}

@ARTICLE{li_he_graphite,
       author = {{Li}, Haixin and {He}, Chao and {Wang}, Sai and {Yang}, Zhengbo and {Liu}, Yu and {Wang}, Yingjian and {Luo}, Xiao'ou and {Moran}, Sarah E. and {Pesciotta}, Cara and {H{\"o}rst}, Sarah M. and {Moses}, Julianne I. and {Vuitton}, V{\'e}ronique},
        title = "{The Impact of Organic Hazes and Graphite on the Observation of CO$_{2}$-rich Sub-Neptune Atmospheres}",
      journal = {\apjl},
         year = 2025,
        month = sep,
       volume = {990},
       number = {2},
          eid = {L66},
        pages = {L66},
          doi = {10.3847/2041-8213/adfa87},
archivePrefix = {arXiv},
       eprint = {2508.07161},
 primaryClass = {astro-ph.EP},
       adsurl = {https://ui.adsabs.harvard.edu/abs/2025ApJ...990L..66L}
}

@ARTICLE{moran_virga,
       author = {{Moran}, Sarah E. and {Lodge}, Matt G. and {Batalha}, Natasha E. and {Ohno}, Kazumasa and {Vahidinia}, Sanaz and {Marley}, Mark S. and {Wakeford}, Hannah R. and {Leinhardt}, Z{\"o}e M.},
        title = "{Fractal Aggregate Aerosols in the Virga Cloud Code I: Model Description and Application to a Benchmark Cloudy Exoplanet}",
      journal = {arXiv e-prints},
         year = 2025,
        month = sep,
          eid = {arXiv:2509.06708},
        pages = {arXiv:2509.06708},
          doi = {10.48550/arXiv.2509.06708},
archivePrefix = {arXiv},
       eprint = {2509.06708},
 primaryClass = {astro-ph.EP},
       adsurl = {https://ui.adsabs.harvard.edu/abs/2025arXiv250906708M}
}

@ARTICLE{chao_he_sulfur, 
       author = {{He}, Chao and {H{\"o}rst}, Sarah M. and {Lewis}, Nikole K. and {Yu}, Xinting and {Moses}, Julianne I. and {McGuiggan}, Patricia and {Marley}, Mark S. and {Kempton}, Eliza M.-R. and {Moran}, Sarah E. and {Morley}, Caroline V. and {Vuitton}, V{\'e}ronique},
        title = "{Sulfur-driven haze formation in warm CO$_{2}$-rich exoplanet atmospheres}",
      journal = {Nature Astronomy},
         year = 2020,
        month = apr,
       volume = {4},
        pages = {986-993},
          doi = {10.1038/s41550-020-1072-9},
archivePrefix = {arXiv},
       eprint = {2004.02728},
 primaryClass = {astro-ph.EP},
       adsurl = {https://ui.adsabs.harvard.edu/abs/2020NatAs...4..986H}
}

@ARTICLE{gao_sulfur_haze,
       author = {{Gao}, Peter and {Marley}, Mark S. and {Zahnle}, Kevin and {Robinson}, Tyler D. and {Lewis}, Nikole K.},
        title = "{Sulfur Hazes in Giant Exoplanet Atmospheres: Impacts on Reflected Light Spectra}",
      journal = {\aj},
         year = 2017,
        month = mar,
       volume = {153},
       number = {3},
          eid = {139},
        pages = {139},
          doi = {10.3847/1538-3881/aa5fab},
archivePrefix = {arXiv},
       eprint = {1701.00318},
 primaryClass = {astro-ph.EP},
       adsurl = {https://ui.adsabs.harvard.edu/abs/2017AJ....153..139G}
}

@ARTICLE{ohno_diamond,
       author = {{Ohno}, Kazumasa},
        title = "{Photochemical Hazes in Exoplanetary Skies with Diamonds: Microphysical Modeling of Haze Composition Evolution via Chemical Vapor Deposition}",
      journal = {\apj},
         year = 2024,
        month = dec,
       volume = {977},
       number = {2},
          eid = {188},
        pages = {188},
          doi = {10.3847/1538-4357/ad8e67},
archivePrefix = {arXiv},
       eprint = {2410.10197},
 primaryClass = {astro-ph.EP},
       adsurl = {https://ui.adsabs.harvard.edu/abs/2024ApJ...977..188O}
}

@ARTICLE{brande_2024,
       author = {{Brande}, Jonathan and {Crossfield}, Ian J.~M. and {Kreidberg}, Laura and {Morley}, Caroline V. and {Barman}, Travis and {Benneke}, Bj{\"o}rn and {Christiansen}, Jessie L. and {Dragomir}, Diana and {Fortney}, Jonathan J. and {Greene}, Thomas P. and {Hardegree-Ullman}, Kevin K. and {Howard}, Andrew W. and {Knutson}, Heather A. and {Lothringer}, Joshua D. and {Mikal-Evans}, Thomas},
        title = "{Clouds and Clarity: Revisiting Atmospheric Feature Trends in Neptune-size Exoplanets}",
      journal = {\apjl},
         year = 2024,
        month = jan,
       volume = {961},
       number = {1},
          eid = {L23},
        pages = {L23},
          doi = {10.3847/2041-8213/ad1b5c},
archivePrefix = {arXiv},
       eprint = {2310.07714},
 primaryClass = {astro-ph.EP},
       adsurl = {https://ui.adsabs.harvard.edu/abs/2024ApJ...961L..23B}
}

@ARTICLE{crossfield_kreidberg,
       author = {{Crossfield}, Ian J.~M. and {Kreidberg}, Laura},
        title = "{Trends in Atmospheric Properties of Neptune-size Exoplanets}",
      journal = {\aj},
         year = 2017,
        month = dec,
       volume = {154},
       number = {6},
          eid = {261},
        pages = {261},
          doi = {10.3847/1538-3881/aa9279},
archivePrefix = {arXiv},
       eprint = {1708.00016},
 primaryClass = {astro-ph.EP},
       adsurl = {https://ui.adsabs.harvard.edu/abs/2017AJ....154..261C}
}

@INPROCEEDINGS{xinting_titan_hetnucleation,
       author = {{Yu}, Xinting and {Yu}, Yue and {Garver}, Julia and {Zhang}, Xi},
        title = "{Cloud Formation on Titan through Heterogenous Nucleation}",
    booktitle = {AGU Fall Meeting Abstracts},
         year = 2024,
       series = {AGU Fall Meeting Abstracts},
       volume = {2024},
        month = dec,
          eid = {P41B-08},
        pages = {P41B-08},
       adsurl = {https://ui.adsabs.harvard.edu/abs/2024AGUFMP41B...08Y}
}

@ARTICLE{ohno_24_gj1214b,
       author = {{Ohno}, Kazumasa and {Schlawin}, Everett and {Bell}, Taylor J. and {Murphy}, Matthew M. and {Beatty}, Thomas G. and {Welbanks}, Luis and {Greene}, Thomas P. and {Fortney}, Jonathan J. and {Parmentier}, Vivien and {Edelman}, Isaac R. and {Mehta}, Nishil and {Rieke}, Marcia J.},
        title = "{A Possible Metal-dominated Atmosphere below the Thick Aerosols of GJ 1214 b Suggested by Its JWST Panchromatic Transmission Spectrum}",
      journal = {\apjl},
         year = 2025,
        month = jan,
       volume = {979},
       number = {1},
          eid = {L7},
        pages = {L7},
          doi = {10.3847/2041-8213/ada02c},
archivePrefix = {arXiv},
       eprint = {2410.10186},
 primaryClass = {astro-ph.EP},
       adsurl = {https://ui.adsabs.harvard.edu/abs/2025ApJ...979L...7O}
}

@ARTICLE{gao_2020,
       author = {{Gao}, Peter and {Thorngren}, Daniel P. and {Lee}, Elspeth K.~H. and {Fortney}, Jonathan J. and {Morley}, Caroline V. and {Wakeford}, Hannah R. and {Powell}, Diana K. and {Stevenson}, Kevin B. and {Zhang}, Xi},
        title = "{Aerosol composition of hot giant exoplanets dominated by silicates and hydrocarbon hazes}",
      journal = {Nature Astronomy},
         year = 2020,
        month = may,
       volume = {4},
        pages = {951-956},
          doi = {10.1038/s41550-020-1114-3},
archivePrefix = {arXiv},
       eprint = {2005.11939},
 primaryClass = {astro-ph.EP},
       adsurl = {https://ui.adsabs.harvard.edu/abs/2020NatAs...4..951G}
}

@ARTICLE{kempton_miri,
       author = {{Kempton}, Eliza M. -R. and {Zhang}, Michael and {Bean}, Jacob L. and {Steinrueck}, Maria E. and {Piette}, Anjali A.~A. and {Parmentier}, Vivien and {Malsky}, Isaac and {Roman}, Michael T. and {Rauscher}, Emily and {Gao}, Peter and {Bell}, Taylor J. and {Xue}, Qiao and {Taylor}, Jake and {Savel}, Arjun B. and {Arnold}, Kenneth E. and {Nixon}, Matthew C. and {Stevenson}, Kevin B. and {Mansfield}, Megan and {Kendrew}, Sarah and {Zieba}, Sebastian and {Ducrot}, Elsa and {Dyrek}, Achr{\`e}ne and {Lagage}, Pierre-Olivier and {Stassun}, Keivan G. and {Henry}, Gregory W. and {Barman}, Travis and {Lupu}, Roxana and {Malik}, Matej and {Kataria}, Tiffany and {Ih}, Jegug and {Fu}, Guangwei and {Welbanks}, Luis and {McGill}, Peter},
        title = "{A reflective, metal-rich atmosphere for GJ 1214b from its JWST phase curve}",
      journal = {\nat},
         year = 2023,
        month = aug,
       volume = {620},
       number = {7972},
        pages = {67-71},
          doi = {10.1038/s41586-023-06159-5},
archivePrefix = {arXiv},
       eprint = {2305.06240},
 primaryClass = {astro-ph.EP},
       adsurl = {https://ui.adsabs.harvard.edu/abs/2023Natur.620...67K}
}

@ARTICLE{pymiescat,
       author = {{Sumlin}, Benjamin J. and {Heinson}, William R. and {Chakrabarty}, Rajan K.},
        title = "{Retrieving the aerosol complex refractive index using PyMieScatt: A Mie computational package with visualization capabilities}",
      journal = {\jqsrt},
         year = 2018,
        month = jan,
       volume = {205},
        pages = {127-134},
          doi = {10.1016/j.jqsrt.2017.10.012},
archivePrefix = {arXiv},
       eprint = {1710.05288},
 primaryClass = {physics.optics},
       adsurl = {https://ui.adsabs.harvard.edu/abs/2018JQSRT.205..127S}
}

@ARTICLE{gao23,
       author = {{Gao}, Peter and {Piette}, Anjali A.~A. and {Steinrueck}, Maria E. and {Nixon}, Matthew C. and {Zhang}, Michael and {Kempton}, Eliza M. -R. and {Bean}, Jacob L. and {Rauscher}, Emily and {Parmentier}, Vivien and {Batalha}, Natasha E. and {Savel}, Arjun B. and {Arnold}, Kenneth E. and {Roman}, Michael T. and {Malsky}, Isaac and {Taylor}, Jake},
        title = "{The Hazy and Metal-rich Atmosphere of GJ 1214 b Constrained by Near- and Mid-infrared Transmission Spectroscopy}",
      journal = {\apj},
         year = 2023,
        month = jul,
       volume = {951},
       number = {2},
          eid = {96},
        pages = {96},
          doi = {10.3847/1538-4357/acd16f},
archivePrefix = {arXiv},
       eprint = {2305.05697},
 primaryClass = {astro-ph.EP},
       adsurl = {https://ui.adsabs.harvard.edu/abs/2023ApJ...951...96G}
}

@ARTICLE{maria_sn_hazes,
       author = {{Steinrueck}, Maria E. and {Parmentier}, Vivien and {Kreidberg}, Laura and {Gao}, Peter and {Kempton}, Eliza M. -R. and {Zhang}, Michael and {Stevenson}, Kevin B. and {Malsky}, Isaac and {Roman}, Michael T. and {Rauscher}, Emily and {Malik}, Matej and {Lupu}, Roxana and {Kataria}, Tiffany and {Piette}, Anjali A.~A. and {Bean}, Jacob L. and {Nixon}, Matthew C.},
        title = "{The radiative effects of photochemical hazes on the atmospheric circulation and phase curves of sub-Neptunes}",
      journal = {arXiv e-prints},
         year = 2025,
        month = mar,
          eid = {arXiv:2503.22642},
        pages = {arXiv:2503.22642},
          doi = {10.48550/arXiv.2503.22642},
archivePrefix = {arXiv},
       eprint = {2503.22642},
 primaryClass = {astro-ph.EP},
       adsurl = {https://ui.adsabs.harvard.edu/abs/2025arXiv250322642S}
}

@ARTICLE{diana_silicate,
       author = {{Powell}, Diana and {Zhang}, Xi and {Gao}, Peter and {Parmentier}, Vivien},
        title = "{Formation of Silicate and Titanium Clouds on Hot Jupiters}",
      journal = {\apj},
         year = 2018,
        month = jun,
       volume = {860},
       number = {1},
          eid = {18},
        pages = {18},
          doi = {10.3847/1538-4357/aac215},
archivePrefix = {arXiv},
       eprint = {1805.01468},
 primaryClass = {astro-ph.EP},
       adsurl = {https://ui.adsabs.harvard.edu/abs/2018ApJ...860...18P}
}

@ARTICLE{diana2D,
       author = {{Powell}, Diana and {Zhang}, Xi},
        title = "{Two-dimensional Models of Microphysical Clouds on Hot Jupiters. I. Cloud Properties}",
      journal = {\apj},
         year = 2024,
        month = jul,
       volume = {969},
       number = {1},
          eid = {5},
        pages = {5},
          doi = {10.3847/1538-4357/ad3de4},
archivePrefix = {arXiv},
       eprint = {2404.08759},
 primaryClass = {astro-ph.EP},
       adsurl = {https://ui.adsabs.harvard.edu/abs/2024ApJ...969....5P}
}

@ARTICLE{Malsky_2025,
       author = {{Malsky}, Isaac and {Rauscher}, Emily and {Stevenson}, Kevin and {Savel}, Arjun B. and {Steinrueck}, Maria E. and {Gao}, Peter and {Kempton}, Eliza M. -R. and {Roman}, Michael T. and {Bean}, Jacob L. and {Zhang}, Michael and {Parmentier}, Vivien and {Piette}, Anjali A.~A. and {Kataria}, Tiffany},
        title = "{Clouds and Hazes in GJ 1214 b's Metal-rich Atmosphere}",
      journal = {\aj},
         year = 2025,
        month = apr,
       volume = {169},
       number = {4},
          eid = {221},
        pages = {221},
          doi = {10.3847/1538-3881/adb7e8},
archivePrefix = {arXiv},
       eprint = {2503.22608},
 primaryClass = {astro-ph.EP},
       adsurl = {https://ui.adsabs.harvard.edu/abs/2025AJ....169..221M}
}

@ARTICLE{lavvas2024,
       author = {{Lavvas}, Panayotis and {Paraskevaidou}, Sophia and {Arfaux}, Anthony},
        title = "{Atmospheric characterisation of GJ1214b from transit and eclipse observations}",
      journal = {arXiv e-prints},
         year = 2024,
        month = oct,
          eid = {arXiv:2410.09981},
        pages = {arXiv:2410.09981},
          doi = {10.48550/arXiv.2410.09981},
archivePrefix = {arXiv},
       eprint = {2410.09981},
 primaryClass = {astro-ph.EP},
       adsurl = {https://ui.adsabs.harvard.edu/abs/2024arXiv241009981L}
}

@ARTICLE{arfaux_cloudhaze,
       author = {{Arfaux}, Anthony and {Lavvas}, Panayotis},
        title = "{Coupling haze and cloud microphysics in WASP-39b's atmosphere based on JWST observations}",
      journal = {\mnras},
         year = 2024,
        month = may,
       volume = {530},
       number = {1},
        pages = {482-500},
          doi = {10.1093/mnras/stae826},
archivePrefix = {arXiv},
       eprint = {2311.07365},
 primaryClass = {astro-ph.EP},
       adsurl = {https://ui.adsabs.harvard.edu/abs/2024MNRAS.530..482A}
}

@ARTICLE{parmentier2013,
       author = {{Parmentier}, Vivien and {Showman}, Adam P. and {Lian}, Yuan},
        title = "{3D mixing in hot Jupiters atmospheres. I. Application to the day/night cold trap in HD 209458b}",
      journal = {\aap},
         year = 2013,
        month = oct,
       volume = {558},
          eid = {A91},
        pages = {A91},
          doi = {10.1051/0004-6361/201321132},
archivePrefix = {arXiv},
       eprint = {1301.4522},
 primaryClass = {astro-ph.EP},
       adsurl = {https://ui.adsabs.harvard.edu/abs/2013A&A...558A..91P}
}

@ARTICLE{gao_benneke,
       author = {{Gao}, Peter and {Benneke}, Bj{\"o}rn},
        title = "{Microphysics of KCl and ZnS Clouds on GJ 1214 b}",
      journal = {\apj},
         year = 2018,
        month = aug,
       volume = {863},
       number = {2},
          eid = {165},
        pages = {165},
          doi = {10.3847/1538-4357/aad461},
archivePrefix = {arXiv},
       eprint = {1807.04924},
 primaryClass = {astro-ph.EP},
       adsurl = {https://ui.adsabs.harvard.edu/abs/2018ApJ...863..165G}
}

@ARTICLE{steinrueck_2021,
       author = {{Steinrueck}, Maria E. and {Showman}, Adam P. and {Lavvas}, Panayotis and {Koskinen}, Tommi and {Tan}, Xianyu and {Zhang}, Xi},
        title = "{3D simulations of photochemical hazes in the atmosphere of hot Jupiter HD 189733b}",
      journal = {\mnras},
         year = 2021,
        month = jun,
       volume = {504},
       number = {2},
        pages = {2783-2799},
          doi = {10.1093/mnras/stab1053},
archivePrefix = {arXiv},
       eprint = {2011.14022},
 primaryClass = {astro-ph.EP},
       adsurl = {https://ui.adsabs.harvard.edu/abs/2021MNRAS.504.2783S}
}

@ARTICLE{charnay_mixing,
       author = {{Charnay}, B. and {Meadows}, V. and {Leconte}, J.},
        title = "{3D Modeling of GJ1214b's Atmosphere: Vertical Mixing Driven by an Anti-Hadley Circulation}",
      journal = {\apj},
         year = 2015,
        month = nov,
       volume = {813},
       number = {1},
          eid = {15},
        pages = {15},
          doi = {10.1088/0004-637X/813/1/15},
archivePrefix = {arXiv},
       eprint = {1509.06814},
 primaryClass = {astro-ph.EP},
       adsurl = {https://ui.adsabs.harvard.edu/abs/2015ApJ...813...15C}
}

@ARTICLE{lavvas_2019,
       author = {{Lavvas}, Panayotis and {Koskinen}, Tommi and {Steinrueck}, Maria E. and {Garc{\'\i}a Mu{\~n}oz}, Antonio and {Showman}, Adam P.},
        title = "{Photochemical Hazes in Sub-Neptunian Atmospheres with a Focus on GJ 1214b}",
      journal = {\apj},
         year = 2019,
        month = jun,
       volume = {878},
       number = {2},
          eid = {118},
        pages = {118},
          doi = {10.3847/1538-4357/ab204e},
archivePrefix = {arXiv},
       eprint = {1905.02976},
 primaryClass = {astro-ph.EP},
       adsurl = {https://ui.adsabs.harvard.edu/abs/2019ApJ...878..118L}
}

@ARTICLE{kawashima_ikoma_2019,
       author = {{Kawashima}, Yui and {Ikoma}, Masahiro},
        title = "{Theoretical Transmission Spectra of Exoplanet Atmospheres with Hydrocarbon Haze: Effect of Creation, Growth, and Settling of Haze Particles. II. Dependence on UV Irradiation Intensity, Metallicity, C/O Ratio, Eddy Diffusion Coefficient, and Temperature}",
      journal = {\apj},
         year = 2019,
        month = jun,
       volume = {877},
       number = {2},
          eid = {109},
        pages = {109},
          doi = {10.3847/1538-4357/ab1b1d},
       adsurl = {https://ui.adsabs.harvard.edu/abs/2019ApJ...877..109K}
}

@ARTICLE{lavvas_koskinen,
       author = {{Lavvas}, P. and {Koskinen}, T.},
        title = "{Aerosol Properties of the Atmospheres of Extrasolar Giant Planets}",
      journal = {\apj},
         year = 2017,
        month = sep,
       volume = {847},
       number = {1},
          eid = {32},
        pages = {32},
          doi = {10.3847/1538-4357/aa88ce},
archivePrefix = {arXiv},
       eprint = {1708.09257},
 primaryClass = {astro-ph.EP},
       adsurl = {https://ui.adsabs.harvard.edu/abs/2017ApJ...847...32L}
}

@ARTICLE{ohno_kawashima_super_rayleigh,
       author = {{Ohno}, Kazumasa and {Kawashima}, Yui},
        title = "{Super-Rayleigh Slopes in Transmission Spectra of Exoplanets Generated by Photochemical Haze}",
      journal = {\apjl},
         year = 2020,
        month = jun,
       volume = {895},
       number = {2},
          eid = {L47},
        pages = {L47},
          doi = {10.3847/2041-8213/ab93d7},
archivePrefix = {arXiv},
       eprint = {2005.08880},
 primaryClass = {astro-ph.EP},
       adsurl = {https://ui.adsabs.harvard.edu/abs/2020ApJ...895L..47O}
}

@ARTICLE{gao_zhang_superpuff,
       author = {{Gao}, Peter and {Zhang}, Xi},
        title = "{Deflating Super-puffs: Impact of Photochemical Hazes on the Observed Mass-Radius Relationship of Low-mass Planets}",
      journal = {\apj},
         year = 2020,
        month = feb,
       volume = {890},
       number = {2},
          eid = {93},
        pages = {93},
          doi = {10.3847/1538-4357/ab6a9b},
archivePrefix = {arXiv},
       eprint = {2001.00055},
 primaryClass = {astro-ph.EP},
       adsurl = {https://ui.adsabs.harvard.edu/abs/2020ApJ...890...93G}
}

@article{raymond_cloud_activation,
author = {Raymond, Timothy M. and Pandis, Spyros N.},
title = {Cloud activation of single-component organic aerosol particles},
journal = {Journal of Geophysical Research: Atmospheres},
volume = {107},
number = {D24},
pages = {AAC 16-1-AAC 16-8},
doi = {https://doi.org/10.1029/2002JD002159},
url = {https://agupubs.onlinelibrary.wiley.com/doi/abs/10.1029/2002JD002159},
eprint = {https://agupubs.onlinelibrary.wiley.com/doi/pdf/10.1029/2002JD002159},
year = {2002}
}

@ARTICLE{yu_tholin_surface_energy,
       author = {{Yu}, Xinting and {H{\"o}rst}, Sarah M. and {He}, Chao and {McGuiggan}, Patricia and {Kristiansen}, Kai and {Zhang}, Xi},
        title = "{Surface Energy of the Titan Aerosol Analog ``Tholin''}",
      journal = {\apj},
         year = 2020,
        month = dec,
       volume = {905},
       number = {2},
          eid = {88},
        pages = {88},
          doi = {10.3847/1538-4357/abc55d},
archivePrefix = {arXiv},
       eprint = {2010.13885},
 primaryClass = {astro-ph.EP},
       adsurl = {https://ui.adsabs.harvard.edu/abs/2020ApJ...905...88Y}
}

@ARTICLE{kawashima_2018,
       author = {{Kawashima}, Yui and {Ikoma}, Masahiro},
        title = "{Theoretical Transmission Spectra of Exoplanet Atmospheres with Hydrocarbon Haze: Effect of Creation, Growth, and Settling of Haze Particles. I. Model Description and First Results}",
      journal = {\apj},
         year = 2018,
        month = jan,
       volume = {853},
       number = {1},
          eid = {7},
        pages = {7},
          doi = {10.3847/1538-4357/aaa0c5},
archivePrefix = {arXiv},
       eprint = {1712.02808},
 primaryClass = {astro-ph.EP},
       adsurl = {https://ui.adsabs.harvard.edu/abs/2018ApJ...853....7K}
}

@Article{kleinheins_seasalt_ccn,
AUTHOR = {Kleinheins, J. and Shardt, N. and Lohmann, U. and Marcolli, C.},
TITLE = {The surface tension and cloud condensation nuclei (CCN) activation of sea spray aerosol particles},
JOURNAL = {Atmospheric Chemistry and Physics},
VOLUME = {25},
YEAR = {2025},
NUMBER = {2},
PAGES = {881--903},
URL = {https://acp.copernicus.org/articles/25/881/2025/},
DOI = {10.5194/acp-25-881-2025}
}

@ARTICLE{mukherjee_picaso,
       author = {{Mukherjee}, Sagnick and {Batalha}, Natasha E. and {Fortney}, Jonathan J. and {Marley}, Mark S.},
        title = "{PICASO 3.0: A One-dimensional Climate Model for Giant Planets and Brown Dwarfs}",
      journal = {\apj},
         year = 2023,
        month = jan,
       volume = {942},
       number = {2},
          eid = {71},
        pages = {71},
          doi = {10.3847/1538-4357/ac9f48},
archivePrefix = {arXiv},
       eprint = {2208.07836},
 primaryClass = {astro-ph.EP},
       adsurl = {https://ui.adsabs.harvard.edu/abs/2023ApJ...942...71M}
}

@ARTICLE{irwin_hazy_blue_worlds,
       author = {{Irwin}, P.~G.~J. and {Teanby}, N.~A. and {Fletcher}, L.~N. and {Toledo}, D. and {Orton}, G.~S. and {Wong}, M.~H. and {Roman}, M.~T. and {P{\'e}rez-Hoyos}, S. and {James}, A. and {Dobinson}, J.},
        title = "{Hazy Blue Worlds: A Holistic Aerosol Model for Uranus and Neptune, Including Dark Spots}",
      journal = {Journal of Geophysical Research (Planets)},
         year = 2022,
        month = jun,
       volume = {127},
       number = {6},
          eid = {e07189},
        pages = {e07189},
          doi = {10.1029/2022JE007189},
archivePrefix = {arXiv},
       eprint = {2201.04516},
 primaryClass = {astro-ph.EP},
       adsurl = {https://ui.adsabs.harvard.edu/abs/2022JGRE..12707189I}
}

@ARTICLE{carlson_giant_planet_microphysics,
       author = {{Carlson}, Barbara E. and {Rossow}, William B. and {Orton}, Glenn S.},
        title = "{Cloud microphysics of the giant planets.}",
      journal = {Journal of the Atmospheric Sciences},
         year = 1988,
        month = jul,
       volume = {45},
        pages = {2066-2081},
          doi = {10.1175/1520-0469(1988)045<2066:CMOTGP>2.0.CO;2},
       adsurl = {https://ui.adsabs.harvard.edu/abs/1988JAtS...45.2066C}
}

@ARTICLE{arras_dust_accretion, 
       author = {{Arras}, Phil and {Wilson}, Megan and {Pryal}, Matthew and {Baker}, Jordan},
        title = "{Dust Accretion onto Exoplanets}",
      journal = {\apj},
         year = 2022,
        month = jun,
       volume = {932},
       number = {2},
          eid = {90},
        pages = {90},
          doi = {10.3847/1538-4357/ac625e},
archivePrefix = {arXiv},
       eprint = {2206.09093},
 primaryClass = {astro-ph.EP},
       adsurl = {https://ui.adsabs.harvard.edu/abs/2022ApJ...932...90A}
}

@ARTICLE{fan_pluto_bimodal,
       author = {{Fan}, Siteng and {Gao}, Peter and {Zhang}, Xi and {Adams}, Danica J. and {Kutsop}, Nicholas W. and {Bierson}, Carver J. and {Liu}, Chao and {Yang}, Jiani and {Young}, Leslie A. and {Cheng}, Andrew F. and {Yung}, Yuk L.},
        title = "{A bimodal distribution of haze in Pluto's atmosphere}",
      journal = {Nature Communications},
         year = 2022,
        month = jan,
       volume = {13},
          eid = {240},
        pages = {240},
          doi = {10.1038/s41467-021-27811-6},
archivePrefix = {arXiv},
       eprint = {2201.04392},
 primaryClass = {astro-ph.EP},
       adsurl = {https://ui.adsabs.harvard.edu/abs/2022NatCo..13..240F}
}

@ARTICLE{west_smith_1991_aggregates,
       author = {{West}, R.~A. and {Smith}, P.~H.},
        title = "{Evidence for aggregate particles in the atmospheres of Titan and Jupiter}",
      journal = {\icarus},
         year = 1991,
        month = apr,
       volume = {90},
       number = {2},
        pages = {330-333},
          doi = {10.1016/0019-1035(91)90113-8},
       adsurl = {https://ui.adsabs.harvard.edu/abs/1991Icar...90..330W}
}

@ARTICLE{xi_aerosol_influence,
       author = {{Zhang}, Xi and {West}, Robert A. and {Irwin}, Patrick G.~J. and {Nixon}, Conor A. and {Yung}, Yuk L.},
        title = "{Aerosol influence on energy balance of the middle atmosphere of Jupiter}",
      journal = {Nature Communications},
         year = 2015,
        month = dec,
       volume = {6},
          eid = {10231},
        pages = {10231},
          doi = {10.1038/ncomms10231},
       adsurl = {https://ui.adsabs.harvard.edu/abs/2015NatCo...610231Z}
}

@ARTICLE{lodge_2024a,
       author = {{Lodge}, M.~G. and {Wakeford}, H.~R. and {Leinhardt}, Z.~M.},
        title = "{Aerosols are not spherical cows: using discrete dipole approximation to model the properties of fractal particles}",
      journal = {\mnras},
         year = 2024,
        month = feb,
       volume = {527},
       number = {4},
        pages = {11113-11137},
          doi = {10.1093/mnras/stad3743},
archivePrefix = {arXiv},
       eprint = {2312.02301},
 primaryClass = {astro-ph.EP},
       adsurl = {https://ui.adsabs.harvard.edu/abs/2024MNRAS.52711113L}
}

@ARTICLE{jordan_albedo,
       author = {{Jordan}, Sean and {Shorttle}, Oliver and {Quanz}, Sascha P.},
        title = "{Planetary Albedo Is Limited by the Above-cloud Atmosphere: Implications for Sub-Neptune Climates}",
      journal = {\apj},
         year = 2025,
        month = nov,
       volume = {993},
       number = {1},
          eid = {86},
        pages = {86},
          doi = {10.3847/1538-4357/ae0192},
archivePrefix = {arXiv},
       eprint = {2504.12030},
 primaryClass = {astro-ph.EP},
       adsurl = {https://ui.adsabs.harvard.edu/abs/2025ApJ...993...86J}
}

@ARTICLE{gao_marley_ackerman,
       author = {{Gao}, Peter and {Marley}, Mark S. and {Ackerman}, Andrew S.},
        title = "{Sedimentation Efficiency of Condensation Clouds in Substellar Atmospheres}",
      journal = {\apj},
         year = 2018,
        month = mar,
       volume = {855},
       number = {2},
          eid = {86},
        pages = {86},
          doi = {10.3847/1538-4357/aab0a1},
archivePrefix = {arXiv},
       eprint = {1802.06241},
 primaryClass = {astro-ph.EP},
       adsurl = {https://ui.adsabs.harvard.edu/abs/2018ApJ...855...86G}
}

@ARTICLE{moses_2011,
       author = {{Moses}, Julianne I. and {Visscher}, C. and {Fortney}, J.~J. and {Showman}, A.~P. and {Lewis}, N.~K. and {Griffith}, C.~A. and {Klippenstein}, S.~J. and {Shabram}, M. and {Friedson}, A.~J. and {Marley}, M.~S. and {Freedman}, R.~S.},
        title = "{Disequilibrium Carbon, Oxygen, and Nitrogen Chemistry in the Atmospheres of HD 189733b and HD 209458b}",
      journal = {\apj},
         year = 2011,
        month = aug,
       volume = {737},
       number = {1},
          eid = {15},
        pages = {15},
          doi = {10.1088/0004-637X/737/1/15},
archivePrefix = {arXiv},
       eprint = {1102.0063},
 primaryClass = {astro-ph.EP},
       adsurl = {https://ui.adsabs.harvard.edu/abs/2011ApJ...737...15M}
}

@ARTICLE{xi_zhang_global_mean_2,
       author = {{Zhang}, Xi and {Showman}, Adam P.},
        title = "{Global-mean Vertical Tracer Mixing in Planetary Atmospheres. II. Tidally Locked Planets}",
      journal = {\apj},
         year = 2018,
        month = oct,
       volume = {866},
       number = {1},
          eid = {2},
        pages = {2},
          doi = {10.3847/1538-4357/aada7c},
archivePrefix = {arXiv},
       eprint = {1808.05365},
 primaryClass = {astro-ph.EP},
       adsurl = {https://ui.adsabs.harvard.edu/abs/2018ApJ...866....2Z}
}

@ARTICLE{lodge_2024b,
       author = {{Lodge}, M.~G. and {Wakeford}, H.~R. and {Leinhardt}, Z.~M.},
        title = "{MANTA-Ray: Supercharging Speeds for Calculating the Optical Properties of Fractal Aggregates in the Long-wavelength Limit}",
      journal = {\mnras},
         year = 2024,
        month = dec,
       volume = {535},
       number = {2},
        pages = {1964-1978},
          doi = {10.1093/mnras/stae2451},
archivePrefix = {arXiv},
       eprint = {2410.21400},
 primaryClass = {astro-ph.EP},
       adsurl = {https://ui.adsabs.harvard.edu/abs/2024MNRAS.535.1964L}
}

@ARTICLE{lodge_virga,
       author = {{Lodge}, Matt G. and {Moran}, Sarah E. and {Wakeford}, Hannah R. and {Leinhardt}, Zo{\"e} M. and {Marley}, Mark S.},
        title = "{Fractal Aggregate Aerosols in the Virga Cloud Code. II. Exploring the Effects of Key Cloud Parameters in Warm Neptune, Hot Jupiter and Brown Dwarf Atmospheres}",
      journal = {\apj},
         year = 2026,
        month = feb,
       volume = {997},
       number = {2},
          eid = {317},
        pages = {317},
          doi = {10.3847/1538-4357/ae2752},
archivePrefix = {arXiv},
       eprint = {2512.04186},
 primaryClass = {astro-ph.EP},
       adsurl = {https://ui.adsabs.harvard.edu/abs/2026ApJ...997..317L}
}

@ARTICLE{curtis_2008_lab,
       author = {{Curtis}, Daniel B. and {Hatch}, Courtney D. and {Hasenkopf}, Christa A. and {Toon}, Owen B. and {Tolbert}, Margaret A. and {McKay}, Christopher P. and {Khare}, Bishun N.},
        title = "{Laboratory studies of methane and ethane adsorption and nucleation onto organic particles: Application to Titan's clouds}",
      journal = {\icarus},
         year = 2008,
        month = jun,
       volume = {195},
       number = {2},
        pages = {792-801},
          doi = {10.1016/j.icarus.2008.02.003},
       adsurl = {https://ui.adsabs.harvard.edu/abs/2008Icar..195..792C}
}

@article{SJOGREN2007157,
title = {Hygroscopic growth and water uptake kinetics of two-phase aerosol particles consisting of ammonium sulfate, adipic and humic acid mixtures},
journal = {Journal of Aerosol Science},
volume = {38},
number = {2},
pages = {157-171},
year = {2007},
issn = {0021-8502},
doi = {https://doi.org/10.1016/j.jaerosci.2006.11.005},
url = {https://www.sciencedirect.com/science/article/pii/S0021850206002096},
author = {S. Sjogren and M. Gysel and E. Weingartner and U. Baltensperger and M.J. Cubison and H. Coe and A.A. Zardini and C. Marcolli and U.K. Krieger and T. Peter}
}

@Article{ice_pores,
AUTHOR = {Marcolli, C.},
TITLE = {Pre-activation of aerosol particles by ice preserved in pores},
JOURNAL = {Atmospheric Chemistry and Physics},
VOLUME = {17},
YEAR = {2017},
NUMBER = {3},
PAGES = {1595--1622},
URL = {https://acp.copernicus.org/articles/17/1595/2017/},
DOI = {10.5194/acp-17-1595-2017}
}

@ARTICLE{kiefer_heterogeneous,
       author = {{Kiefer}, S. and {Samra}, D. and {Lewis}, D.~A. and {Schneider}, A.~D. and {Min}, M. and {Carone}, L. and {Decin}, L. and {Helling}, Ch.},
        title = "{Why heterogeneous cloud particles matter: Iron-bearing species and cloud particle morphology affect exoplanet transmission spectra}",
      journal = {\aap},
         year = 2024,
        month = oct,
       volume = {690},
          eid = {A244},
        pages = {A244},
          doi = {10.1051/0004-6361/202450526},
archivePrefix = {arXiv},
       eprint = {2409.01121},
 primaryClass = {astro-ph.EP},
       adsurl = {https://ui.adsabs.harvard.edu/abs/2024A&A...690A.244K}
}

@ARTICLE{hamill_polarimetry,
       author = {{Hamill}, Colin D. and {Johnson}, Alexandria V. and {Gao}, Peter},
        title = "{Light Scattering Measurements of KCl Particles as an Exoplanet Cloud Analog}",
      journal = {\psj},
         year = 2024,
        month = aug,
       volume = {5},
       number = {8},
          eid = {186},
        pages = {186},
          doi = {10.3847/PSJ/ad6569},
archivePrefix = {arXiv},
       eprint = {2411.00952},
 primaryClass = {astro-ph.EP},
       adsurl = {https://ui.adsabs.harvard.edu/abs/2024PSJ.....5..186H}
}

@ARTICLE{gao_aerosol_composition,
       author = {{Gao}, Peter and {Thorngren}, Daniel P. and {Lee}, Elspeth K.~H. and {Fortney}, Jonathan J. and {Morley}, Caroline V. and {Wakeford}, Hannah R. and {Powell}, Diana K. and {Stevenson}, Kevin B. and {Zhang}, Xi},
        title = "{Aerosol composition of hot giant exoplanets dominated by silicates and hydrocarbon hazes}",
      journal = {Nature Astronomy},
         year = 2020,
        month = may,
       volume = {4},
        pages = {951-956},
          doi = {10.1038/s41550-020-1114-3},
archivePrefix = {arXiv},
       eprint = {2005.11939},
 primaryClass = {astro-ph.EP},
       adsurl = {https://ui.adsabs.harvard.edu/abs/2020NatAs...4..951G}
}

@ARTICLE{pesciotta_hazes,
       author = {{Pesciotta}, Cara and {H{\"o}rst}, Sarah M. and {Radke}, Michael J. and {Moran}, Sarah E. and {He}, Chao and {Vuitton}, V{\'e}ronique},
        title = "{Hydrolyzed Hazes on Water-rich Exoplanets: Optical Constants and Detectability}",
      journal = {arXiv e-prints},
         year = 2026,
        month = apr,
          eid = {arXiv:2604.07498},
        pages = {arXiv:2604.07498},
          doi = {10.48550/arXiv.2604.07498},
archivePrefix = {arXiv},
       eprint = {2604.07498},
 primaryClass = {astro-ph.EP},
       adsurl = {https://ui.adsabs.harvard.edu/abs/2026arXiv260407498P}
}

@ARTICLE{corrales_2023_hazes_c/o_ratio,
       author = {{Corrales}, L{\'\i}a and {Gavilan}, Lisseth and {Teal}, D.~J. and {Kempton}, Eliza M.-R.},
        title = "{Photochemical Hazes Can Trace the C/O Ratio in Exoplanet Atmospheres}",
      journal = {\apjl},
         year = 2023,
        month = feb,
       volume = {943},
       number = {2},
          eid = {L26},
        pages = {L26},
          doi = {10.3847/2041-8213/acaf86},
archivePrefix = {arXiv},
       eprint = {2301.01093},
 primaryClass = {astro-ph.EP},
       adsurl = {https://ui.adsabs.harvard.edu/abs/2023ApJ...943L..26C}
}

@ARTICLE{huseby_hazes_uv,
       author = {{Huseby}, Lori and {Moran}, Sarah E. and {Pearson}, Neil and {Kataria}, Tiffany and {He}, Chao and {Pesciotta}, Cara and {H{\"o}rst}, Sarah M. and {Haenecour}, Pierre and {Barman}, Travis and {Reddy}, Vishnu and {Lewis}, Nikole K. and {Vuitton}, V{\'e}ronique},
        title = "{Effects of Ultraviolet Radiation on Sub-Neptune Exoplanet Hazes through Laboratory Experiments}",
      journal = {\psj},
         year = 2025,
        month = jun,
       volume = {6},
       number = {6},
          eid = {145},
        pages = {145},
          doi = {10.3847/PSJ/adda4a},
archivePrefix = {arXiv},
       eprint = {2505.13692},
 primaryClass = {astro-ph.EP},
       adsurl = {https://ui.adsabs.harvard.edu/abs/2025PSJ.....6..145H}
}

@article{powellTransitSignaturesInhomogeneous2019,
    title = {Transit {Signatures} of {Inhomogeneous} {Clouds} on {Hot} {Jupiters}: {Insights} {From} {Microphysical} {Cloud} {Modeling}},
    volume = {887},
    issn = {0004-637X, 1538-4357},
    shorttitle = {Transit {Signatures} of {Inhomogeneous} {Clouds} on {Hot} {Jupiters}},
    url = {http://arxiv.org/abs/1910.07527},
    doi = {10.3847/1538-4357/ab55d9},
    language = {en},
    number = {2},
    urldate = {2025-04-11},
    journal = {The Astrophysical Journal},
    author = {Powell, Diana and Louden, Tom and Kreidberg, Laura and Zhang, Xi and Gao, Peter and Parmentier, Vivien},
    month = dec,
    year = {2019},
    note = {arXiv:1910.07527 [astro-ph]},
    pages = {170},
}

@ARTICLE{poseidon_joss,
       author = {{MacDonald}, Ryan J.},
        title = "{POSEIDON: A Multidimensional Atmospheric Retrieval Code for Exoplanet Spectra}",
      journal = {The Journal of Open Source Software},
         year = 2023,
        month = jan,
       volume = {8},
          eid = {4873},
        pages = {4873},
          doi = {10.21105/joss.04873},
archivePrefix = {arXiv},
       eprint = {2410.18181},
 primaryClass = {astro-ph.IM},
       adsurl = {https://ui.adsabs.harvard.edu/abs/2023JOSS....8.4873M}
}

@BOOK{HWO_pathways,
       author = {{National Academies of Sciences, Engineering, and Medicine}},
        title = "{Pathways to Discovery in Astronomy and Astrophysics for the 2020s}",
         year = 2021,
          doi = {10.17226/26141},
       adsurl = {https://ui.adsabs.harvard.edu/abs/2021pdaa.book.....N}
}

@ARTICLE{karyu_haze_cosmic_dust,
       author = {{Karyu}, Hiroki and {Kuroda}, Takeshi and {M{\"a}{\"a}tt{\"a}nen}, Anni and {Mahieux}, Arnaud and {Viscardy}, S{\'e}bastien and {Terada}, Naoki and {Robert}, S{\'e}verine and {Vandaele}, Ann Carine and {Crucifix}, Michel},
        title = "{A cosmic origin of Venus' lower haze}",
      journal = {Nature Astronomy},
         year = 2026,
        month = apr,
          doi = {10.1038/s41550-026-02843-4},
       adsurl = {https://ui.adsabs.harvard.edu/abs/2026NatAs.tmp...80K}
}

@ARTICLE{wallack_toi836c,
       author = {{Wallack}, Nicole L. and {Batalha}, Natasha E. and {Alderson}, Lili and {Scarsdale}, Nicholas and {Adams Redai}, Jea I. and {Aguichine}, Artyom and {Alam}, Munazza K. and {Gao}, Peter and {Wolfgang}, Angie and {Batalha}, Natalie M. and {Kirk}, James and {L{\'o}pez-Morales}, Mercedes and {Moran}, Sarah E. and {Teske}, Johanna and {Wakeford}, Hannah R. and {Wogan}, Nicholas F.},
        title = "{JWST COMPASS: A NIRSpec/G395H Transmission Spectrum of the Sub-Neptune TOI-836c}",
      journal = {\aj},
         year = 2024,
        month = aug,
       volume = {168},
       number = {2},
          eid = {77},
        pages = {77},
          doi = {10.3847/1538-3881/ad3917},
archivePrefix = {arXiv},
       eprint = {2404.01264},
 primaryClass = {astro-ph.EP},
       adsurl = {https://ui.adsabs.harvard.edu/abs/2024AJ....168...77W}
}

@ARTICLE{wallack_hd15337,
       author = {{Wallack}, Nicole L. and {Gao}, Peter and {Greklek-McKeon}, Michael and {Meech}, Annabella and {Aguichine}, Artyom and {Alam}, Munazza K. and {Alderson}, Lili and {Batalha}, Natasha E. and {Batalha}, Natalie M. and {Gagnebin}, Anna and {Gordon}, Tyler A. and {Kirk}, James and {L{\'o}pez-Morales}, Mercedes and {Moran}, Sarah E. and {Redai}, Jea Iyanla and {Scarsdale}, Nicholas and {Teske}, Johanna and {Wakeford}, Hannah R. and {Wogan}, Nicholas F. and {Wolfgang}, Angie},
        title = "{JWST COMPASS: NIRSpec/G395H Transmission Observations of the Sub-Neptune HD 15337 c}",
      journal = {\aj},
         year = 2026,
        month = mar,
       volume = {171},
       number = {3},
          eid = {180},
        pages = {180},
          doi = {10.3847/1538-3881/ae2d12},
archivePrefix = {arXiv},
       eprint = {2602.22327},
 primaryClass = {astro-ph.EP},
       adsurl = {https://ui.adsabs.harvard.edu/abs/2026AJ....171..180W}
}

@article{jenson_charlson_nucleation_scavenging,
author = {JENSEN, JØRGEN B. and CHARLSON, ROBERT J.},
title = {On the efficiency of nucleation scavenging},
journal = {Tellus B},
volume = {36B},
number = {5},
pages = {367-375},
doi = {https://doi.org/10.1111/j.1600-0889.1984.tb00255.x},
url = {https://onlinelibrary.wiley.com/doi/abs/10.1111/j.1600-0889.1984.tb00255.x},
eprint = {https://onlinelibrary.wiley.com/doi/pdf/10.1111/j.1600-0889.1984.tb00255.x},
year = {1984}
}

@article{svenningson_nucleation_scavenging,
title = {Cloud droplet nucleation scavenging in relation to the size and hygroscopic behaviour of aerosol particles},
journal = {Atmospheric Environment},
volume = {31},
number = {16},
pages = {2463-2475},
year = {1997},
note = {The Great Dun Fell Cloud Experiment 1993, Eurotrac sub-project Ground-based Cloud Experiment (GCE)},
issn = {1352-2310},
doi = {https://doi.org/10.1016/S1352-2310(96)00179-3},
url = {https://www.sciencedirect.com/science/article/pii/S1352231096001793},
author = {Birgitta Svenningsson and Hans-Christen Hansson and Bengt Martinsson and Alfred Wiedensohler and Erik Swietlicki and Sven-Inge Cederfelt and Manfred Wendisch and Keith N. Bower and Tom W. Choularton and Roy N. Colvile}
}

@ARTICLE{lavvas_huygens,
       author = {{Lavvas}, P. and {Yelle}, R.~V. and {Griffith}, C.~A.},
        title = "{Titan{\textquoteright}s vertical aerosol structure at the Huygens landing site: Constraints on particle size, density, charge, and refractive index}",
      journal = {\icarus},
         year = 2010,
        month = dec,
       volume = {210},
       number = {2},
        pages = {832-842},
          doi = {10.1016/j.icarus.2010.07.025},
       adsurl = {https://ui.adsabs.harvard.edu/abs/2010Icar..210..832L}
}

@ARTICLE{cloutier_gj1214,
       author = {{Cloutier}, Ryan and {Charbonneau}, David and {Deming}, Drake and {Bonfils}, Xavier and {Astudillo-Defru}, Nicola},
        title = "{A More Precise Mass for GJ 1214 b and the Frequency of Multiplanet Systems Around Mid-M Dwarfs}",
      journal = {\aj},
         year = 2021,
        month = nov,
       volume = {162},
       number = {5},
          eid = {174},
        pages = {174},
          doi = {10.3847/1538-3881/ac1584},
archivePrefix = {arXiv},
       eprint = {2107.14732},
 primaryClass = {astro-ph.EP},
       adsurl = {https://ui.adsabs.harvard.edu/abs/2021AJ....162..174C}
}

@ARTICLE{goumans_stardust_silicates,
       author = {{Goumans}, T.~P.~M. and {Bromley}, Stefan T.},
        title = "{Efficient nucleation of stardust silicates via heteromolecular homogeneous condensation}",
      journal = {\mnras},
         year = 2012,
        month = mar,
       volume = {420},
       number = {4},
        pages = {3344-3349},
          doi = {10.1111/j.1365-2966.2011.20255.x},
       adsurl = {https://ui.adsabs.harvard.edu/abs/2012MNRAS.420.3344G}
}

@article{ahrer_escaping_2025,
	title = {Escaping {Helium} and a {Highly} {Muted} {Spectrum} {Suggest} a {Metal}-enriched {Atmosphere} on {Sub}-{Neptune} {GJ} 3090 b from {JWST} {Transit} {Spectroscopy}},
	volume = {985},
	issn = {0004-637X},
	url = {https://ui.adsabs.harvard.edu/abs/2025ApJ...985L..10A},
	doi = {10.3847/2041-8213/add010},
	urldate = {2025-10-05},
	journal = {The Astrophysical Journal},
	author = {Ahrer, Eva-Maria and Radica, Michael and Piaulet-Ghorayeb, Caroline and Raul, Eshan and Wiser, Lindsey and Welbanks, Luis and Acuña, Lorena and Allart, Romain and Coulombe, Louis-Philippe and Louca, Amy and MacDonald, Ryan and Saidel, Morgan and Evans-Soma, Thomas M. and Benneke, Björn and Christie, Duncan and Beatty, Thomas G. and Cadieux, Charles and Cloutier, Ryan and Doyon, René and Fortney, Jonathan J. and Gagnebin, Anna and Gapp, Cyril and Innes, Hamish and Knutson, Heather A. and Komacek, Thaddeus and Krissansen-Totton, Joshua and Miguel, Yamila and Pierrehumbert, Raymond and Roy, Pierre-Alexis and Schlichting, Hilke E.},
	month = may,
	year = {2025},
	note = {Publisher: IOP
ADS Bibcode: 2025ApJ...985L..10A},
	pages = {L10},
}

@ARTICLE{powell_depletion_of_co,
       author = {{Powell}, Diana and {Gao}, Peter and {Murray-Clay}, Ruth and {Zhang}, Xi},
        title = "{Depletion of gaseous CO in protoplanetary disks by surface-energy-regulated ice formation}",
      journal = {Nature Astronomy},
         year = 2022,
        month = aug,
       volume = {6},
        pages = {1147-1155},
          doi = {10.1038/s41550-022-01741-9},
archivePrefix = {arXiv},
       eprint = {2208.13806},
 primaryClass = {astro-ph.EP},
       adsurl = {https://ui.adsabs.harvard.edu/abs/2022NatAs...6.1147P}
}

@ARTICLE{ros_nucleation_disks,
       author = {{Ros}, Katrin and {Johansen}, Anders and {Riipinen}, Ilona and {Schlesinger}, Daniel},
        title = "{Effect of nucleation on icy pebble growth in protoplanetary discs}",
      journal = {\aap},
         year = 2019,
        month = sep,
       volume = {629},
          eid = {A65},
        pages = {A65},
          doi = {10.1051/0004-6361/201834331},
archivePrefix = {arXiv},
       eprint = {1907.08471},
 primaryClass = {astro-ph.EP},
       adsurl = {https://ui.adsabs.harvard.edu/abs/2019A&A...629A..65R}
}

@ARTICLE{seki_hasegawa,
       author = {{Seki}, J. and {Hasegawa}, H.},
        title = "{The Heterogeneous Condensation of Interstellar Ice Grains}",
      journal = {\apss},
         year = 1983,
        month = jul,
       volume = {94},
       number = {1},
        pages = {177-189},
          doi = {10.1007/BF00651770},
       adsurl = {https://ui.adsabs.harvard.edu/abs/1983Ap&SS..94..177S}
}

@ARTICLE{bernatowicz_grains_within_grains,
       author = {{Bernatowicz}, Thomas J. and {Amari}, Sachiko and {Zinner}, Ernst K. and {Lewis}, Roy S.},
        title = "{Interstellar Grains within Interstellar Grains}",
      journal = {\apjl},
         year = 1991,
        month = jun,
       volume = {373},
        pages = {L73},
          doi = {10.1086/186054},
       adsurl = {https://ui.adsabs.harvard.edu/abs/1991ApJ...373L..73B}
}

@ARTICLE{tielens_dust_review,
       author = {{Tielens}, A.~G.~G.~M.},
        title = "{Dust Formation in Astrophysical Environments: The Importance of Kinetics}",
      journal = {Frontiers in Astronomy and Space Sciences},
         year = 2022,
        month = may,
       volume = {9},
          eid = {908217},
        pages = {908217},
          doi = {10.3389/fspas.2022.908217},
archivePrefix = {arXiv},
       eprint = {2206.01548},
 primaryClass = {astro-ph.EP},
       adsurl = {https://ui.adsabs.harvard.edu/abs/2022FrASS...9.8217T}
}

@ARTICLE{astropy2022,
       author = {{Astropy Collaboration} and {Price-Whelan}, Adrian M. and {Lim}, Pey Lian and {Earl}, Nicholas and {Starkman}, Nathaniel and {Bradley}, Larry and {Shupe}, David L. and {Patil}, Aarya A. and {Corrales}, Lia and {Brasseur}, C.~E. and {N{\"o}the}, Maximilian and {Donath}, Axel and {Tollerud}, Erik and {Morris}, Brett M. and {Ginsburg}, Adam and {Vaher}, Eero and {Weaver}, Benjamin A. and {Tocknell}, James and {Jamieson}, William and {van Kerkwijk}, Marten H. and {Robitaille}, Thomas P. and {Merry}, Bruce and {Bachetti}, Matteo and {G{\"u}nther}, H. Moritz and {Aldcroft}, Thomas L. and {Alvarado-Montes}, Jaime A. and {Archibald}, Anne M. and {B{\'o}di}, Attila and {Bapat}, Shreyas and {Barentsen}, Geert and {Baz{\'a}n}, Juanjo and {Biswas}, Manish and {Boquien}, M{\'e}d{\'e}ric and {Burke}, D.~J. and {Cara}, Daria and {Cara}, Mihai and {Conroy}, Kyle E. and {Conseil}, Simon and {Craig}, Matthew W. and {Cross}, Robert M. and {Cruz}, Kelle L. and {D'Eugenio}, Francesco and {Dencheva}, Nadia and {Devillepoix}, Hadrien A.~R. and {Dietrich}, J{\"o}rg P. and {Eigenbrot}, Arthur Davis and {Erben}, Thomas and {Ferreira}, Leonardo and {Foreman-Mackey}, Daniel and {Fox}, Ryan and {Freij}, Nabil and {Garg}, Suyog and {Geda}, Robel and {Glattly}, Lauren and {Gondhalekar}, Yash and {Gordon}, Karl D. and {Grant}, David and {Greenfield}, Perry and {Groener}, Austen M. and {Guest}, Steve and {Gurovich}, Sebastian and {Handberg}, Rasmus and {Hart}, Akeem and {Hatfield-Dodds}, Zac and {Homeier}, Derek and {Hosseinzadeh}, Griffin and {Jenness}, Tim and {Jones}, Craig K. and {Joseph}, Prajwel and {Kalmbach}, J. Bryce and {Karamehmetoglu}, Emir and {Ka{\l}uszy{\'n}ski}, Miko{\l}aj and {Kelley}, Michael S.~P. and {Kern}, Nicholas and {Kerzendorf}, Wolfgang E. and {Koch}, Eric W. and {Kulumani}, Shankar and {Lee}, Antony and {Ly}, Chun and {Ma}, Zhiyuan and {MacBride}, Conor and {Maljaars}, Jakob M. and {Muna}, Demitri and {Murphy}, N.~A. and {Norman}, Henrik and {O'Steen}, Richard and {Oman}, Kyle A. and {Pacifici}, Camilla and {Pascual}, Sergio and {Pascual-Granado}, J. and {Patil}, Rohit R. and {Perren}, Gabriel I. and {Pickering}, Timothy E. and {Rastogi}, Tanuj and {Roulston}, Benjamin R. and {Ryan}, Daniel F. and {Rykoff}, Eli S. and {Sabater}, Jose and {Sakurikar}, Parikshit and {Salgado}, Jes{\'u}s and {Sanghi}, Aniket and {Saunders}, Nicholas and {Savchenko}, Volodymyr and {Schwardt}, Ludwig and {Seifert-Eckert}, Michael and {Shih}, Albert Y. and {Jain}, Anany Shrey and {Shukla}, Gyanendra and {Sick}, Jonathan and {Simpson}, Chris and {Singanamalla}, Sudheesh and {Singer}, Leo P. and {Singhal}, Jaladh and {Sinha}, Manodeep and {Sip{\H{o}}cz}, Brigitta M. and {Spitler}, Lee R. and {Stansby}, David and {Streicher}, Ole and {{\v{S}}umak}, Jani and {Swinbank}, John D. and {Taranu}, Dan S. and {Tewary}, Nikita and {Tremblay}, Grant R. and {de Val-Borro}, Miguel and {Van Kooten}, Samuel J. and {Vasovi{\'c}}, Zlatan and {Verma}, Shresth and {de Miranda Cardoso}, Jos{\'e} Vin{\'\i}cius and {Williams}, Peter K.~G. and {Wilson}, Tom J. and {Winkel}, Benjamin and {Wood-Vasey}, W.~M. and {Xue}, Rui and {Yoachim}, Peter and {Zhang}, Chen and {Zonca}, Andrea and {Astropy Project Contributors}},
        title = "{The Astropy Project: Sustaining and Growing a Community-oriented Open-source Project and the Latest Major Release (v5.0) of the Core Package}",
      journal = {\apj},
         year = 2022,
        month = aug,
       volume = {935},
       number = {2},
          eid = {167},
        pages = {167},
          doi = {10.3847/1538-4357/ac7c74},
archivePrefix = {arXiv},
       eprint = {2206.14220},
 primaryClass = {astro-ph.IM},
       adsurl = {https://ui.adsabs.harvard.edu/abs/2022ApJ...935..167A}
}

@ARTICLE{astropy2013,
       author = {{Astropy Collaboration} and {Robitaille}, Thomas P. and {Tollerud}, Erik J. and {Greenfield}, Perry and {Droettboom}, Michael and {Bray}, Erik and {Aldcroft}, Tom and {Davis}, Matt and {Ginsburg}, Adam and {Price-Whelan}, Adrian M. and {Kerzendorf}, Wolfgang E. and {Conley}, Alexander and {Crighton}, Neil and {Barbary}, Kyle and {Muna}, Demitri and {Ferguson}, Henry and {Grollier}, Fr{\'e}d{\'e}ric and {Parikh}, Madhura M. and {Nair}, Prasanth H. and {Unther}, Hans M. and {Deil}, Christoph and {Woillez}, Julien and {Conseil}, Simon and {Kramer}, Roban and {Turner}, James E.~H. and {Singer}, Leo and {Fox}, Ryan and {Weaver}, Benjamin A. and {Zabalza}, Victor and {Edwards}, Zachary I. and {Azalee Bostroem}, K. and {Burke}, D.~J. and {Casey}, Andrew R. and {Crawford}, Steven M. and {Dencheva}, Nadia and {Ely}, Justin and {Jenness}, Tim and {Labrie}, Kathleen and {Lim}, Pey Lian and {Pierfederici}, Francesco and {Pontzen}, Andrew and {Ptak}, Andy and {Refsdal}, Brian and {Servillat}, Mathieu and {Streicher}, Ole},
        title = "{Astropy: A community Python package for astronomy}",
      journal = {\aap},
         year = 2013,
        month = oct,
       volume = {558},
          eid = {A33},
        pages = {A33},
          doi = {10.1051/0004-6361/201322068},
archivePrefix = {arXiv},
       eprint = {1307.6212},
 primaryClass = {astro-ph.IM},
       adsurl = {https://ui.adsabs.harvard.edu/abs/2013A&A...558A..33A}
}

@ARTICLE{numpy2020, 
       author = {{Harris}, Charles R. and {Millman}, K. Jarrod and {van der Walt}, St{\'e}fan J. and {Gommers}, Ralf and {Virtanen}, Pauli and {Cournapeau}, David and {Wieser}, Eric and {Taylor}, Julian and {Berg}, Sebastian and {Smith}, Nathaniel J. and {Kern}, Robert and {Picus}, Matti and {Hoyer}, Stephan and {van Kerkwijk}, Marten H. and {Brett}, Matthew and {Haldane}, Allan and {del R{\'\i}o}, Jaime Fern{\'a}ndez and {Wiebe}, Mark and {Peterson}, Pearu and {G{\'e}rard-Marchant}, Pierre and {Sheppard}, Kevin and {Reddy}, Tyler and {Weckesser}, Warren and {Abbasi}, Hameer and {Gohlke}, Christoph and {Oliphant}, Travis E.},
        title = "{Array programming with NumPy}",
      journal = {\nat},
         year = 2020,
        month = sep,
       volume = {585},
       number = {7825},
        pages = {357-362},
          doi = {10.1038/s41586-020-2649-2},
archivePrefix = {arXiv},
       eprint = {2006.10256},
 primaryClass = {cs.MS},
       adsurl = {https://ui.adsabs.harvard.edu/abs/2020Natur.585..357H}
}

@ARTICLE{scipy2020,
       author = {{Virtanen}, Pauli and {Gommers}, Ralf and {Oliphant}, Travis E. and {Haberland}, Matt and {Reddy}, Tyler and {Cournapeau}, David and {Burovski}, Evgeni and {Peterson}, Pearu and {Weckesser}, Warren and {Bright}, Jonathan and {van der Walt}, St{\'e}fan J. and {Brett}, Matthew and {Wilson}, Joshua and {Millman}, K. Jarrod and {Mayorov}, Nikolay and {Nelson}, Andrew R.~J. and {Jones}, Eric and {Kern}, Robert and {Larson}, Eric and {Carey}, C.~J. and {Polat}, {\.I}lhan and {Feng}, Yu and {Moore}, Eric W. and {VanderPlas}, Jake and {Laxalde}, Denis and {Perktold}, Josef and {Cimrman}, Robert and {Henriksen}, Ian and {Quintero}, E.~A. and {Harris}, Charles R. and {Archibald}, Anne M. and {Ribeiro}, Ant{\^o}nio H. and {Pedregosa}, Fabian and {van Mulbregt}, Paul and {SciPy 1.  0 Contributors}},
        title = "{SciPy 1.0: fundamental algorithms for scientific computing in Python}",
      journal = {Nature Medicine},
         year = 2020,
        month = feb,
       volume = {17},
        pages = {261-272},
          doi = {10.1038/s41592-019-0686-2},
archivePrefix = {arXiv},
       eprint = {1907.10121},
 primaryClass = {cs.MS},
       adsurl = {https://ui.adsabs.harvard.edu/abs/2020NatMe..17..261V}
}

@ARTICLE{mang_picaso4,
       author = {{Mang}, James and {Batalha}, Natasha E. and {Morley}, Caroline V. and {Wogan}, Nicholas F. and {Mukherjee}, Sagnick and {Visscher}, Channon and {Marley}, Mark S. and {Fortney}, Jonathan J. and {Chubb}, Katy L. and {Gao}, Peter and {Malsky}, Isaac},
        title = "{PICASO 4.0: Clouds and Photochemistry in Climate Models of Brown Dwarfs and Exoplanets}",
      journal = {\apj},
         year = 2026,
        month = mar,
       volume = {1000},
       number = {1},
          eid = {98},
        pages = {98},
          doi = {10.3847/1538-4357/ae47ff},
archivePrefix = {arXiv},
       eprint = {2602.22468},
 primaryClass = {astro-ph.EP},
       adsurl = {https://ui.adsabs.harvard.edu/abs/2026ApJ..1000...98M}
}
\bibliographystyle{aasjournalv7}



\end{document}